\documentclass[english]{article}
\pdfoutput=1

\usepackage[T1]{fontenc}
\usepackage[latin9]{inputenc}
\usepackage{geometry}
\usepackage{amsmath}
\usepackage{amssymb}
\usepackage{graphicx}
\usepackage{esint}

\makeatletter
\numberwithin{equation}{section}
\numberwithin{figure}{section}

\@ifundefined{date}{}{\date{}}
\usepackage{url}
\usepackage{cite}

\makeatother

\usepackage{babel}

\newcommand{\etalchar}[1]{$^{#1}$}

\begin{document}
\global\long\def\SLE{\mathrm{SLE}}
 \global\long\def\SLEk{\mathrm{SLE}_{\kappa}}
 \global\long\def\SLEkappa#1{\mathrm{SLE}_{#1}}
 \global\long\def\SLEkapparho#1#2{\mathrm{SLE}_{#1}(#2)}

\global\long\def\PR{\mathsf{P}}
 \global\long\def\EX{\mathsf{E}}

\global\long\def\bR{\mathbb{R}}
 \global\long\def\bZ{\mathbb{Z}}
 \global\long\def\bN{\mathbb{N}}
 \global\long\def\bQ{\mathbb{Q}}

\global\long\def\bC{\mathbb{C}}
 \global\long\def\Rsphere{\overline{\bC}}
 \global\long\def\re{\Re\mathfrak{e}}
 \global\long\def\im{\Im\mathfrak{m}}
 \global\long\def\arg{\mathrm{arg}}
 \global\long\def\ii{\mathfrak{i}}

\global\long\def\bD{\mathbb{D}}
 \global\long\def\bH{\mathbb{H}}

\global\long\def\dist{\mathrm{dist}}
 \global\long\def\reg{\mathrm{reg}}

\global\long\def\eps{\varepsilon}
 \global\long\def\const{\mathrm{const.}}
 \global\long\def\half{\frac{1}{2}}

\global\long\def\domain{\Lambda}
\global\long\def\bdry{\partial}
 \global\long\def\cl#1{\overline{#1}}

\global\long\def\Ampl{\zeta}
 \global\long\def\Corr{\chi}
 \global\long\def\lft{-}
 \global\long\def\rgt{+}
 \global\long\def\rgtlft{\pm}

\global\long\def\ud{\mathrm{d}}
 \global\long\def\der#1{\frac{\ud}{\ud#1}}
 \global\long\def\pder#1{\frac{\partial}{\partial#1}}

\global\long\def\set#1{\left\{  #1\right\}  }
 \global\long\def\setcond#1#2{\left\{  #1\;\big|\;#2\right\}  }

\global\long\def\FWint#1{F_{#1}}
 \global\long\def\RealInt#1{I_{#1}}
 \global\long\def\OrdInt#1{R_{#1}}

\global\long\def\braid{\sigma}

\global\long\def\Uqsltwo{\mathcal{U}_{q}(\mathfrak{sl}_{2})}
 \global\long\def\Hcp{\Delta}

\global\long\def\qnum#1{\left[#1\right] }
 \global\long\def\qfact#1{\left[#1\right]! }
 \global\long\def\qbin#1#2{\left[\begin{array}{c}
 #1\\
#2 
\end{array}\right]}

\global\long\def\Hom{\mathrm{Hom}}
 \global\long\def\End{\mathrm{End}}
 \global\long\def\Aut{\mathrm{Aut}}
 \global\long\def\Rad{\mathrm{Rad}}
 \global\long\def\Ext{\mathrm{Ext}}

\global\long\def\Kern{\mathrm{Ker}}
 \global\long\def\Imag{\mathrm{Im}}

\global\long\def\dmn{\mathrm{dim}}
 \global\long\def\spn{\mathrm{span}}
 \global\long\def\tens{\otimes}
 \global\long\def\Mat{\mathrm{Mat}}
 \global\long\def\unitmat{\mathbb{I}}
 \global\long\def\id{\mathrm{id}}
 \global\long\def\isom{\cong}

\global\long\def\SymmGrp{\mathfrak{S}}

\global\long\def\bra{\langle}
 \global\long\def\ket{\rangle}
 \global\long\def\bravec#1{|#1\rangle}
 \global\long\def\ketvec#1{|#1\rangle}

\global\long\def\Wd{M}
 \global\long\def\Wbas{e}
 \global\long\def\Tbas{\tau}

\global\long\def\chamber{\mathfrak{X}}
 \global\long\def\Wchamber{\mathfrak{W}}
 \global\long\def\sR{\mathcal{R}}
 \global\long\def\FKcone{L\mathrm{-cone}}

\global\long\def\FWint#1{F_{#1}}
 \global\long\def\RealInt#1{I_{#1}}
 \global\long\def\OrdInt#1{R_{#1}}
 \global\long\def\anchor{z_{0}}

\global\long\def\FWintegrand{f^{\Supset}}

\title{\textbf{\huge{SLE boundary visits}}}

\author{Niko Jokela%
\footnote{\emph{Current address:} Department of Physics and Helsinki Institute of Physics, POB 64, 00014 University of Helsinki, Finland.}\\
{\normalsize{\url{niko.jokela@helsinki.fi}}}\\
{\normalsize{Departamento de Física de Partículas, Universidade de
Santiago de Compostela}}\\
{\normalsize{and}}\\
{\normalsize{Instituto Galego de Física de Altas Enerxías (IGFAE)}}\\
{\normalsize{E-15782, Santiago de Compostela, Spain}}\bigskip{}
\bigskip{}
\\
Matti Järvinen%
\footnote{\emph{Current address:} Laboratoire de Physique Théorique, École Normale Supérieure \& Institut de Physique Théorique Philippe Meyer, 24 rue Lhomond, 75231 Paris, France.}\\
{\normalsize{\url{jarvinen@lpt.ens.fr}}}\\
{\normalsize{Crete Center for Theoretical Physics, Department of Physics}}\\
{\normalsize{University of Crete, 71003 Heraklion, Greece}}\bigskip{}
\bigskip{}
\\
Kalle Kytölä%
\footnote{\emph{Current address:} Department of Mathematics and Systems Analysis,
POB 11100, 00076 Aalto University, Finland.}\\
{\normalsize{\url{kalle.kytola@aalto.fi}}}\\
{\normalsize{Department of Mathematics and Statistics}}\\
{\normalsize{P.O. Box 68, 00014 University of Helsinki, Finland}}}

\begingroup
\begin{flushright}
CCTP-2013-14
\par\end{flushright}
\let\newpage\relax%
\maketitle
\endgroup

\begin{abstract}
We study the probabilities with which chordal Schramm-Loewner Evolutions (SLE) 
visit small neighborhoods of boundary points. We find 
formulas for general chordal SLE boundary visiting probability amplitudes,
also known as SLE boundary zig-zags or order refined SLE multi-point
Green's functions on the boundary. Remarkably, an exact answer can
be found to this important SLE question for an arbitrarily large number
of marked points. The main technique employed is a spin chain~-~Coulomb
gas correspondence between tensor product representations of a quantum
group and functions given by Dotsenko-Fateev type integrals. We show
how to express these integral formulas in terms of regularized real
integrals, and we discuss their numerical evaluation.

The results are universal in the sense that apart from an overall
multiplicative constant the same formula gives the amplitude for many
different formulations of the SLE boundary visit problem. The formula
also applies to renormalized boundary visit probabilities for interfaces
in critical lattice models of statistical mechanics: we compare the
results with numerical simulations of percolation, loop-erased random
walk, and Fortuin-Kasteleyn random cluster models at $Q=2$ and $Q=3$,
and find good agreement.

\vfill{}

\end{abstract}

\newpage{}

\tableofcontents{}

\section{Introduction\label{sec: Introduction}}

\subsection{SLE curves}

\begin{figure}
\begin{centering}
\includegraphics[width=0.4\textwidth]{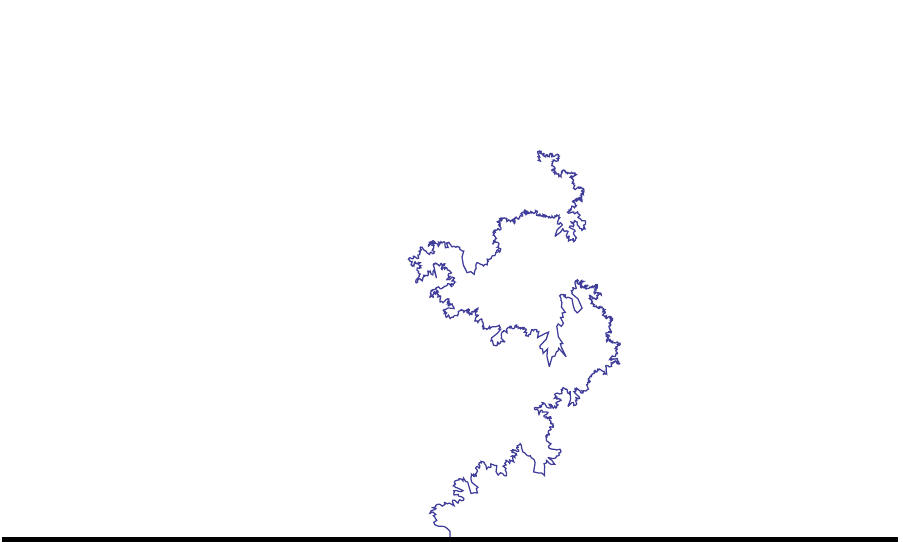}\quad{}\enskip{}\includegraphics[width=0.55\textwidth]{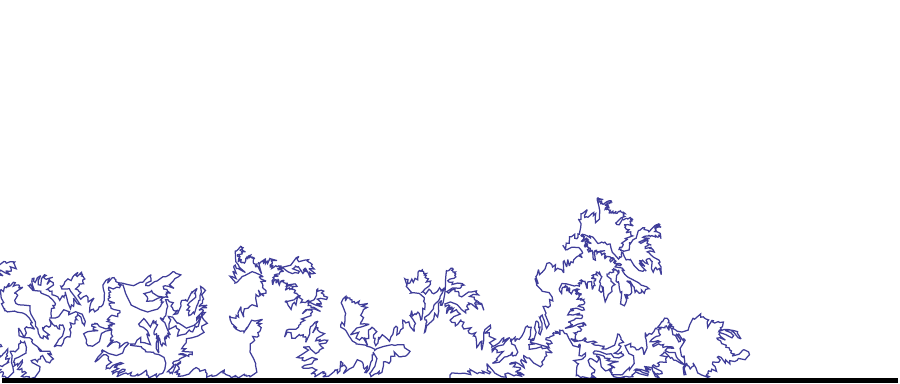}
\par\end{centering}

\caption{\emph{Chordal $\SLEk$ is a random fractal curve. For $\kappa\leq4$
the curve is simple and does not touch boundary, and for $4<\kappa<8$
the curve has double points and touches the boundary on a random Cantor
set. The two pictures show chordal $\SLEk$ in the upper half-plane
$\bH$ from $0$ to $\infty$ --- in the left picture $\kappa=3$,
and the right picture $\kappa=6$.\label{fig: SLEs}}}
\end{figure}

Schramm-Loewner evolutions (SLE) are conformally invariant random
fractal curves in the plane, whose most important characteristics
are determined by one parameter $\kappa>0$. They were introduced
by Oded Schramm \cite{Schramm-LERW_and_UST} as the only plausible
candidates for the scaling limits of random interfaces in statistical
mechanics models that are expected to display conformal invariance,
with different models corresponding to different values of the parameter
$\kappa$.%
\footnote{Figure \ref{fig: SLEs} shows two SLE curves. Examples of interfaces
in lattice models are shown in Figures \ref{fig: LERW}, \ref{fig: perco},
and \ref{fig: FK model}, on pages \pageref{fig: LERW}, \pageref{fig: perco},
and \pageref{fig: FK model}, respectively.%
} Proofs that interfaces in various critical lattice models do converge
to SLEs in the scaling limit have been obtained for example in \cite{Smirnov-critical_percolation,LSW-LERW_and_UST,SS-harmonic_explorer,Smirnov-towards_conformal_invariance,CN-critical_percolation_exploration_path,Zhan-scaling_limits_of_planar_LERW,Smirnov-conformal_invariance_in_RCM_1,HK-Ising_interfaces_and_free_boundary_conditions,Izyurov-critical_Ising_interfaces_in_multiply_connected_domains,CDHKS-convergence_of_Ising_interfaces_to_SLE}.

The fundamental example of SLEs is the chordal $\SLEkappa{\kappa}$
\cite{LSW-intersection_exponents_1,RS-basic_properties}. For a given
simply connected domain $\domain\subset\bC$ with two marked boundary
points $a,b\in\bdry\domain$, the chordal $\SLEkappa{\kappa}$ in
$\domain$ from $a$ to $b$ is an oriented but unparametrized random
curve $\gamma$ in the closure of $\domain$ starting from $a$ and
ending at $b$. Its two characterizing properties are conformal invariance
and domain Markov property:
\begin{itemize}
\item Conformal invariance states that the image of a chordal SLE under
a conformal map is a chordal SLE in the image domain.
\item Domain Markov property states that given an initial segment of a chordal
SLE, the conditional law of the continuation is a chordal SLE in the
remaining subdomain.
\end{itemize}
\noindent 

Some features of SLEs vary continuously in $\kappa$, notably the
Hausdorff dimension of the fractal curve is given by $d_{{\rm H}}(\gamma)=1+\frac{\kappa}{8}$
for $0<\kappa\leq8$ \cite{Beffara-dimension_of_the_SLE_curves}.
On the other hand, some qualitative properties of SLEs show abrupt
phase transitions with respect to the parameter $\kappa$. For the
present purposes, it is important to distinguish the following three
phases \cite{RS-basic_properties}:
\begin{description}
\item [{$0<\kappa\leq4$:}] The chordal $\SLEk$ is a simple curve, i.e.,
the curve does not have double points, see Figure~\ref{fig: SLEs}~(left).
The curve does not touch the boundary $\bdry\domain$ of the domain
except at the starting point $a$ and the end point $b$. The curve
avoids any given point $z\in\domain$ of the domain with probability
one.
\item [{$4<\kappa<8$:}] The chordal $\SLEk$ is a non self-traversing
curve with double points, see Figure~\ref{fig: SLEs}~(right). The
intersection of the curve with the boundary $\bdry\domain$ of the
domain is a random Cantor set. The curve still avoids any given point
$z\in\overline{\domain}\setminus\set{a,b}$ of the domain or of its
boundary with probability one.
\item [{$8\leq\kappa$:}] The chordal $\SLEk$ is a space-filling curve;
any point $z\in\domain$ of the domain is on the curve.
\end{description}
The behavior in the case $\kappa\geq8$ is somewhat pathological.
No interfaces in statistical mechanics models are expected to correspond
to $\kappa>8$.%
\footnote{In the borderline case $\kappa=8$, the (space-filling) chordal $\SLE_{8}$
curve is the scaling limit of the Peano curve of the uniform spanning
tree \cite{LSW-LERW_and_UST}.%
} In this article we restrict our attention to the cases $0<\kappa<8$.

\subsection{Chordal SLE boundary visits}

The main goal in this article is to find formulas for the probabilities with which the
chordal SLE visits small neighborhoods of given boundary points. Partial
answers to similar questions have been obtained
in \cite{BB-zig_zag,SZ-boundary_proximity_of_SLE,AS-Hausdorff_dimension_of_SLE_real_line,
AS-covariant_measure_of_SLE_on_boundary,Lawler-Minkowski_SLE_real_line}.

It is easiest to illustrate the question in the upper half-plane
\begin{align*}
\bH=\; & \setcond{z\in\bC}{\text{\ensuremath{\im}}(z)>0},
\end{align*}
with the chordal $\SLEk$ curve $\gamma$ starting from the origin
and ending at infinity.
We will briefly recall the precise definition of chordal $\SLEk$ in $\bH$ from $0$ to $\infty$
in Section~\ref{sub: def chordal SLE}, and we refer the reader to \cite{RS-basic_properties}
for more thorough background.

\begin{figure}
\begin{centering}
\includegraphics[width=0.95\textwidth]{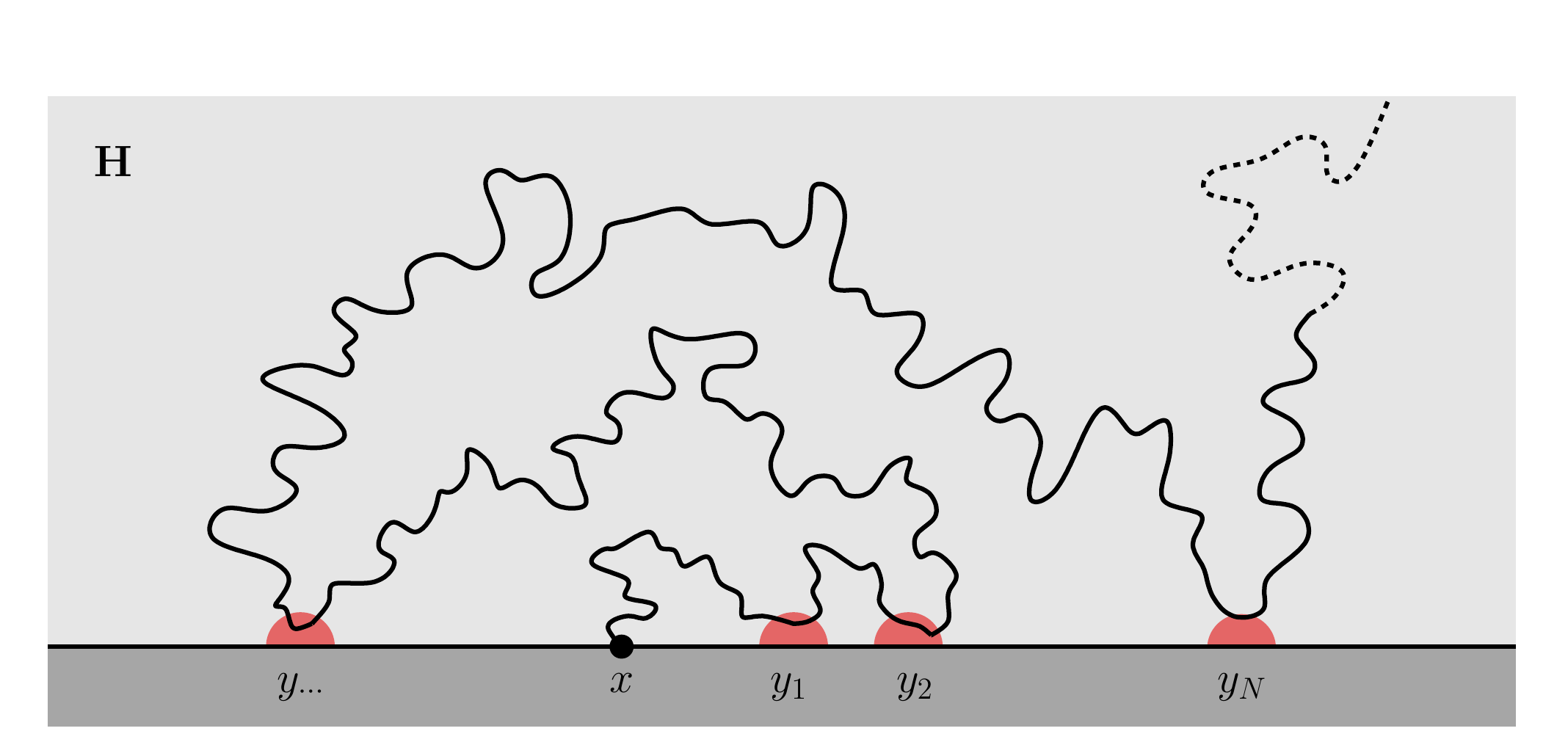}
\par\end{centering}

\caption{\emph{A schematic illustration of the boundary zig-zag studied in
this article: the chordal $\SLEk$ curve in the upper half-plane $\bH$
starts from $x$ and visits small neighborhoods of boundary points
$y_{1},y_{2},\ldots,y_{N}$.\label{fig: SLE boundary zig-zag}}}
\end{figure}

Denote the half-disk of radius $\eps>0$ centered
at a boundary point $y\in\bR\subset\bdry\bH$ by 
\begin{align*}
B_{\eps}(y)=\; & \setcond{z\in\bH}{|z-y|<\eps}.
\end{align*}
Given points $y_{1},y_{2},\ldots,y_{N}\in\bR$ and radii $\eps_{1},\eps_{2},\ldots,\eps_{N}>0$,
the probability that the curve $\gamma$ visits all of $B_{\eps_{j}}(y_{j})$,
$j=1,2,\ldots,N$, tends to zero as a power law as the radius
$\eps_{j}$ is taken small. 
More precisely, the scaling exponent of the power law is
\begin{align}
h=\; & \frac{8-\kappa}{\kappa}\label{eq: h13}
\end{align}
(see Appendix \ref{sec: derivations of the value of h}), and we are
interested in the limit%
\footnote{The existence of the limit has been proved in \cite{Lawler-Minkowski_SLE_real_line}.}
\begin{align}
C_{(\bH;0,\infty)}^{(N)}(y_{1},y_{2},\ldots,y_{N})=\; & \lim_{\eps_{1},\ldots,\eps_{N}\searrow0}\;\frac{1}{\eps_{1}^{h}\cdots\eps_{N}^{h}}\,\PR\left[\gamma\cap B_{\eps_{j}}(y_{j})\neq\emptyset\quad\text{for }j=1,2,\ldots,N\right]\label{eq: full correlation with balls}\\
\nonumber 
\end{align}
of probabilities of events illustrated schematically in Figure \ref{fig: SLE boundary zig-zag}.
In the spirit of \cite{Lawler-fractal_and_multifractal_properties_of_SLE,
LS-natural_parametrization_of_SLE,AKL-the_Greens_function_for_radial_SLE,
LW-multi_point_Greens_functions_for_SLE,LZ-SLE_curves_and_natural_parametrization,
LR-Minkowski_content_and_natural_parametrization_for_SLE},
it is appropriate to call the limit \eqref{eq: full correlation with balls}
an SLE boundary Green's function. We emphasize that one could in principle choose
to define a boundary visit of SLE differently, for example for $\kappa>4$ one could
ask the curve $\gamma$ to touch a boundary segment of length $\eps$, or one could choose
the neighborhood shape to be something other than a half-disk.
Yet, independently of the precise formulation, the limit remains universal apart from
a multiplicative constant which depends on the details of the chosen formulation.%
\footnote{Compare also with the proof \cite{LR-Minkowski_content_and_natural_parametrization_for_SLE}
that the SLE Green's function defined using conformal radius differs by a multiplicative
constant (whose explicit value is not known) from the SLE Green's function defined using Euclidean distance.}
Different formulations and universality will be discussed in
Section~\ref{sec: interpretations and applications}.

\noindent 

Recalling that $\gamma$ is an oriented curve, we may even specify
the order of the boundary visits, i.e., require that the curve $\gamma$
first reaches the chosen small neighborhood of $y_{1}$, then the
neighborhood of $y_{2}$ and so on until reaching the neighborhood of $y_{N}$.
The order refinement of the SLE boundary Green's function is the limit
\begin{align}
P_{(\bH;0,\infty)}^{(N)}(y_{1},y_{2},\ldots,y_{N})=\; & \lim_{\eps_{1},\ldots,\eps_{N}\searrow0}\;\frac{1}{\eps_{1}^{h}\cdots\eps_{N}^{h}}\,\PR\left[\tau_{y_{1};\eps_{1}}<\tau_{y_{2};\eps_{2}}<\cdots<\tau_{y_{N};\eps_{N}}<\infty\right],\label{eq: zig-zag proba with balls}
\end{align}
where any increasing parametrization $t\mapsto\gamma_{t}$ of the
curve $\gamma$ is chosen, and we denote by 
\begin{align}
\tau_{y_{j};\eps_{j}}=\; & \inf\setcond{t\geq0}{\gamma_{t}\in B_{\eps_{j}}(y_{j})}\label{eq: definition of visit time with balls}
\end{align}
the stopping time at which the curve $\gamma$ first reaches the $\eps_{j}$-neighborhood
of $y_{j}$. Obviously one can recover the complete correlation function
$C_{(\bH;0,\infty)}^{(N)}$ from the ordered ones $P_{(\bH;0,\infty)}^{(N)}$
by summing over all possible orders of visits%
\footnote{In fact in the sum we only need those permutations which respect the
order of positive $y_{j}$'s and reverse the order of negative $y_{j}$'s,
otherwise the curve essentially disconnects its future passage to
a point that it would need to visit later. This will be discussed in
some more detail in Section~\ref{sec: detailed asymptotics}. %
}
\begin{align*}
C_{(\bH;0,\infty)}^{(N)}(y_{1},y_{2},\ldots,y_{N})=\; & \sum_{\sigma\in \SymmGrp_{N}}P_{(\bH;0,\infty)}^{(N)}(y_{\sigma(1)},y_{\sigma(2)},\ldots,y_{\sigma(N)}).
\end{align*}
In the general form with the order of visits specified, the question
of finding the asymptotic amplitudes of the visiting probabilities
of chordal $\SLEk$ was posed in \cite{BB-zig_zag}, where these quantities
were called ``(boundary) zig-zag probabilities''.

Depending on the details of the precise formulation of the boundary
visit question, one would obtain a different non-universal multiplicative
constant in the SLE boundary Green's function \eqref{eq: full correlation with balls}
and its order refinement \eqref{eq: zig-zag proba with balls}. We
therefore prefer to use a generic notation for a quantity of this
type, for which we are free to choose a more convenient multiplicative
normalization. We also prefer to make explicit the dependence of the
question on the starting point $x\in\bR$ of the chordal $\text{\ensuremath{\SLEk}}$
curve, but the end point of the curve will always be kept at infinity.
In the rest of this article, 
\begin{align*}
 & \Ampl^{(N)}(x;y_{1},y_{2},\ldots,y_{N})
\end{align*}
denotes a \textbf{(boundary) zig-zag amplitude}, which is proportional
to any of the interpretations (see Sections \ref{sub: different definitions of boundary visit}
and \ref{sub: applications and universality}) of the order refined
boundary visit question. In particular we have
\begin{align*}
P_{(\bH;0,\infty)}^{(N)}(y_{1},y_{2},\ldots,y_{N})=\; & \const\times\zeta^{(N)}(0;y_{1},y_{2},\ldots,y_{N}).
\end{align*}
Similarly, we denote by 
\begin{align*}
\Corr^{(N)}(x;y_{1},y_{2},\ldots,y_{N})
\end{align*}
a \textbf{complete (boundary) correlation function}, so that in particular

\begin{align*}
C_{(\bH;0,\infty)}^{(N)}(y_{1},y_{2},\ldots,y_{N})=\; & \const\times\Corr^{(N)}(0;y_{1},y_{2},\ldots,y_{N}),
\end{align*}
with the same proportionality constant.

Explicit formulas for the above types of quantities are known in the
following two special cases:
\begin{itemize}
\item The one-point function ($N=1$) behaves simply as a power law, as
follows immediately from the invariance under dilatations $z\mapsto\lambda z$
($\lambda>0$) of the chordal $\SLEk$ in $(\bH;0,\infty)$ 
\begin{align}
\zeta^{(1)}(x;y_{1})=\Corr^{(1)}(x;y_{1})\propto\; & |y_{1}-x|^{-h}=|y_{1}-x|^{1-\frac{8}{\kappa}}.\label{eq: known 1-pt result}
\end{align}

\item The two-point function when $y_{1}$ and $y_{2}$ are on the same
side of the starting point (either $x<y_{1}<y_{2}$ or $y_{2}<y_{1}<x$)
is given by a hypergeometric function \cite{SZ-boundary_proximity_of_SLE}
(see also \cite{BB-zig_zag}) 
\begin{align}
\zeta^{(2)}(x;y_{1},y_{2})=\Corr^{(2)}(x;y_{1},y_{2})\propto\; & |y_{1}-x|^{1-\frac{8}{\kappa}}|y_{2}-y_{1}|^{1-\frac{8}{\kappa}}\times\phantom{}_{2}F_{1}\left(\frac{4}{\kappa},\frac{\kappa-8}{\kappa};\frac{8}{\kappa};\frac{y_{2}-y_{1}}{y_{2}-x}\right).\label{eq: known 2-pt result}
\end{align}

\end{itemize}
In this article we present a method for finding the solutions in the
general case. We write down a system of partial differential equations
(PDEs) motivated by conformal field theory (CFT) for the quantities
of interest, $\Ampl^{(N)}$ and $\Corr^{(N)}$. Our solutions for
them are written in terms of Coulomb gas integrals (Dotsenko-Fateev
integrals \cite{DF-multipoint_correlation_functions}) and are found
by quantum group calculations. This technique is developed in the
present article and in \cite{KP-covariant_boundary_correlations},
we call it the spin chain~-~Coulomb gas correspondence. Our primary
goal here is to find the explicit formulas and show their wide applicability:
the functions $\Ampl^{(N)}$ and $\Corr^{(N)}$ answer various formulations
of boundary visit questions for SLEs as well as for interfaces in
lattice models. We also compare the results to numerical simulations
of various lattice models, and outline a strategy of proof that our
formulas give the (order refined) SLE boundary Green's functions.

We emphasize that it is very rarely possible to find the exact solution
for an SLE problem involving a large number of marked points --- the
few existing solutions to such problems rely on finding tricks that
appear particular to each problem
\cite{Hagendorf-generalization_of_Schramms_formula_for_SLE2,
HD-SLE_on_doubly_connected_domains_and_the_winding_of_LERWs,
SZK-factorization_of_percolation_density_correlation_functions_for_clusters_touching_the_sides_of_a_rectangle,
SK-complete_CFT_solution_of_a_chiral_6pt_correlation_function,
SKFZ-cluster_densities_at_2d_critical_points_in_rectangular_geometries,
BI-proof_of_a_factorization_formula_for_critical_percolation,
AKL-the_Greens_function_for_radial_SLE,FKZ-cluster_pinch_point_densities_in_polygons,
FK-solution_space_for_a_system_of_null_state_PDEs_1}.%
\footnote{In contrast, it is almost routine to answer chordal SLE questions
which involve only two boundary points or one bulk point in addition
to the starting point and end point of the curve. This is so essentially
because the three-dimensional group of conformal automorphisms of
the domain allows to reduce the problem with four real variables to
just one cross ratio, and a standard application of Itô calculus yields
a second order linear ordinary differential equation for the quantity
in question. Boundary conditions then pin down the correct answer
in the two-dimensional space of solutions. For example the known formulas
\eqref{eq: known 1-pt result} and \eqref{eq: known 2-pt result}
were found by such methods. For questions depending on a larger number
of points, such as the one studied in this article, instead of ordinary
differential equations one would need to solve partial differential
equations, and the spaces of solutions become substantially harder
to manage.%
} The key technique that enables us to find the exact solution here
is the spin chain~-~Coulomb gas correspondence. It provides a systematic
method to solve a quite general class of SLE problems.

\subsection{Organization of the article}

The rest of the article is organized as follows.

In Section~\ref{sec: the problem} we formulate the PDE problem which
we solve in the subsequent sections to find the zig-zag amplitudes
$\Ampl^{(N)}$ and the complete correlation functions $\Corr^{(N)}$:
\begin{itemize}
\item The functions $\Ampl^{(N)}$ and $\Corr^{(N)}$ are conformally covariant.
\item The functions $\Ampl^{(N)}$ and $\Corr^{(N)}$ satisfy a second order
PDE and $N$ third order PDEs.
\item The boundary conditions depend on the order of visits: they are written
in terms of asymptotic behaviors of $\Ampl^{(N)}$ and their inhomogeneous
terms involve the $\zeta^{(N-1)}$ in a recursive manner.
\end{itemize}

In Section \ref{sec: quantum group and Coulomb gas} we discuss the
spin chain~-~Coulomb gas correspondence, by which the PDE problem
is translated to a linear problem in representations of a quantum
group:
\begin{itemize}
\item We associate functions defined by Coulomb gas integrals to vectors
in a finite-dimensional tensor product representation of the quantum
group $\Uqsltwo$.
\item The functions associated to highest weight vectors are solutions to
the partial differential equations of Section \ref{sec: the problem},
and for particular highest weights they also have the correct conformal
covariance.
\item Projections to subrepresentations in consecutive tensorands determine
the asymptotic behaviors of the functions.
\item There are unique highest weight vectors of the correct highest weights
whose subrepresentation projections correspond to the boundary conditions
imposed on the zig-zag amplitudes $\Ampl^{(N)}$.
\end{itemize}

In Section \ref{sec: regularized real integrals and evaluation} the
integrals obtained in the spin chain~-~Coulomb gas correspondence
are rewritten as regularized real integrals.
The transformation
to real integrals concretely exhibits the needed closed homology properties
of our solutions.

In Section~\ref{sec: interpretations and applications} we 
discuss basic properties, applications, interpretations,
and universality of the SLE boundary visit question and outline a
strategy of proof.

In Section \ref{sec: lattice models and numerics} we compare our
formula numerically to simulations of lattice models of statistical
mechanics. We study random interfaces in percolation, random cluster
model, and loop-erased random walk. We perform computer simulations
of them and collect frequencies of multi-point boundary visits of
the interfaces, and compare renormalized frequencies to the zig-zag
amplitudes $\Ampl^{(N)}$.

We conclude the article by discussion and outlook in Section \ref{sec: conclusions}.

The article is complemented with several appendices. Appendix \ref{sec: derivations of the value of h}
provides two derivations of the value of the scaling exponent \eqref{eq: h13},
and a derivation of the second order PDE. Appendix \ref{sec: conformal field theory}
contains relevant background on conformal field theory. Our normalization
conventions for some quantum group representations and some explicit
four-point solutions are contained in Appendix \ref{sec: explicit quantum group formulas}.
Numerical evaluation of the integrals of Sections \ref{sec: quantum group and Coulomb gas}
and \ref{sec: regularized real integrals and evaluation} is treated
in Appendix \ref{sub: numerical evaluation}.

\bigskip{}

\section{The problem: partial differential equations and asymptotics\label{sec: the problem}}

We find the boundary visit amplitudes $\Ampl^{(N)}$ and $\Corr^{(N)}$
by solving a PDE problem. The system of partial differential equations
is given below in Section \ref{sub: differential equations}. This
part is the same for $\Corr^{(N)}$ and for $\Ampl^{(N)}$, and moreover
the system is the same for all boundary zig-zag amplitudes corresponding
to different orders of visits to the same set of points. The results
will be different, however, as each of the functions satisfies different
boundary conditions, detailed in Section \ref{sec: boundary condition asymptotics}.

\subsection{Differential equations for boundary visit amplitudes\label{sub: differential equations}}

The linear homogeneous system of PDEs below contains essentially three
different types of partial differential equations --- all of them
can be argued to hold by conformal field theory (see Appendix \ref{sub: singular vectors}),
but from the point of view of SLE analysis, the argument leading to
each of them is different. For $\Ampl^{(N)}$ the system reads:

\begin{align}
\left[\pder x+\sum_{j=1}^{N}\pder{y_{j}}\right]\Ampl^{(N)}(x;y_{1},\ldots,y_{N})=\; & 0\label{eq: translation invariance for zig-zag}\\
\left[x\pder x+\sum_{j=1}^{N}y_{j}\pder{y_{j}}-Nh\right]\Ampl^{(N)}(x;y_{1},\ldots,y_{N})=\; & 0\label{eq: scaling covariance for zig-zag}\\
\left[\frac{\partial^{2}}{\partial x^{2}}-\frac{4}{\kappa}\mathcal{L}_{-2}\right]\Ampl^{(N)}(x;y_{1},\ldots,y_{N})=\; & 0\label{eq: second order differential equation}\\
\left[\frac{\partial^{3}}{\partial y_{j}^{3}}-\frac{16}{\kappa}\mathcal{L}_{-2}^{(j)}\frac{\partial}{\partial y_{j}}+\frac{8(8-\kappa)}{\kappa^{2}}\mathcal{L}_{-3}^{(j)}\right]\Ampl^{(N)}(x;y_{1},\ldots,y_{N})=\; & 0\qquad(j=\;1,2,\ldots,N),\label{eq: third order differential equations}
\end{align}
where
\begin{align*}
\mathcal{L}_{-2}=\; & \sum_{k=1}^{N}\left(\frac{-1}{y_{k}-x}\frac{\partial}{\partial y_{k}}+\frac{h}{(y_{k}-y_{j})^{2}}\right)
\end{align*}
and
\begin{align*}
\mathcal{L}_{-n}^{(j)}=\; & \frac{-1}{(x-y_{j})^{n-1}}\frac{\partial}{\partial x}+\frac{(n-1)\delta}{(x-y_{j})^{n}}+\sum_{k\neq j}\left(\frac{-1}{(y_{k}-y_{j})^{n-1}}\frac{\partial}{\partial y_{k}}+\frac{(n-1)h}{(y_{k}-y_{j})^{n}}\right) ,
\end{align*}
and we have used the parameters $h = \frac{8-\kappa}{\kappa}$ and $\delta=\frac{6-\kappa}{2\kappa}$.

The first order PDEs \eqref{eq: translation invariance for zig-zag}
and \eqref{eq: scaling covariance for zig-zag} express the translation
invariance and homogeneity of the amplitudes. More general conformal
covariance of the answer will be discussed in Section \ref{sub: conformal covariance}
and again from a conformal field theory point of view in Appendix
\ref{sub: conformal covariance from CFT}. The second order PDE \eqref{eq: second order differential equation}
can be interpreted either in terms of the SLE process as the statement
of a local martingale property of the answer, see Appendix \ref{sec: 2nd order PDE},
or in terms of conformal field theory as a conformal Ward identity
associated to a second order degeneracy of the boundary field located
at $x$, as will be discussed in Appendix \ref{sub: singular vectors}.
The $N$ third order PDEs~\eqref{eq: third order differential equations}
are similarly the conformal Ward identities associated to third order
degeneracies of the boundary fields located at $y_{j}$, $j=1,2,\ldots,N$,
see Appendix \ref{sub: singular vectors}. Unlike for the first and
second order equations we do not know how to explain the third order
equations by SLE analysis directly. As a partial justification, however,
we note that Equations~\eqref{eq: third order differential equations}
coincide with the third order partial differential equations \cite{Dubedat-SLE_and_Virasoro_representations_fusion}
derived by Dub\'edat 
for limiting cases of multiple SLE partition functions, which morally
describe the same configurations of curves as our boundary visiting SLEs.
Ultimately, the validity of all of the above equations
for the SLE boundary visit amplitudes would need to be established
by first finding the explicit answer, which is the main task in the
present article, and then proving that it gives the SLE boundary Green's
function following the strategy that will be outlined in Section \ref{sub: conditioning application and proof strategy}.%
\footnote{Given that this proposed route to Equations \eqref{eq: third order differential equations}
is somewhat indirect, one may wonder if more direct hints of these
third order differential equations exist. To this end, recall that
for $N=1$ and $N=2$ the explicit zig-zag amplitudes \eqref{eq: known 1-pt result}
and \eqref{eq: known 2-pt result} can in any case be found by routine
SLE calculations. For these already known functions $\Ampl^{(1)}$
and $\Ampl^{(2)}$, we have by direct calculation verified the validity of the
third order equations, which conformal field theory predicts.%
}

\subsection{Asymptotics for boundary visit amplitudes\label{sec: boundary condition asymptotics}\label{sec: detailed asymptotics}}

The system of differential equations of Section \ref{sub: differential equations}
has a large space of solutions. To pin down the correct solution we
need boundary conditions, which will be specified in the form of asymptotic
behavior of the boundary zig-zag amplitudes. Considerations of the
possible asymptotics allowed by conformal field theory can be found
in Appendix \ref{sec: CFT asymptotics}. The particular requirements
that finally specify the solutions are given below.

Consider the question of visiting the neighborhoods of $y_{1},y_{2},\ldots,y_{N}$
in this order. Some notation and terminology is needed to conveniently
describe the specific asymptotics of $\Ampl^{(N)}$ in this case.
We say that points $y_{j}$ such that $y_{j}<x$ are \emph{on the
left} and points $y_{j}$ such that $x<y_{j}$ are \emph{on the right}.
We say that the points are in an \emph{outwards increasing order}
if for any $y_{j},y_{k}$ on the left we have that $j<k$ implies
$y_{k}<y_{j}$ and for any $y_{j},y_{k}$ on the right we have that
$j<k$ implies $y_{j}<y_{k}$, in other words that among points on
the same side, the point further away from starting point is visited
later.

The boundary visit amplitude vanishes unless the points are in an
outwards increasing order --- a visit to a small neighborhood of a
point further away on the same side almost disconnects the future
passage of the curve to the point that would need to be visited later.%
\footnote{For positive $\eps$ it is in principle possible
for visits to occur in an order that is not outwards increasing,
but these probabilities are suppressed by a higher power of
$\eps$, and as such do not survive the limit~\eqref{eq: zig-zag proba with balls} of $\eps \searrow 0$.
Rigorous estimates of the appropriate SLE probabilities are of the
type considered, e.g., in \cite{Beffara-dimension_of_the_SLE_curves,LW-multi_point_Greens_functions_for_SLE},
although the present situation is somewhat easier.%
}

It is convenient to use a separate ordering for the points on the
left and right. Denote therefore $y_{1}^{\lft},\ldots,y_{L}^{\lft}$
the points on the left in a decreasing order (in the order of visits)
and $y_{1}^{\rgt},\ldots,y_{R}^{\rgt}$ the points on the right in
an increasing order (in the order of visits). The following notation
makes the arguments of the zig-zag amplitude appear in the same order
as they are on the real axis,
\begin{align*}
\Ampl_{\omega}(y_{L}^{-},\ldots,y_{1}^{-};x;y_{1}^{+},\ldots,y_{R}^{+})=\; & \zeta^{(N)}(x;y_{1},y_{2},\ldots,y_{N}),
\end{align*}
where $\omega=(\omega_{1},\omega_{2},\ldots,\omega_{N})\in\set{+,-}^{N}$
is a sequence of ``$\pm$''-symbols specifying the sequence of sides
of the visits in the sense that $\omega_{j}=-$ (resp. $\omega_{j}=+$)
if $y_{j}$ is on the left (resp. on the right). If we fix the number
$L$ of points on the left and the number $R$ of points on the right,
$N=L+R$, then the number of different outwards increasing orders
is ${N \choose L}$, corresponding to the choices of $\omega\in\set{+,-}^{N}$
with $L$ ``$-$''-symbols and $R$ ``$+$''-symbols. The complete
correlation function $\chi^{(N)}$ is the sum of these ${N \choose L}$
zig-zag amplitudes. In the particular case when all the points are
on the same side, the complete correlation function coincides with
the zig-zag amplitude.

The specific asymptotics depend on the order of visits, and to describe
them we need a few separate cases. We call the consecutive points
$y_{m}^{\rgtlft}$ and $y_{m+1}^{\rgtlft}$ on the same side ($\rgtlft$)
\emph{successively visited points on the same side} if for some $j$
we have $y_{m}^{\pm}=y_{j}$ and $y_{m+1}^{\pm}=y_{j+1}$.

We claim that for any outwards increasing order $\omega$ the boundary
zig-zag amplitude $\Ampl_{\omega}$ satisfies the asymptotics conditions
given below%
\footnote{The eventual justification of these requirements would be a proof
of the fact that the SLE boundary Green's function agrees with our formula
obtained by solving the PDE system with these conditions.
A strategy of proof is discussed in Section \ref{sub: conditioning application and proof strategy}.},
and that up to a multiplicative constant these asymptotics
determine all $\Ampl^{(N)}$. The conditions are intuitive in view
of the possibilities listed in Appendix \ref{sec: CFT asymptotics}:
they state that the order of magnitude of the amplitude is larger
if successively visited points are close and smaller if non-successively
visited points are close, and in the former case the leading asymptotic
is proportional to an $(N-1)$-point function, where the two close-by
points are replaced by a single point. Moreover, they state that the
leading behavior when successively visited points are close-by is
given by the $(N-1)$-point function with the two close-by points
replaced by just one.
\begin{itemize}
\item \emph{Asymptotics for successively visited points}: If $y_{j}$ and
$y_{j+1}$ are successively visited points on the same side, then
\begin{align}
\lim_{y_{j},y_{j+1}\rightarrow y'}\;\; & \frac{1}{|y_{j+1}-y_{j}|^{1-\frac{8}{\kappa}}}\Ampl^{(N)}(x;\, y_{1},\ldots,y_{j},y_{j+1},\ldots,y_{N})\label{eq: successively visited}\\
=\; & \const\times\zeta^{(N-1)}(x;\, y_{1},\ldots,y_{j-1},y',y_{j+2},y_{j+3},\ldots,y_{N}).\nonumber 
\end{align}

\item \emph{Asymptotics for non-successively visited points}: If $y_{j}$
and $y_{k}$ are non-successively visited consecutive points on the
same side, then
\begin{align}
\lim_{y_{j},y_{k}\rightarrow y'}\;\frac{1}{|y_{k}-y_{j}|^{1-\frac{8}{\kappa}}}\Ampl^{(N)}(x;\, y_{1},y_{2},\ldots,y_{N})=\; & 0.\label{eq: non-successively visited}
\end{align}

\item \emph{Asymptotics for the first points on the left and right}: For
the first point $y_{1}$ to be visited we have
\begin{align}
\lim_{x,y_{1}\rightarrow x'}\;\frac{1}{|y_{1}-x|^{1-\frac{8}{\kappa}}}\Ampl^{(N)}(x;\, y_{1},y_{2},\ldots,y_{N})=\; & \const\times\zeta^{(N-1)}(x';\, y_{2},y_{3},\ldots,y_{N}).\label{eq: the very first visited point}
\end{align}
For the first point on the opposite side, i.e., for $y_{1}^{\rgtlft}\neq y_{1}$
, we have
\begin{align}
\lim_{x,y_{1}^{\rgtlft}\rightarrow x'}\;\frac{1}{|y_{1}^{\rgtlft}-x|^{1-\frac{8}{\kappa}}}\Ampl^{(N)}(x;\, y_{1},y_{2},\ldots,y_{N})=\; & 0.\label{eq: first opposite visited point}
\end{align}
\end{itemize}
The constants in \eqref{eq: successively visited} and \eqref{eq: the very first visited point}
are different, but for different pairs of successively visited consecutive points,
the constant in \eqref{eq: successively visited} should be the same.\footnote{As a remark,
we have not found any new solutions by relaxing the requirement that the constants
for different pairs are equal --- even with unspecified constants treated as additional variables,
the system of equations forces the correct universal ratios between the constants, at least for small $N$.}
Moreover, the constants should not depend on $N$.

We conjecture that the solution space to the partial differential equations
\eqref{eq: translation invariance for zig-zag},
\eqref{eq: scaling covariance for zig-zag},
\eqref{eq: second order differential equation},
\eqref{eq: third order differential equations}
is finite-dimensional and that its dimension is exactly the multiplicity $m_N$
of a certain irreducible direct summand in a tensor product,
see Section~\ref{sec: existence and uniqueness of quantum group solutions}.
Under this assumption, it could be shown with the techniques introduced in Section~\ref{sec: quantum group and Coulomb gas},
that recursively in $N$ the asymptotics conditions
\eqref{eq: successively visited}, \eqref{eq: non-successively visited},
\eqref{eq: the very first visited point}, \eqref{eq: first opposite visited point}
specify uniquely, up to a multiplicative constant, solutions for all outwards increasing orders of visits $\omega$.

Our choice of normalization of $\Ampl^{(N)}$ and $\Corr^{(N)}$ will
be determined recursively by fixing the constant appearing in Equation
\eqref{eq: the very first visited point}, see Section \ref{sec: quantum group problem}.
Once this natural choice is made, the different $N$-point functions
$\Ampl_{\omega}$ obtain correct relative normalizations, with the
universal ratios referred to in Section \ref{sub: applications and universality}.
In particular, the constant appearing in Equation \eqref{eq: successively visited}
gets automatically fixed as well.

\bigskip{}

\section{Quantum group and integral formulas\label{sec: quantum group and Coulomb gas}}

\subsection{Coulomb gas integrals\label{sec: Coulomb gas integrals}}

The main tool that allows us to solve the PDE problem of Section \ref{sec: the problem}
and therefore to find the explicit formula for the SLE boundary visit
amplitudes is the spin chain~-~Coulomb gas correspondence. In this
article, for the sake of concreteness, we describe only the case relevant
to the problem of boundary visit amplitudes --- a more general treatment
can be found in \cite{KP-covariant_boundary_correlations}.

\subsubsection{Standard Coulomb gas integrals and their properties\label{sec: general Coulomb gas integrals}}

The Coulomb gas formalism of conformal field theory, or Dotsenko-Fateev
integrals \cite{DF-multipoint_correlation_functions}, is a way of
producing solutions to systems of differential equations of the type
of Section \ref{sub: differential equations} by integrating an auxiliary
function, which in our case takes the form
\begin{align}
 & f_{\ell}^{(N)}(x;\; y_{1},y_{2},\ldots,y_{N};\; w_{1},w_{2},\ldots,w_{\ell})\label{eq: Coulomb gas integrand}\\
=\; & \prod_{j=1}^{N}(y_{j}-x)^{\frac{4}{\kappa}}\times\prod_{1\leq j<k\leq N}(y_{k}-y_{j})^{\frac{8}{\kappa}}\times\prod_{s=1}^{\ell}(w_{s}-x)^{-\frac{4}{\kappa}}\times\prod_{j=1}^{N}\prod_{s=1}^{\ell}(w_{s}-y_{j})^{-\frac{8}{\kappa}}\times\prod_{1\leq s<r\leq\ell}(w_{r}-w_{s})^{\frac{8}{\kappa}}.\nonumber 
\end{align}

Consider the function
\begin{align}
F(x;y_{1},\ldots,y_{N})=\; & \int_{\Gamma}f_{\ell}^{(N)}(x;\; y_{1},\ldots,y_{N};w_{1},\ldots,w_{\ell})\,\ud w_{1}\cdots\ud w_{\ell},\label{eq: integral of Coulomb gas type general l}
\end{align}
where $\Gamma$ is a closed $\ell$-surface avoiding the points $x,y_{1},\ldots,y_{N}$.
The integral of course only depends on the homotopy type of the surface
$\Gamma$. The function is defined such that while the contour $\Gamma$
of the $w$-variables may depend on the positions of $x,y_{1},\ldots,y_{N}$,
the choice is locally constant. One then observes:
\begin{itemize}
\item \emph{translation invariance}: $F$ satisfies Equation \eqref{eq: translation invariance for zig-zag}.
\item \emph{scale covariance}: $F$ is homogeneous of degree $\Delta_{N;\ell}=\ell+\frac{4}{\kappa}(N^{2}+\ell^{2}-2\ell-2N\ell)$,
and in particular if $\ell=N$ it satisfies Equation \eqref{eq: scaling covariance for zig-zag}.
\item \emph{second order differential} \emph{equation}: $F$ satisfies Equation
\eqref{eq: second order differential equation}.
\item \emph{third order differential} \emph{equations}: $F$ satisfies Equations
\eqref{eq: third order differential equations}.
\end{itemize}
The translation invariance follows immediately from the translation
invariance of the integrand $f_{\ell}^{(N)}$ by considering a shift
of the variables $x,y_{1},\ldots,y_{N}$ small enough so that the
integration contour $\Gamma$ can be kept constant, and then the same
shift of the integration contour, which now does not change the homotopy
type. The scaling covariance is shown similarly, starting with scaling
close enough to identity. The relevant scaling covariance of the integrand
reads
\begin{align*}
f_{\ell}^{(N)}(\lambda x;\lambda y_{1},\ldots;\lambda w_{1},\ldots)=\; & \lambda^{\frac{4}{\kappa}(N^{2}+\ell^{2}-2\ell-2N\ell)}\; f_{\ell}^{(N)}(x;y_{1},\ldots;w_{1},\ldots)
\end{align*}
and an extra factor $\lambda^{\ell}$ comes from the change of variables
in the integration --- the formal proofs can be found in \cite[Lemma 3.3 and Theorem 4.17]{KP-covariant_boundary_correlations}.

The second and third order differential equations rely more crucially
on the fact that the integration surface $\Gamma$ is closed. One
again starts from a property satisfied by the integrand alone. Starting
from the second order equation, let
\begin{align*}
\mathcal{D}_{1,2}=\; & \frac{\kappa}{2}\frac{\partial^{2}}{\partial x^{2}}+\sum_{j=1}^{N}\left(\frac{2}{y_{j}-x}\pder{y_{j}}-\frac{2h}{(y_{j}-x)^{2}}\right)
\end{align*}
be the differential operator we want to show annihilates $F$. It
is a matter of straightforward verification to see that the integrand
satisfies
\begin{align*}
\left[\mathcal{D}_{1,2}+\sum_{s=1}^{\ell}\left(\frac{2}{w_{s}-x}\pder{w_{s}}-\frac{2}{(w_{s}-x)^{2}}\right)\right]f_{\ell}^{(N)}(x;y_{1},\ldots;w_{1},\ldots)=\; & 0
\end{align*}
and to notice that this can also be read as
\begin{align*}
\mathcal{D}_{1,2}\, f_{\ell}^{(N)}(x;y_{1},\ldots;w_{1},\ldots)=\; & -2\sum_{s=1}^{\ell}\pder{w_{s}}\left(\frac{1}{w_{s}-x}\times f_{\ell}^{(N)}(x;y_{1},\ldots;w_{1},\ldots)\right).
\end{align*}
Thus when acting on $F$ by the differential operator $\mathcal{D}_{1,2}$,
we may take the operator inside the integral, and rewrite the integrand
as a sum of total derivatives. The integral of these vanish because
the contour was assumed to be closed. Hence one gets the second order
differential equation for $F$. The third order differential equations
are shown to hold similarly --- the formal proof of a more general statement can be found in
\cite[Proposition 4.12 and Theorem 4.17]{KP-covariant_boundary_correlations}.

\subsubsection{Spin chain - Coulomb gas basis functions\label{sec: spin chain Coulomb gas basis}}

\begin{figure}
\begin{centering}
\includegraphics[width=1\textwidth]{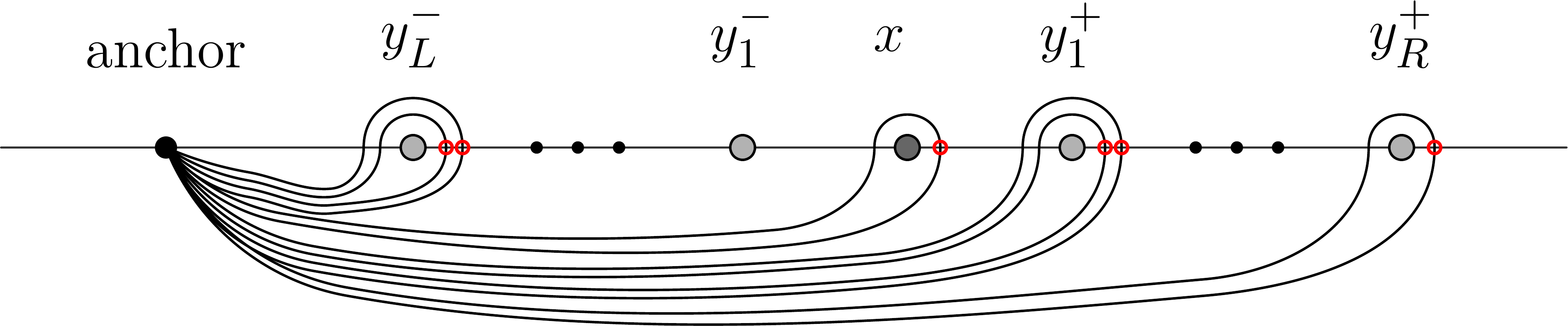}
\par\end{centering}

\caption{\emph{The integration contours of the $w_{j}$-variables in }$\Gamma_{t_{L}^{\lft},\ldots,t_{2}^{\lft},t_{1}^{\lft};d;t_{1}^{\rgt},t_{2}^{\rgt},\ldots,t_{R}^{\rgt}}$
\emph{and the point (marked by red circles) where the integrand $\FWintegrand$ is
rephased to be positive.\label{fig: FW integration contour}}}
\end{figure}

Our solution will eventually be of the form \eqref{eq: integral of Coulomb gas type general l},
with $\ell=N$. As in \cite{KP-covariant_boundary_correlations,KP-pure_partition_functions_of_multiple_SLEs},
we need to unveil an underlying quantum group structure, which will
be useful for calculations, and in particular crucial for dealing
with the asymptotics. For this purpose, we introduce the functions
\begin{align*}
 & \varphi_{t_{L}^{\lft},\ldots,t_{2}^{\lft},t_{1}^{\lft};d;t_{1}^{\rgt},t_{2}^{\rgt},\ldots,t_{R}^{\rgt}}(y_{L}^{-},\ldots,y_{2}^{-},y_{1}^{-};x;y_{1}^{+},y_{2}^{+},\ldots,y_{R}^{+})
\end{align*}
indexed by $t_{j}^{\pm}\in\set{0,1,2}$ and $d\in\set{0,1}$, which
are defined by the integrals
\begin{align}
\varphi_{t_{L}^{\lft},\ldots,t_{R}^{\rgt}}(y_{L}^{-},\ldots,y_{R}^{+})=\; & \int_{\Gamma_{t_{L}^{\lft},\ldots,t_{R}^{\rgt}}}\FWintegrand_{t_{L}^{\lft},\ldots,t_{R}^{\rgt}}(y_{L}^{-},\ldots,y_{R}^{+};w_{1},\ldots,w_{\ell})\,\ud w_{1}\cdots\ud w_{\ell},\label{eq: basis functions}
\end{align}
where:
\begin{itemize}
\item The integration surface $\Gamma_{t_{L}^{\lft},\ldots,t_{2}^{\lft},t_{1}^{\lft};d;t_{1}^{\rgt},t_{2}^{\rgt},\ldots,t_{R}^{\rgt}}$
is shown in Figure \ref{fig: FW integration contour}. The dimension
of the integration surface, i.e., the number of integration variables
$w_{s}$, is $\ell=d+\sum_{j=1}^{L}t_{j}^{-}+\sum_{j=1}^{R}t_{j}^{+}$.
In the functions appearing in our final answer this will always be
$\ell=N$. The contour of each integration variable $w_{s}$ is a
loop based at an anchor point $\anchor$ to the left of all of the
variables, and the loop encircles one of the points in the positive
direction. The loops of the first $t_{L}^{-}$ variables encircle
the point $y_{L}^{-}$, the next $t_{L-1}^{-}$ variables encircle
the point $y_{L-1}^{-}$ and so on. The loops encircling the same
point are nested. The loops encircling different points avoid each
other so that the contours to a point further on the right go below.
\item The integrand $\FWintegrand_{t_{L}^{\lft},\ldots,t_{1}^{\lft};d;t_{1}^{\rgt},\ldots,t_{R}^{\rgt}}$
is a rephased branch of the integrand $f_{\ell}^{(N)}$ defined in
Equation \eqref{eq: Coulomb gas integrand}: we multiply by a suitable
complex number of modulus one to make $\FWintegrand_{t_{L}^{\lft},\ldots,t_{R}^{\rgt}}$
real and positive at the point where each of the integration variables
is on the real axis to the right of the point it encircles, see Figure
\ref{fig: FW integration contour}. 
\end{itemize}

We make the following remarks about the role and properties of the
above functions:
\begin{itemize}
\item Individually the surfaces $\Gamma_{t_{L}^{\lft},\ldots,t_{R}^{\rgt}}$
are not closed, but our solution will be a linear combination which
is closed in the appropriate homology \cite{FW-topological_representation_of_Uqsl2}.
\item The individual functions $\varphi_{t_{L}^{\lft},\ldots,t_{R}^{\rgt}}$
depend also on the point $\anchor$ where the loops in $\Gamma_{t_{L}^{\lft},\ldots,,t_{R}^{\rgt}}$
are anchored. This dependence will cancel in the final answer ---
the cancellation will be shown concretely in Section \ref{sec: regularized real integrals and evaluation},
and a proof of this property in a general setup is given in
\cite[Proposition 4.5 and Theorem 4.17]{KP-covariant_boundary_correlations}.
\end{itemize}
In the spin chain~-~Coulomb gas correspondence defined in Section
\ref{sec: definition of correspondence}, we will make basis vectors
in a quantum group representation correspond to the functions $\varphi_{t_{L}^{\lft},\ldots,t_{2}^{\lft},t_{1}^{\lft};d;t_{1}^{\rgt},t_{2}^{\rgt},\ldots,t_{R}^{\rgt}}$.
In Sections \ref{sec: asymptotics via correspondence} and \ref{sec: closed homology classes}
we explain how straightforward quantum group calculations will allow
us to decide about the asymptotics of the functions as well as the
closedness of the surfaces in an appropriate homology --- see also
\cite{FW-topological_representation_of_Uqsl2,KP-covariant_boundary_correlations}.

\subsection{Quantum group\label{sec: quantum group}}

We need to recall some facts and fix some notation for the quantum
group $\Uqsltwo$. It should be thought of as a deformation of (the
universal enveloping algebra of) the Lie algebra $\mathfrak{sl}_{2}$,
with a deformation parameter $q$ --- with a suitable normalization
when $q\rightarrow1$ one recovers $\mathfrak{sl}_{2}$ from the definitions
we give below.

We let $q=e^{4\pi\ii/\kappa}$, and assume that $\kappa$ is generic
in the sense that $\kappa\notin\bQ$.%
\footnote{For irrational $\kappa$ the parameter $q$ is not a root of unity,
and the representation theory of the quantum group is semisimple.
To obtain the SLE boundary visit amplitudes in general, we may in the end use
continuity in the parameter $\kappa$.%
} We define the $q$-integers $\qnum m$ (for $m\in\bZ$) 
\begin{align*}
\qnum m\,:=\; & \frac{q^{m}-q^{-m}}{q-q^{-1}}
\end{align*}
Since we assume $\kappa\notin\bQ$, all $q$-integers $\qnum m$ with
$m\neq0$ are non-zero.

\subsubsection{Definition of the quantum group\label{sec: quantum group definition}}

The quantum group $\Uqsltwo$ is the algebra over $\bC$ with generators
$E,F,K,K^{-1}$ and relations 
\begin{align*}
KK^{-1}=\; & 1=K^{-1}K,\qquad KE=q^{2}EK,\qquad KF=q^{-2}FK,\\
EF-FE=\; & \frac{1}{q-q^{-1}}\left(K-K^{-1}\right).
\end{align*}
Moreover, $\Uqsltwo$ is equipped with the unique Hopf algebra structure
such that the coproducts of the generators are
\begin{align*}
\Hcp(K)=\; & K\tens K,\qquad & \Hcp(E)=\; & E\tens K+1\tens E,\qquad & \Hcp(F)=\; & F\tens1+K^{-1}\tens F.
\end{align*}
The coproduct $\Hcp$ determines the action of the quantum group in
tensor product $\Wd\tens\Wd'$ of two representations $\Wd$ and $\Wd'$,
for example $E.(v\tens v')=E.v\tens K.v'+v\tens E.v'$. The tensor
product of representations is then associative but not commutative:
multiple tensor products are well defined, for example $(\Wd\tens\Wd')\tens\Wd''\isom\Wd\tens(\Wd'\tens\Wd'')$,
but the order of the tensorands is important.

\subsubsection{Representations of the quantum group\label{sec: irreducible representations}}

The quantum group $\Uqsltwo$ is semisimple (for $q$ not a root of
unity) in the sense that any finite dimensional representation is
the direct sum of its irreducible subrepresentations. In fact, the
representation theory essentially just deforms that of $\mathfrak{sl}_{2}$.
We recall the following standard facts, the proofs of which  can be found in,
e.g., \cite[Lemmas 2.3 and 2.4]{KP-covariant_boundary_correlations}.

For any $d\in\bN$, there exists a $d$-dimensional irreducible representation
$\Wd_{d}$ with a basis $\Wbas_{0},\Wbas_{1},\Wbas_{2},\ldots,\Wbas_{d-1}$
such that the action of the generators on the basis vectors is given
by 
\begin{align*}
K.\Wbas_{j}=\; & q^{d-1-2j}\;\Wbas_{j}\\
F.\Wbas_{j}=\; & \Wbas_{j+1} & \text{(with interpretation \ensuremath{\Wbas_{d}=0})}\\
E.\Wbas_{j}=\; & \qnum j\qnum{d-j}\,\Wbas_{j-1} & \text{(with interpretation \ensuremath{\Wbas_{-1}=0})} & \,.
\end{align*}
This representation $\Wd_{d}$ is the appropriate deformation of the
$d$-dimensional irreducible representation of $\mathfrak{sl}_{2}$ (``the spin-$\frac{d-1}{2}$
representation''). The tensor products of $\Wd_{d}$ decompose according to the 
formula
\begin{align*}
\Wd_{d_{2}}\otimes\Wd_{d_{1}}\isom\; & \Wd_{d_{1}+d_{2}-1}\oplus\Wd_{d_{1}+d_{2}-3}\oplus\cdots\oplus\Wd_{|d_{1}-d_{2}|+1}.
\end{align*}
Our calculations will require some specific cases of such
(quantum) Clebsch-Gordan decompositions
to be made explicit. Formulas for those cases are given in Appendix \ref{sec: explicit submodules}.

The one-dimensional irreducible $\Wd_{1}\isom\bC$ is the \emph{trivial
representation}, it acts as a neutral element of the tensor products:
for any representation $\Wd$ we have the isomorphisms $\Wd_{1}\tens\Wd\isom\Wd\isom\Wd\tens\Wd_{1}$.
This allows us to omit $\Wd_{1}$ in tensor products, when needed.

\subsection{Spin chain - Coulomb gas correspondence\label{sec: correspondence}}

\subsubsection{Definition of the correspondence\label{sec: definition of correspondence}}

With the above preparations we can now define the correspondence.
The spin chain~-~Coulomb gas correspondence linearly associates
to vectors
\begin{align*}
v\in\; & \Wd_{3}^{\tens R}\tens\Wd_{2}\tens\Wd_{3}^{\tens L}
\end{align*}
in a tensor product of representations of $\Uqsltwo$ a function,
so that for the natural tensor product basis vectors the associated
functions are those defined in Section \ref{sec: spin chain Coulomb gas basis}:
\begin{align*}
\Wbas_{t_{R}^{+}}\tens\cdots\tens\Wbas_{t_{1}^{+}}\tens\Wbas_{d}\tens\Wbas_{t_{1}^{-}}\tens\cdots\tens\Wbas_{t_{L}^{-}}\;\mapsto\;\; & \varphi_{t_{L}^{\lft},\ldots,t_{1}^{\lft};d;t_{1}^{\rgt},\ldots,t_{R}^{\rgt}}.
\end{align*}

Note that in our convention, the order of the variables of the function
is the reverse of the order of the corresponding factors in the tensor
product.

\subsubsection{Asymptotics via the correspondence\label{sec: asymptotics via correspondence}}

A key property of the spin chain~-~Coulomb gas correspondence is
that the asymptotics of the functions can be straightforwardly read
from the projections to subrepresentations of the corresponding vectors
in $\Wd_{3}^{\tens R}\tens\Wd_{2}\tens\Wd_{3}^{\tens L}$.
For these projections, we use below the notation and normalization conventions of Appendix~\ref{sec: explicit submodules}.

Let $v\in\Wd_{3}^{\tens R}\tens\Wd_{2}\tens\Wd_{3}^{\tens L}$ and
let $\varphi$ be the function associated to $v$ by the correspondence
of Section \ref{sec: definition of correspondence}. The correspondence
of asymptotics and subrepresentations is stated precisely in the following:
\begin{itemize}
\item Consider two consecutive points $y_{m}^{\pm},y_{m+1}^{\pm}$ on the
right or left (superscript ``$+$'' or ``$-$'', respectively).
For $d \in \set{1,3,5}$, denote accordingly by $\pi^{(d)}_{\pm;m}$
the projection to $d$-dimensional subrepresentation of $\Wd_3 \tens \Wd_3$
acting on the $m$:th and $m+1$:st components on the appropriate side.
\begin{itemize}
\item Suppose that $v$ is in the singlet of the components corresponding
to $y_{m}^{\pm},y_{m+1}^{\pm}$, that is $v=\pi_{\pm;m}^{(1)}(v)$.
Then as $y_{m}^{\pm},y_{m+1}^{\pm}\rightarrow y'$, we have 
\begin{align*}
\varphi(x;y_{1},\ldots,y_{N})\sim\; & B_{1}\times|y_{m+1}^{\rgtlft}-y_{m}^{\rgtlft}|^{2-\frac{16}{\kappa}}\times\varphi^{(1)}(x;y_{1},\ldots,y_{N}),
\end{align*}
where the variables $y_{m}^{\pm},y_{m+1}^{\pm}$ have been removed
from the right hand side, the function $\varphi^{(1)}$ is the function
of two variables less associated to the vector $\hat{\pi}_{\pm;m}^{(1)}(v)$
interpreted as a vector in either $\Wd_{3}^{\tens(R-2)}\tens\Wd_{2}\tens\Wd_{3}^{\tens L}$
or $\Wd_{3}^{\tens R}\tens\Wd_{2}\tens\Wd_{3}^{\tens(L-2)}$, and
the constant is the generalized beta-function%
\begin{align*}
B_1 = \; & \frac{\Gamma (\frac{\kappa -8}{\kappa})^2 \; \Gamma (\frac{\kappa -4}{\kappa})^2 \; \Gamma (\frac{\kappa +8}{\kappa})}
{2 \; \Gamma (2-\frac{8}{\kappa }) \; \Gamma (2 \frac{\kappa -6}{\kappa }) \; \Gamma (\frac{\kappa +4}{\kappa } )} .
\end{align*}
\item Suppose that $v$ is in the triplet of the components corresponding
to $y_{m}^{\pm},y_{m+1}^{\pm}$, that is $v=\pi_{\pm;m}^{(3)}(v)$.
Then as $y_{m}^{\pm},y_{m+1}^{\pm}\rightarrow y'$, we have 
\begin{align*}
\varphi(x;y_{1},\ldots,y_{N})\sim\; & B_{3}\times|y_{m+1}^{\rgtlft}-y_{m}^{\rgtlft}|^{1-\frac{8}{\kappa}}\times\varphi^{(3)}(x;y_{1},\ldots,y',\ldots,y_{N}),
\end{align*}
where on the right hand side the two variables $y_{m}^{\pm},y_{m+1}^{\pm}$
have been removed and replaced by one $y'$, the function $\varphi^{(3)}$
is the function of one variable less associated to the vector $\hat{\pi}_{\pm;m}^{(3)}(v)$
interpreted as a vector in either $\Wd_{3}^{\tens(R-1)}\tens\Wd_{2}\tens\Wd_{3}^{\tens L}$
or $\Wd_{3}^{\tens R}\tens\Wd_{2}\tens\Wd_{3}^{\tens(L-1)}$ and the
constant is the beta-function
\begin{align*}
B_{3}=\; & \frac{\Gamma(\frac{\kappa-8}{\kappa})^{2}}{\Gamma(2\frac{\kappa-8}{\kappa})} .
\end{align*}
\item Suppose that $v$ is in the quintuplet of the components corresponding
to $y_{m}^{\pm},y_{m+1}^{\pm}$, that is $v=\pi_{\pm;m}^{(5)}(v)$.
Then as $y_{m}^{\pm},y_{m+1}^{\pm}\rightarrow y'$, we have 
\begin{align*}
\varphi(x;y_{1},\ldots,y_{N})\sim\; & |y_{m+1}^{\rgtlft}-y_{m}^{\rgtlft}|^{\frac{8}{\kappa}}\times\varphi^{(5)}(x;y_{1},\ldots,y',\ldots,y_{N}),
\end{align*}
where on the right hand side the two variables $y_{m}^{\pm},y_{m+1}^{\pm}$
have been removed and replaced by one $y'$.
We will not need any properties of the function $\varphi^{(5)}$,
but we nevertheless remark that with a generalization of the present method it becomes in principle explicit
(see \cite[Proposition 4.4]{KP-covariant_boundary_correlations} for details).
\end{itemize}
\item Consider the point $x$ and the first point $y_{1}^{\pm}$ on the
right or left (superscript ``$+$'' or ``$-$'', respectively).
For $d \in \set{2,4}$, denote accordingly by $\pi^{(d)}_{\pm}$
the projection to $d$-dimensional subrepresentation of ${\Wd_3 \tens \Wd_2}$ or
$\Wd_2 \tens \Wd_3$ acting on the middle factor $\Wd_2$ and the $\Wd_3$ on the appropriate side of it.
\begin{itemize}
\item Suppose that $v$ is in the doublet of the components corresponding
to $x,y_{1}^{\pm}$, that is $v=\pi_{\pm}^{(2)}(v)$. Then as $x,y_{1}^{\pm}\rightarrow x'$,
we have 
\begin{align*}
\varphi(x;y_{1},\ldots,y_{N})\sim\; & B_{2}\times|y_{1}^{\rgtlft}-x|^{1-\frac{8}{\kappa}}\times\varphi^{(2)}(x';\ldots,y_{N}),
\end{align*}
where on the right hand side the two variables $x,y_{1}^{\pm}$ have
been removed and replaced by one $x'$, the function $\varphi^{(2)}$
is the function of one variable less associated to the vector $\hat{\pi}_{\pm}^{(2)}(v)$
interpreted as a vector in either $\Wd_{3}^{\tens(R-1)}\tens\Wd_{2}\tens\Wd_{3}^{\tens L}$
or $\Wd_{3}^{\tens R}\tens\Wd_{2}\tens\Wd_{3}^{\tens(L-1)}$ and the
constant is the beta-function
\begin{align}
B_{2}=\; & \frac{\Gamma(\frac{\kappa-4}{\kappa})\Gamma(\frac{\kappa-8}{\kappa})}{\Gamma(2\frac{\kappa-6}{\kappa})} . \label{eq: B2 formula}
\end{align}
\item Suppose that $v$ is in the quadruplet of the components corresponding
to $x,y_{1}^{\pm}$, that is $v=\pi_{\pm}^{(4)}(v)$. Then as $x,y_{1}^{\pm}\rightarrow x'$,
we have 
\begin{align*}
\varphi(x;y_{1},\ldots,y_{N})\sim\; & |y_{1}^{\rgtlft}-x|^{\frac{4}{\kappa}}\times\varphi^{(4)}(x';\ldots,y_{N}),
\end{align*}
where on the right hand side the two variables $x,y_{1}^{\pm}$ have
been removed and replaced by one $x'$. We will not need any
properties of the function $\varphi^{(4)}$, although it
could also be written explicitly (see \cite[Proposition 4.4]{KP-covariant_boundary_correlations}
for details).
\end{itemize}
\end{itemize}
For a general $v\in\Wd_{3}^{\tens R}\tens\Wd_{2}\tens\Wd_{3}^{\tens L}$
the asymptotics of $\varphi$ are obtained by the above formulas and
linearity.

The statements are proved by straightforward manipulations of the
integrals, which are done in a more general setup
in \cite[Lemmas 4.2, 4.3, 3.11, Proposition 4.4]{KP-covariant_boundary_correlations}.
Indeed, when the vector $v$ is of the supposed form, we know from
Appendix \ref{sec: explicit submodules} explicitly how its two consecutive
tensor components must be related. Considering the different possibilities
for $y_{m}^{\pm},y_{m+1}^{\pm}$ namely $v=\pi_{\pm;m}^{(5)}(v)$,
$v=\pi_{\pm;m}^{(3)}(v)$, or $v=\pi_{\pm;m}^{(1)}(v)$, one manages
to rearrange zero, one, or two integration variables on contours between
the points $y_{m}^{\pm}$ and $y_{m+1}^{\pm}$ so that the contours
of the rest of the integration variables remain away from these points.
Then extracting the asymptotics becomes easy: first of all there is
a factor $|y_{m+1}^{\pm}-y_{m}^{\pm}|^{\frac{8}{\kappa}}$ in the
integrand, and secondly the integral over the contours between the
points $y_{m}^{\pm}$ and $y_{m+1}^{\pm}$ can be rescaled to produce
(modulo error terms that can be neglected in the limit $y_{m}^{\pm},y_{m+1}^{\pm}\rightarrow y'$)
a generalized beta-function
\begin{align*}
B_{1} = \; &\int_{0}^{1}\ud w_{1}\int_{w_{1}}^{1}\ud w_{2}\; w_{1}^{-\frac{8}{\kappa}}w_{2}^{-\frac{8}{\kappa}}(w_{2}-w_{1})^{\frac{8}{\kappa}}(1-w_{1})^{-\frac{8}{\kappa}}(1-w_{2})^{-\frac{8}{\kappa}}
 \quad \text{ or } \\
 B_{3} = \; & \int_{0}^{1}\ud w\; w^{-\frac{8}{\kappa}}(1-w)^{-\frac{8}{\kappa}} 
 \quad \text{ or } \quad 
  B_5 = 1
\end{align*}
times a power law $|y_{m+1}^{\pm}-y_{m}^{\pm}|^{\Delta_{l}}$
with $\Delta_{l}=l+\frac{8}{\kappa}(\frac{(l-1)l}{2}-2l)$ according
to the number $l=2,1,0$ of integration variables on contours between
the points $y_{m}^{\pm}$ and $y_{m+1}^{\pm}$. For the rest of the
integrations, we may combine the factors in the integrand containing
the variables $y_{m}^{\pm},y_{m+1}^{\pm}$ or any of the integration
variables between them, and we get a function of the same type, with
fewer variables. The different possibilities for $x,y_{1}^{\pm}$
are treated in an entirely parallel fashion.

The multiplicative constants $B_1$, $B_3$, and $B_2$
appear in the derivation as integrals which are a priori convergent only for $\kappa>8$.
The assertions nevertheless remain true by analytic continuation also in the cases of
interest $0 < \kappa < 8$, and we have given the constants $B_1$, $B_3$, and $B_2$
explicitly as generalized beta functions which are well defined, non-zero, and analytic in
$\kappa$ apart from certain rational values of $\kappa$.

\subsubsection{Highest weight vectors and closed integration surfaces\label{sec: closed homology classes}}

For fundamental properties of the Dotsenko-Fateev functions in Section
\ref{sec: general Coulomb gas integrals}, it was important that the
integration surface $\Gamma$ was closed in an appropriate homology
related to the multivalued integrand \eqref{eq: Coulomb gas integrand},
see \cite{FW-topological_representation_of_Uqsl2}. Our basis functions
$\varphi_{t_{L}^{-},\ldots,t_{1}^{-},d,t_{1}^{+},\ldots,t_{R}^{+}}$
for the spin chain~-~Coulomb gas correspondence, introduced in Section
\ref{sec: spin chain Coulomb gas basis}, are obtained by integrals
along the contours $\Gamma_{t_{L}^{-},\ldots,t_{1}^{-},d,t_{1}^{+},\ldots,t_{R}^{+}}$
of Figure \ref{fig: FW integration contour}, which do not constitute
a closed surface. Remarkably, however, Felder and Wieczerkowski \cite{FW-topological_representation_of_Uqsl2}
showed that if the vector $v$ is annihilated by the quantum group
generator $E$, i.e., if $v$ is a sum of highest weight vectors of
subrepresentations of $\Wd_3^{\tens R} \tens \Wd_2 \tens \Wd_3^{\tens L}$,
then the homology class of the associated linear
combination of $\Gamma_{t_{L}^{-},\ldots,t_{R}^{+}}$ is closed.
Less abstractly, as in
\cite[Proposition 4.5 and Corollary 4.8]{KP-covariant_boundary_correlations},
this can be viewed as a generalization of the manipulations
of the integrals we described in the end of Section \ref{sec: asymptotics via correspondence},
and we exhibit this property very concretely by transforming the integrals
to integrals along the real axis in Section \ref{sec: transformation to real integrals}.

Importantly, if $v\in\Wd_{3}^{\tens R}\tens\Wd_{2}\tens\Wd_{3}^{\tens L}$
satisfies $E.v=0$, then the associated function $\varphi$ has the
following properties:
\begin{itemize}
\item The function $\varphi$ does not depend on the choice of the anchor
point $z_{0}$ of the contours $\Gamma_{t_{L}^{-},\ldots,t_{R}^{+}}$.
\item The function $\varphi$ satisfies the second order differential equation
\eqref{eq: second order differential equation}.
\item The function $\varphi$ satisfies the third order differential equations
\eqref{eq: third order differential equations}.
\end{itemize}
Generalizations and formal proofs are given in
\cite[Propositions 4.5 and 4.12]{KP-covariant_boundary_correlations}.

\subsection{Linear problem in quantum group representations\label{sec: quantum group problem}}

Recall that we are looking for solutions to the partial differential
equations \eqref{eq: translation invariance for zig-zag}, \eqref{eq: scaling covariance for zig-zag},
\eqref{eq: second order differential equation}, \eqref{eq: third order differential equations},
with boundary conditions specified in terms of the asymptotics \eqref{eq: successively visited},
\eqref{eq: non-successively visited}, \eqref{eq: the very first visited point},
\eqref{eq: first opposite visited point}. We will produce the solution
by the spin chain~-~Coulomb gas correspondence of Section \ref{sec: definition of correspondence}:
we will find a vector $v$ so that the associated function $\varphi$
solves the problem.

More precisely, for all order specifications $\omega\in\set{+,-}^{N}$
with $R$ ``$+$''-symbols and $L$ ``$-$''-symbols, we want
vectors
\begin{align*}
v_{\omega}^{(N)}\in\; & \Wd_{3}^{\tens R}\tens\Wd_{2}\tens\Wd_{3}^{\tens L}
\end{align*}
such that the function associated to $v_{\omega}$ by the spin chain~-~Coulomb
gas correspondence is the boundary zig-zag
amplitude $\Ampl_{\omega}(y_{L}^{-},\ldots,y_{1}^{-};x;y_{1}^{+},\ldots,y_{R}^{+})$.
This will be achieved if the vectors $v_{\omega}$ satisfy the following
conditions, written in terms of the projections $\pi$, $\hat{\pi}$ defined in
Appendix~\ref{sec: explicit submodules}:
\begin{itemize}
\item \emph{Highest weight vector of a doublet subrepresentation}: 
\begin{align}
E.v_{\omega}^{(N)}=\; & 0\label{eq: highest weight vector equation}\\
K.v_{\omega}^{(N)}=\; & q\, v_{\omega}^{(N)}.\label{eq: K eigenvalue equation}
\end{align}

\item \emph{Projections to singlet and triplet for successively visited
points}: If $y_{m}^{\rgtlft}$ and $y_{m+1}^{\rgtlft}$ are successively
visited points on the same side ($y_{m}^{\rgtlft}=y_{j}$ and $y_{m+1}^{\rgtlft}=y_{j+1}$),
then
\begin{align}
\pi_{\pm;m}^{(1)}(v_{\omega}^{(N)})=\; & 0\label{eq: QG successively visited}\\
\hat{\pi}_{\pm;m}^{(3)}(v_{\omega}^{(N)})=\; & \const\times v_{\omega'}^{(N-1)},\nonumber 
\end{align}
where $\omega'=(\omega_{1},\omega_{2},\ldots,\omega_{j-1},\omega_{j},\omega_{j+2},\omega_{j+3},\ldots,\omega_{N})$.
\item \emph{Projections to singlet and triplet for non-successively visited
points}: If $y_{m}^{\rgtlft}$ and $y_{m+1}^{\rgtlft}$ are non-successively
visited consecutive points on the same side ($y_{m}^{\rgtlft}=y_{j}$
and $y_{m+1}^{\rgtlft}=y_{k}$ with $k-j>1$), then
\begin{align}
\pi_{\pm;m}^{(1)}(v_{\omega}^{(N)})=\; & 0\label{eq: QG non-successively visited}\\
\pi_{\pm;m}^{(3)}(v_{\omega}^{(N)})=\; & 0.\nonumber 
\end{align}

\item \emph{Projections to doublet for the first points on the left and
right}: Let $\rgtlft$ denote the side of the first visited point,
$y_{1}=y_{1}^{\rgtlft}$, and $\mp$ the opposite side. For the first
visited point the condition is 
\begin{align}
\hat{\pi}_{\pm}^{(2)}(v_{\omega}^{(N)})=\; & \const\times v_{\omega'}^{(N-1)},\label{eq: QG the very first visited point}
\end{align}
where $\omega'=(\omega_{2},\omega_{3},\ldots,\omega_{N})$. For the
first point on the opposite side the condition is
\begin{align}
\pi_{\mp}^{(2)}(v_{\omega}^{(N)})=\; & 0.\label{eq: QG first opposite visited point}
\end{align}

\end{itemize}
By the closed integration surface considerations of Section \ref{sec: closed homology classes},
Equation \eqref{eq: highest weight vector equation} guarantees that
the function associated to $v_{\omega}$ is independent of the anchor
point and satisfies the PDEs \eqref{eq: second order differential equation},
\eqref{eq: third order differential equations}. The translation invariance
\eqref{eq: translation invariance for zig-zag} is then obvious. Equation
\eqref{eq: K eigenvalue equation} guarantees that the associated
function is a linear combination of $\varphi_{t_{L}^{-},\ldots,t_{1}^{-},d,t_{1}^{+},\ldots,t_{R}^{+}}$
with $d+\sum_{j}t_{j}^{-}+\sum_{j}t_{j}^{+}=N$, and therefore by
the results of Section \ref{sec: general Coulomb gas integrals},
it has the correct scaling covariance \eqref{eq: scaling covariance for zig-zag}.
Finally, by the asymptotics properties of Section \ref{sec: asymptotics via correspondence},
we see that Equations \eqref{eq: QG successively visited}, \eqref{eq: QG non-successively visited},
\eqref{eq: QG the very first visited point}, \eqref{eq: QG first opposite visited point},
guarantee \eqref{eq: successively visited}, \eqref{eq: non-successively visited},
\eqref{eq: the very first visited point}, \eqref{eq: first opposite visited point},
respectively.

As for the choice of multiplicative normalization, we first make an
explicit choice for the cases $N=1$ in Section \ref{sec: 1pt quantum group solution}.
The rest of the multiplicative factors are fixed recursively in $N$
by requiring that the constant appearing on the right hand side of
Equation \eqref{eq: QG the very first visited point} is equal to
one. This corresponds to fixing the multiplicative constant in Equation
\eqref{eq: the very first visited point} to the value $B_{2}$ given
in \eqref{eq: B2 formula}.

\subsection{Solutions in terms of quantum group representations\label{sec: quantum group solution}}

A priori, the system of equations \eqref{eq: highest weight vector equation},
\eqref{eq: K eigenvalue equation}, \eqref{eq: QG successively visited},
\eqref{eq: QG non-successively visited}, \eqref{eq: QG the very first visited point},
\eqref{eq: QG first opposite visited point} given in Section \ref{sec: quantum group problem}
is a linear algebra problem in the $2\times3^{N}$-dimensional tensor
product space $\Wd_{3}^{\tens R}\tens\Wd_{2}\tens\Wd_{3}^{\tens L}$.
The first two equations \eqref{eq: highest weight vector equation},
\eqref{eq: K eigenvalue equation} reduce this ambient dimension in
a well understood way: their meaning is that $v_{\omega}$ is a highest
weight vector of a subrepresentation of dimension two in the tensor
product. We have
\begin{align*}
\dmn\Big(\Kern(E)\cap\Kern(K-q)\Big)=\; & m_{N},
\end{align*}
where $m_{N}$ is the multiplicity of $\Wd_{2}$ in the semisimple
decomposition of the tensor product, determined recursively by the
formula of Section \ref{sec: irreducible representations}. 
Explicitly for small $N$ and asymptotically as $N \to \infty$,
the multiplicities $m_N$ are
\begin{align*}
 & \begin{array}{c|ccccccccccccc}
N & 1 & 2 & 3 & 4 & 5 & 6 & 7 & 8 & 9 & 10 &  \cdots\\
\hline m_{N} & 1 & 2 & 4 & 9 & 21 & 51 & 127 & 323 & 835 & 2188 &  \cdots
\end{array}\;,
& m_{N}\sim\; & 3\sqrt{\frac{3}{4\pi}}\times N^{-\frac{3}{2}}\,3^{N} .
\end{align*}

Superficially the system of Section \ref{sec: quantum group problem}
still seems overdetermined, but we find that in each case the solution
space is one-dimensional, so up to multiplicative normalizations the
solutions are unique.

Next we give the explicit solutions to the system of equations for
a few small values of $N$.

\subsubsection{One-point solutions\label{sec: 1pt quantum group solution}}

There are two separate states that we need to solve, $v_{\lft}^{(1)}\in\Wd_{2}\tens\Wd_{3}$
for a visit on the left ($y_{1}<x$), and $v_{\rgt}^{(1)}\in\Wd_{3}\tens\Wd_{2}$
for a visit on the right ($x<y_{1}$). The solutions, unique up to
normalization, are
\begin{align}
v_{\lft}^{(1)}=\; & \frac{q^{4}}{1-q^{4}}\;\Wbas_{0}\tens\Wbas_{1}-\frac{q}{1-q^{2}}\;\Wbas_{1}\tens\Wbas_{0}\label{eq: 1pt vector L}\\
v_{\rgt}^{(1)}=\; & \frac{q^{2}}{1-q^{2}}\;\Wbas_{0}\tens\Wbas_{1}-\frac{q^{2}}{1-q^{4}}\;\Wbas_{1}\tens\Wbas_{0}.\label{eq: 1pt vector R}
\end{align}
The normalization above has been chosen such that the corresponding
functions both are equal to
\begin{align*}
\zeta^{(1)}(x;y_{1})=\; & B_{2}\,|y_{1}-x|^{1-\frac{8}{\kappa}},
\end{align*}
where the constant $B_{2}$ is given by \eqref{eq: B2 formula} (in
particular both functions take positive real values). The calculation
of the corresponding integrals is discussed in more detail in Section~\ref{sec: 1pt real integrals}.

\subsubsection{Two-point solutions\label{sec: 2pt quantum group solution}}

There are four separate states that we need to solve for,
\begin{align*}
v_{\lft\lft}^{(2)}\in\Wd_{2}\tens\Wd_{3}\tens\Wd_{3} & \qquad\qquad(y_{2}<y_{1}<x)\\
v_{\lft\rgt}^{(2)}\in\Wd_{3}\tens\Wd_{2}\tens\Wd_{3} & \qquad\qquad(y_{1}<x<y_{2})\\
v_{\rgt\lft}^{(2)}\in\Wd_{3}\tens\Wd_{2}\tens\Wd_{3} & \qquad\qquad(y_{2}<x<y_{1})\\
v_{\rgt\rgt}^{(2)}\in\Wd_{3}\tens\Wd_{3}\tens\Wd_{2} & \qquad\qquad(x<y_{1}<y_{2}).
\end{align*}
For the normalization of the states, we use the asymptotics as $|y_{1}-x|\rightarrow0$,
i.e., we fix the constant in either \eqref{eq: the very first visited point}
or \eqref{eq: QG the very first visited point}.

The solutions, unique with the chosen normalization, read
\begin{align*}
v_{\rgt\rgt}^{(2)}=\; & \frac{q^{4}(1+q^{2}+q^{4})}{(1-q^{4})^{2}(1+q^{4})}\Big((q^{2}+q^{4})\Wbas_{011}-\Wbas_{020}-(1+q^{2})\Wbas_{101}-(1-q^{2})\Wbas_{110}+\Wbas_{200}\Big)\\
v_{\lft\lft}^{(2)}=\; & \frac{q^{3}(1+q^{2}+q^{4})}{(1-q^{4})^{2}(1+q^{4})}\Big(q^{3}\Wbas_{002}+(q^{5}-q^{3})\Wbas_{011}-q^{3}\Wbas_{020}-q^{2}\Wbas_{101}-q^{4}\Wbas_{101}+(1+q^{2})\Wbas_{110}\Big)\\
v_{\rgt\lft}^{(2)}=\; & \frac{q^{3}(1+q^{2}+q^{4})}{(1-q^{4})^{2}(1+q^{4})}\Big(\frac{q^{4}}{1+q^{2}}\Wbas_{002}+q^{5}\Wbas_{011}-q^{3}\Wbas_{101}-q^{4}\Wbas_{110}+\frac{1+q^{2}+q^{4}}{1+q^{2}}\Wbas_{200}\Big)\\
v_{\lft\rgt}^{(2)}=\; & \frac{q^{3}(1+q^{2}+q^{4})}{(1-q^{4})^{2}(1+q^{4})}\Big(\frac{q^{2}(1+q^{2}+q^{4})}{1+q^{2}}\Wbas_{002}-q\Wbas_{011}-q^{3}\Wbas_{101}+\Wbas_{110}+\frac{q^{2}}{1+q^{2}}\Wbas_{200}\Big),
\end{align*}
where we use the shorthand notation $\Wbas_{t_{2}t_{1}d}=\Wbas_{t_{2}}\tens\Wbas_{t_{1}}\tens\Wbas_{d}\in\Wd_{3}\tens\Wd_{3}\tens\Wd_{2}$
in the first case, and similarly for the rest.

\subsubsection{Three-point solutions\label{sec: 3pt quantum group solution}}

For $N=3$ there are eight separate states that we need to solve for.
For brevity, in the formulas below, we factor out the constant

\begin{align*}
C_{3}=\; & \frac{q^{5}\left(q^{4}+q^{2}+1\right)^{2}}{\left(q^{4}-1\right)^{3}\left(q^{12}+q^{10}+2q^{8}+2q^{6}+2q^{4}+q^{2}+1\right)}.
\end{align*}
Then, with a shorthand notation similar to above, the unique normalized
solutions are
\begin{align*}
v_{\rgt+\rgt}^{(3)}=\; & C_{3}\times\Bigg(-\left(q^{6}+2q^{4}+2q^{2}+1\right)q^{3}\Wbas_{0021}-\left(q^{2}+1\right)\left(q^{6}+q^{4}-1\right)q^{3}\Wbas_{0111}+\left(q^{2}+1\right)^{2}q^{3}\Wbas_{0120}\\
 & \qquad+\left(q^{2}+1\right)^{2}q^{3}\Wbas_{0201}+\left(q^{6}-q^{2}-1\right)q\Wbas_{0210}+\left(q^{3}+q\right)^{3}\Wbas_{1011}+\left(q^{6}-q^{2}-1\right)q\Wbas_{1020}\\
 & \qquad+\left(q^{2}+1\right)\left(q^{6}-q^{2}-1\right)q\Wbas_{1101}+\left(q^{9}-q^{7}-2q^{5}-q^{3}+q\right)\Wbas_{1110}-\left(q^{6}+q^{4}-1\right)q\Wbas_{1200}\\
 & \qquad-\left(q^{6}+2q^{4}+2q^{2}+1\right)q\Wbas_{2001}-\left(q^{6}+q^{4}-1\right)q\Wbas_{2010}+\left(q^{2}+1\right)^{2}q\Wbas_{2100}\Bigg)
\end{align*}
\begin{align*}
v_{\rgt+-}^{(3)}=\; & C_{3}\times\Bigg(-\left(q^{4}+q^{2}+1\right)q^{4}\Wbas_{0012}+\left(-q^{8}+q^{2}+\frac{1}{q^{2}+1}-1\right)\Wbas_{0102}-\left(q^{6}+q^{4}-1\right)q^{5}\Wbas_{0111}\\
 & \qquad+\left(q^{7}+q^{5}\right)\Wbas_{0201}+\left(q^{8}+q^{6}\right)\Wbas_{0210}+\left(q^{6}+q^{4}\right)\Wbas_{1002}+\left(q^{2}+1\right)^{2}q^{5}\Wbas_{1011}\\
 & \qquad+\left(q^{6}-q^{2}-1\right)q^{3}\Wbas_{1101}+\left(q^{6}-q^{2}-1\right)q^{4}\Wbas_{1110}-\left(q^{4}+q^{2}+1\right)q^{4}\Wbas_{1200}\\
 & \qquad-\left(q^{4}+q^{2}+1\right)q^{3}\Wbas_{2001}-\left(q^{4}+q^{2}+1\right)q^{4}\Wbas_{2010}+\frac{\left(q^{4}+q^{2}+1\right)^{2}\Wbas_{2100}}{q^{2}+1}\Bigg)
\end{align*}

\begin{align*}
v_{\rgt-+}^{(3)}=\; & C_{3}\times\Bigg(-\left(q^{4}+1\right)\left(q^{4}+q^{2}+1\right)q^{2}\Wbas_{0012}+\left(q^{4}+\frac{q^{2}}{q^{2}+1}\right)\Wbas_{0102}+\left(q^{7}+q^{5}+q^{3}\right)\Wbas_{0111}\\
 & \qquad-q\Wbas_{0201}-q^{2}\Wbas_{0210}+\left(q^{6}+q^{2}+\frac{1}{q^{2}+1}-1\right)\Wbas_{1002}+\left(q^{9}+q^{7}+q^{5}\right)\Wbas_{1011}\\
 & \qquad-\left(q^{2}+1\right)q^{3}\Wbas_{1101}-\left(q^{2}+1\right)q^{4}\Wbas_{1110}+\left(q^{2}+\frac{1}{q^{2}+1}\right)\Wbas_{1200}-q^{5}\Wbas_{2001}\\
 & \qquad+q^{6}\left(-\Wbas_{2010}\right)+\left(q^{4}+\frac{q^{2}}{q^{2}+1}\right)\Wbas_{2100}\Bigg)
\end{align*}
\begin{align*}
v_{\rgt--}^{(3)}=\; & C_{3}\times\Bigg(-\left(q^{2}+1\right)q^{6}\Wbas_{0012}+\frac{\left(-q^{10}+q^{6}+q^{4}\right)\Wbas_{0021}}{q^{2}+1}-\left(q^{2}+1\right)q^{7}\Wbas_{0102}\\
 & \qquad+\left(-q^{11}+q^{7}+q^{5}\right)\Wbas_{0111}+\left(q^{9}+q^{7}+q^{5}\right)\Wbas_{0120}+\left(q^{7}+q^{5}+q^{3}\right)\Wbas_{1002}\\
 & \qquad+\left(q^{6}+q^{4}-1\right)q^{3}\Wbas_{1011}-\left(q^{2}+1\right)q^{3}\Wbas_{1020}+\left(q^{6}+q^{4}-1\right)q^{4}\Wbas_{1101}-\left(q^{2}+1\right)^{2}q^{4}\Wbas_{1110}\\
 & \qquad-\frac{\left(q^{5}+q^{3}+q\right)^{2}\Wbas_{2001}}{q^{2}+1}+\left(q^{4}+q^{2}+1\right)\Wbas_{2010}+\left(q^{5}+q^{3}+q\right)\Wbas_{2100}\Bigg)
\end{align*}
\begin{align*}
v_{-\rgt+}^{(3)}=\; & C_{3}\times\Bigg(-\left(q^{4}+q^{2}+1\right)q^{4}\Wbas_{0012}-\left(q^{4}+q^{2}+1\right)q^{6}\Wbas_{0102}+\left(q^{2}+1\right)^{2}q^{3}\Wbas_{0111}\\
 & \qquad+\left(q^{7}+q^{5}\right)\Wbas_{0201}+\left(-q^{4}-q^{2}-1\right)\Wbas_{0210}+\frac{\left(q^{5}+q^{3}+q\right)^{2}\Wbas_{1002}}{q^{2}+1}+\left(q^{6}-q^{2}-1\right)q\Wbas_{1011}\\
 & \qquad+\left(q^{6}-q^{2}-1\right)q^{3}\Wbas_{1101}+\left(-q^{6}-q^{4}+1\right)\Wbas_{1110}+\left(\frac{q^{2}}{q^{2}+1}-q^{6}\right)\Wbas_{1200}\\
 & \qquad-\left(q^{4}+q^{2}+1\right)q^{3}\Wbas_{2001}+\left(q^{2}+1\right)\Wbas_{2010}+\left(q^{4}+q^{2}\right)\Wbas_{2100}\Bigg)
\end{align*}
\begin{align*}
v_{-+-}^{(3)}=\; & C_{3}\times\Bigg(-\frac{\left(q^{10}+q^{8}+q^{6}\right)\Wbas_{0012}}{q^{2}+1}-\frac{\left(q^{4}+q^{2}+1\right)q^{8}\Wbas_{0021}}{q^{2}+1}+q^{3}\Wbas_{0102}\\
 & \qquad+\left(q^{7}+q^{5}\right)\Wbas_{0111}+q^{7}\Wbas_{0120}+q^{5}\Wbas_{1002}+\left(q^{9}+q^{7}\right)\Wbas_{1011}+q^{9}\Wbas_{1020}\\
 & \qquad-\left(q^{4}+q^{2}+1\right)q^{2}\Wbas_{1101}-\left(q^{4}+q^{2}+1\right)q^{4}\Wbas_{1110}-\frac{\left(q^{8}+q^{6}+q^{4}\right)\Wbas_{2001}}{q^{2}+1}\\
 & \qquad-\frac{\left(q^{10}+q^{8}+q^{6}\right)\Wbas_{2010}}{q^{2}+1}+\left(q^{7}+q^{5}+2q^{3}+q+\frac{1}{q}\right)\Wbas_{2100}\Bigg)
\end{align*}
\begin{align*}
v_{--+}^{(3)}=\; & C_{3}\times\Bigg(-\frac{\left(q^{4}+q^{2}+1\right)^{2}q^{4}\Wbas_{0012}}{q^{2}+1}+\left(q^{6}+q^{4}+q^{2}\right)\Wbas_{0021}+\left(q^{5}+q^{3}+q\right)\Wbas_{0102}\\
 & \qquad+\left(q^{6}+q^{4}-1\right)q\Wbas_{0111}-\left(q^{2}+1\right)q\Wbas_{0120}+\left(q^{7}+q^{5}+q^{3}\right)\Wbas_{1002}+\left(q^{6}+q^{4}-1\right)q^{3}\Wbas_{1011}\\
 & \qquad-\left(q^{2}+1\right)q^{3}\Wbas_{1020}-\left(q^{3}+q\right)^{2}\Wbas_{1101}+\left(-q^{6}+q^{2}+1\right)\Wbas_{1110}-\left(q^{2}+1\right)q^{4}\Wbas_{2001}\\
 & \qquad+\frac{\left(-q^{8}+q^{4}+q^{2}\right)\Wbas_{2010}}{q^{2}+1}+\left(q^{5}+q^{3}+q\right)\Wbas_{2100}\Bigg)
\end{align*}
\begin{align*}
v_{---}^{(3)}=\; & C_{3}\times\Bigg(-\left(q^{2}+1\right)^{2}q^{5}\Wbas_{0012}+\left(-q^{9}+q^{5}+q^{3}\right)\Wbas_{0021}+\left(-q^{9}+q^{5}+q^{3}\right)\Wbas_{0102}\\
 & \qquad+\left(-q^{8}+q^{6}+2q^{4}+q^{2}-1\right)q^{3}\Wbas_{0111}+\left(q^{6}+q^{4}-1\right)q^{3}\Wbas_{0120}+\left(q^{6}+q^{4}-1\right)q^{3}\Wbas_{0201}\\
 & \qquad-\left(q^{2}+1\right)^{2}q^{3}\Wbas_{0210}+\left(q^{2}+1\right)\left(q^{4}+q^{2}+1\right)q^{2}\Wbas_{1002}+\left(q^{2}+1\right)\left(q^{6}+q^{4}-1\right)q^{2}\Wbas_{1011}\\
 & \qquad-\left(q^{3}+q\right)^{2}\Wbas_{1020}-\left(q^{2}+1\right)^{3}q^{2}\Wbas_{1101}-\left(q^{2}+1\right)\left(q^{6}-q^{2}-1\right)\Wbas_{1110}+\left(q^{2}+1\right)\left(q^{4}+q^{2}+1\right)\Wbas_{1200}\Bigg)
\end{align*}

\subsubsection{Four-point solutions\label{sec: 4pt quantum group solution}}

For $N=4$ there are sixteen separate states that we need to solve
for. The solutions are again unique (with the chosen normalization).
In Appendix \ref{sub: 4-point QG solutions} we include the results
for those vectors that have been used in the plots of Figure \ref{fig: perco 4pt data}.

\subsubsection{Well-posedness of the problem\label{sec: existence and uniqueness of quantum group solutions}}

The linear problem of Section \ref{sec: quantum group problem} is
well-posed: one always finds solutions and they are unique (with the
chosen normalization). Up to $N=4$ this was explicitly stated above.

The uniqueness of solutions is checked by considering the homogeneous
equations for $N$-point vectors, where the inhomogeneous terms coming
from the $(N-1)$-point vectors on the right hand sides of Equations
\eqref{eq: QG successively visited} and \eqref{eq: QG the very first visited point}
are omitted, that is
\[ (K-q).v = 0 , \quad E.v =0 , \quad \pi^{(1)}_{\pm;m}(v) = 0 , \quad \pi^{(3)}_{\pm;m}(v) = 0 , \quad \pi^{(2)}_{\pm}(v) = 0 \qquad
\text{ for } v \in \Wd_3^{\tens R} \tens \Wd_2 \tens \Wd_3^{\tens L} . \]
The projection conditions here, i.e., the homogeneous versions of Equations \eqref{eq: QG successively visited},
\eqref{eq: QG non-successively visited}, \eqref{eq: QG the very first visited point},
\eqref{eq: QG first opposite visited point}, force the vector to
lie in the unique subrepresentation of the highest spin $\Wd_{2N+2} \subset \Wd_3^{\tens R} \tens \Wd_2 \tens \Wd_3^{\tens L}$.
On the other hand, the first two equations, i.e., Equations \eqref{eq: highest weight vector equation}, \eqref{eq: K eigenvalue equation},
force the solution to lie in a doublet $\Wd_{2} \subset \Wd_3^{\tens R} \tens \Wd_2 \tens \Wd_3^{\tens L}$.
The doublet subrepresentation and the subrepresentation of highest spin intersect only at zero.
The homogeneous problem therefore has no non-zero solutions, which shows uniqueness.

The easiest way to prove the existence of solutions for all $N$ seems
to be by exhibiting an algorithm, which recovers the solutions to
our problem from the solutions 
to a slightly simpler similar problem related to multiple SLEs.
This is done in detail in \cite[Section 5]{KP-pure_partition_functions_of_multiple_SLEs}.

\bigskip{}

\section{Regularized real integrals and evaluation of the formulas\label{sec: regularized real integrals and evaluation}}

\subsection{Transformation to real integration contours\label{sec: transformation to real integrals}}

Let us then analyze further the integrals $\varphi_{t_{L}^{\lft},\ldots,t_{2}^{\lft},t_{1}^{\lft};d;t_{1}^{\rgt},t_{2}^{\rgt},\ldots,t_{R}^{\rgt}}$
given by the spin chain~-~Coulomb gas correspondence. Recall that
the integral was defined in Section~\ref{sec: spin chain Coulomb gas basis},
where the integration surface $\Gamma$ consists of non-intersecting
loop contours for each of the integration variables $w_{s}$ as depicted
in Figure~\ref{fig: FW integration contour}.

First we shall describe a transformation of the contours which makes
the integrands explicitly real in general, and examples will follow
below. 
The procedure is, in principle, straightforward. We assume that the
anchor point~$z_{0}$ of the loop integrals lies on the real axis
left of the points $x$ and $y_{L}^{\lft}$. (As stated in Section~\ref{sec: closed homology classes}
and as we shall see below, the integrals of interest to us in the
end are independent of this anchor point.) We can then deform the
loop-shaped contours so that they follow the real line, starting from
the innermost loops on the left and proceeding towards right.

There is, however, a complication as the integrals along the real
axis may become singular. Notice that as any of the integration variables
$w_{i}$ approaches any of the points $y_{j}$, the integrand behaves
as $\sim|w_{i}-y_{j}|^{1-8/\kappa}$. Thus the resulting integrals
will be convergent if $\kappa>8$.
For simplicity let us therefore first assume
that $\kappa>8$, although
for the application to SLE boundary visit amplitudes we are
ultimately interested in $\kappa < 8$.
We will discuss the divergences and the needed regularization for $\kappa<8$
in Section~\ref{sec: divergences of integrals}.

When $\kappa>8$, a loop contour enclosing, for example, $y_{k}^{\lft}$
can be divided into $2(L-k+1)$ subcontours on the real line. We get
two contours (one from both the lower and the upper edges of the loop)
between the base point and $y_{L}^{\lft}$ as well as between all
consecutive pairs $\{y_{j}^{\lft},y_{j+1}^{\lft}\}$ with $j=k,\ldots,L-1$.
The corresponding (one-dimensional) integral thus becomes a sum of
integrals over the real line. Extending this procedure to the loops
enclosing $x$ and the points $y_{k}^{\rgt}$ right of $x$, each
integral $\varphi_{t_{L}^{\lft},\ldots,t_{2}^{\lft},t_{1}^{\lft};d;t_{1}^{\rgt},t_{2}^{\rgt},\ldots,t_{R}^{\rgt}}$
can be written as a linear combination of integrals having all integration
contours on the real line.

\begin{figure}
\begin{centering}
\includegraphics[width=1\textwidth]{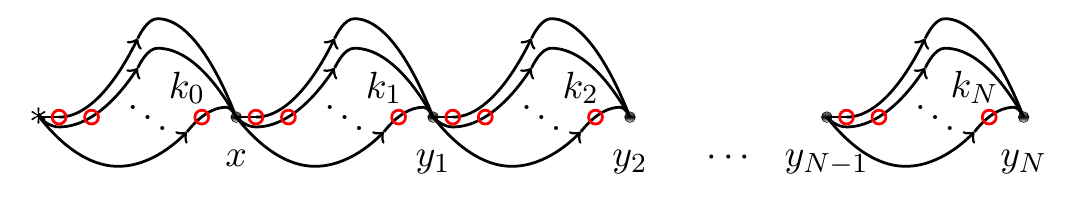}
\par\end{centering}

\caption{\emph{The integration contours of the $w_{j}$-variables for }$\hat{\rho}_{;k;k_{1}^{+},k_{2}^{\rgt},\ldots,k_{N}^{\rgt}}$\emph{
and the point (marked by red circles) where the integrand is rephased
to be positive. The contours in this figure have been deformed away
from the real axis, for the sake of clarity of the phase convention.\label{fig: multiple interval integration contour}}}
\end{figure}

In order to obtain the explicit linear combination of the integrals,
the remaining and most non-trivial task is to calculate the phase
factors which arise as the integrand is a multi-valued function. The
phase convention for the integrand $\FWintegrand_{t_{L}^{\lft},\ldots,t_{R}^{\rgt}}$
of \eqref{eq: basis functions} for the loop contours was defined
by the red circles in Figure~\ref{fig: FW integration contour},
and this convention leads to rather impractical branch choices for
the integrand as the contours are transformed. We shall choose the
phases for the contours along the real line as depicted by the red
circles in Figure~\ref{fig: multiple interval integration contour},
where the integration contours have been deformed away from the real
line in order to make their multiplicity and the phase convention
visible. Let us denote these integrals by $\hat{\rho}_{k_{L}^{\lft},\ldots,k_{2}^{\lft},k_{1}^{\lft};k;k_{1}^{\rgt},k_{2}^{\rgt},\ldots,k_{R}^{\rgt}}$,
when the number of variables integrated from the anchor to $y_{L}^{-}$
is $k_{L}^{-}$, the number of variables integrated from $y_{L}^{-}$
to $y_{L-1}^{-}$ is $k_{L-1}^{-}$ and so on (we thus choose to index
the integrals in terms of the rightmost points of the integration
intervals). It is not worthwhile to write down a general formula for
the phase factors which appear when expressing each $\varphi$ as
a sum of the integrals $\hat{\rho}$, but it is straightforward to
calculate them case by case as seen in the examples below. As the
phase factors reflect the branch choices of the integrand in~\eqref{eq: integral of Coulomb gas type general l},
they will be integer powers of $q=\exp(4\pi\ii/\kappa)$, possibly
multiplied by $-1$ if the direction of integration needs to be reversed.

As the final step, we arrange the integration over each interval such
that the integration variables have a fixed order. The natural phase
convention in this case is that the integrand is real and positive.
We denote these integrals by $\rho_{k_{L}^{\lft},\ldots,k_{2}^{\lft},k_{1}^{\lft};k;k_{1}^{\rgt},k_{2}^{\rgt},\ldots,k_{R}^{\rgt}}$.
The integrals are over products of simplexes of dimensions $k_{L}^{\lft},\ldots,,k_{1}^{\lft};k;k_{1}^{\rgt},,\ldots,k_{R}^{\rgt}$,
for example when $L=0$, $R=N$ and $K=k+\sum_{j=1}^{N}k_{j}^{+}$
we have
\begin{align*}
\rho_{;k;k_{1}^{+},k_{2}^{\rgt},\ldots,k_{N}^{\rgt}}(;x;y_{1},\ldots,y_{N})=\; & \underset{\substack{x<w_{1}<w_{2}<w_{3}<\cdots<w_{k}<y_{1}\\
y_{1}<w_{k+1}<\cdots<w_{k+k_{1}}<y_{2}\\
\vdots\qquad\qquad\qquad\vdots\\
y_{N-1}<w_{K-k_{N}+1}<\cdots<w_{K}<y_{N}
}
}{\idotsint}\ud w_{1}\ud w_{2}\cdots\ud w_{K}\;\big|f_{N}^{(N)}(x;y_{1},\ldots,y_{N};w_{1},\ldots,w_{N})\big|.
\end{align*}
The reordering gives a factor of $\qnum k!\, q^{-k(k-1)/2}$
for each interval with $k$ integrations, where $\qnum k!=\prod_{m=1}^{k}\qnum m$
is a $q$-factorial (see \cite[Lemma 3.2]{KP-covariant_boundary_correlations} for details). Thus, we have 
\begin{align*}
\hat{\rho}_{k_{L}^{\lft},\ldots,k_{2}^{\lft},k_{1}^{\lft};k;k_{1}^{\rgt},k_{2}^{\rgt},\ldots,k_{R}^{\rgt}}=\; & \prod_{j=1}^{L}\left[k_{j}^{\lft}\right]!\, q^{-k_{j}^{\lft}\left(k_{j}^{\lft}-1\right)/2}\times\left[k\right]!\, q^{-k\left(k-1\right)/2}\\
 & \times\prod_{j=1}^{R}\left[k_{j}^{\rgt}\right]!\, q^{-k_{j}^{\rgt}\left(k_{j}^{\rgt}-1\right)/2}\times\rho_{k_{L}^{\lft},\ldots,k_{2}^{\lft},k_{1}^{\lft};k;k_{1}^{\rgt},k_{2}^{\rgt},\ldots,k_{R}^{\rgt}}.
\end{align*}

\subsection{Solutions in terms of real integrals\label{sec: solutions with real integrals}}

Let us then calculate explicitly the solutions obtained in Section~\ref{sec: quantum group solution}
for low numbers of boundary visits $N$. We shall discuss in detail
the case $N=1$, and list the results for the solutions with a higher
number of points. Case by case, we will check that the obtained solutions
for the boundary zig-zag amplitudes satisfy the following two requirements: 
\begin{itemize}
\item The integration contour $\Gamma$ is closed, and therefore the solution
is independent of the choice of the anchor point of the loop integrals.
When the amplitude is expressed in terms of the integrals $\rho_{k_{L}^{\lft},\ldots,k_{1}^{\lft};k;k_{1}^{\rgt},\ldots,k_{R}^{\rgt}}$
(or a similar $\hat{\rho}$), this will be clear as the solutions
do not depend on the integrals which include integrations starting
from the base point --- we will only have terms with $k_{L}^{-}=0$
(or if $L=0$ then $k=0$). 
\item The solution is real: when expressed in terms of the integrals $\rho$,
all coefficients will be real.
\end{itemize}

\subsubsection{One-point solutions\label{sec: 1pt real integrals}}

We start from the $N=1$ case where the single visit takes place right
of the starting point $x$. In this case we found that the state $v_{\rgt}^{(1)}\in\Wd_{3}\tens\Wd_{2}$
in \eqref{eq: 1pt vector R} which satisfies the constraints is 
\begin{align*}
v_{\rgt}^{(1)}=\;\frac{q^{2}}{1-q^{2}}\;\Wbas_{0}\tens\Wbas_{1}-\frac{q^{2}}{1-q^{4}}\;\Wbas_{1}\tens\Wbas_{0}.
\end{align*}
By the spin chain~-~Coulomb gas correspondence of Section \ref{sec: definition of correspondence},
the zig-zag probability amplitude is given by 
\begin{align*}
\Ampl_{\rgt}^{(1)}(x;y_{1}^{\rgt})=\;\frac{q^{2}}{1-q^{2}}\;\varphi_{;1;0}(x;y_{1}^{\rgt})-\frac{q^{2}}{1-q^{4}}\;\varphi_{;0;1}(x;y_{1}^{\rgt}).
\end{align*}

Let us then do the transformation to the integrals along the real
line. The first term $\varphi_{;1;0}$ has the loop integral encircling
$x$, which can only lead to integrals over the real line between
the base point and $x$, i.e., the integral $\hat{\rho}_{;1;0}$.
The phase factor from the lower edge of the loop is $q=e^{4\pi\ii/\kappa}$
(as the phase conventions of Figures~\ref{fig: FW integration contour}
and~\ref{fig: multiple interval integration contour} differ by a
rotation of the integration variable $w$ around $x$ by the angle
$-\pi$), whereas the phase factor for the upper edge of the loop
is $-q^{-1}$ (where the rotation is in the opposite direction, and
the minus sign arises from reversing the direction of integration).
Together, 
\begin{align*}
\varphi_{;1;0}(x;y_{1}^{\rgt})=\;\left(q-\frac{1}{q}\right)\hat{\rho}_{;1;0}(x;y_{1}^{\rgt}).
\end{align*}

The other loop integral $\varphi_{;0;1}$ breaks into four integrals
along the intervals on the real axis, two integrals between the base
point and $x$, and two integrals between $x$ and $y_{1}^{\rgt}$.
The phase factors can be calculated analogously to the case of $\varphi_{;1;0}$,
and they are integer powers of $q$. We find that 
\begin{align}
\varphi_{;0;1}(x;y_{1}^{\rgt})=\;\left(q^{2}-\frac{1}{q^{2}}\right)\hat{\rho}_{;0;1}(x;y_{1}^{\rgt})+\left(q^{3}-\frac{1}{q}\right)\hat{\rho}_{;1;0}(x;y_{1}^{\rgt}).\label{eq: phi01transf}
\end{align}

Substituting in these results, we get 
\begin{align*}
\Ampl_{\rgt}^{(1)}(x;y_{1}^{\rgt})=\;\hat{\rho}_{;0;1}(x;y_{1}^{\rgt})=\;\rho_{;0;1}(x;y_{1}^{\rgt}).
\end{align*}
In particular, the contributions of the integral $\hat{\rho}_{;1;0}$
cancel. The remaining integral $\hat{\rho}_{;0;1}$ is independent
of the anchor point of the loop contours, which shows that the contour
$\Gamma$ was closed. In this case there is only one integration variable,
so trivially $\hat{\rho}_{;0;1}=\rho_{;0;1}$. From the final expression
we also see that the result is real.

When $N=1$ the resulting integral can be calculated easily. Using
the definitions from~\eqref{eq: integral of Coulomb gas type general l},
\begin{align*}
\zeta_{\rgt}^{(1)}(x;y_{1}^{\rgt})=\;(y_{1}^{\rgt}-x)^{4/\kappa}\int_{x}^{y_{1}^{\rgt}}dw\,(w-x)^{-4/\kappa}(y_{1}^{\rgt}-w)^{-8/\kappa}=\; B_{2}\,(y_{1}^{\rgt}-x)^{1-8/\kappa},
\end{align*}
where the constant is given by the same beta function $B_{2}=B\left(\frac{\kappa-8}{\kappa},\frac{\kappa-4}{\kappa}\right)=\Gamma(\frac{\kappa-4}{\kappa})\Gamma(\frac{\kappa-8}{\kappa})/\Gamma(2\frac{\kappa-6}{\kappa})$
as in Equation~\eqref{eq: B2 formula}.

For comparison, let us also take a look at the case where the visit
takes place left of $x$. The state $v_{\lft}^{(1)}\in W_{2}\tens W_{3}$
was given in \ref{eq: 1pt vector L}, and by the correspondence we
get the probability amplitude 
\begin{align*}
\zeta_{\lft}^{(1)}(y_{1}^{\lft};x)=\; & \frac{q^{4}}{1-q^{4}}\;\varphi_{1;0;}(y_{1}^{\lft};x)-\frac{q}{1-q^{2}}\;\varphi_{0;1;}(y_{1}^{\lft};x).
\end{align*}

The transformations to real integrals read in this case 
\begin{align*}
\varphi_{1;0;}(y_{1}^{\lft};x)=\; & \left(q^{2}-\frac{1}{q^{2}}\right)\hat{\rho}_{1;0;}(y_{1}^{\lft};x)\\
\varphi_{0;1;}(y_{1}^{\lft};x)=\; & \left(q-\frac{1}{q}\right)\hat{\rho}_{0;1;}(y_{1}^{\lft};x)+\left(q^{3}-q\right)\hat{\rho}_{1;0;}(y_{1}^{\lft};x).
\end{align*}
Inserting these gives again a simple result 
\begin{align*}
\zeta_{\lft}^{(1)}(y_{1}^{\lft};x)=\; & \hat{\rho}_{0;1;}(y_{1}^{\lft};x)=\rho_{0;1;}(y_{1}^{\lft};x).
\end{align*}
This evaluates to 
\begin{align*}
\zeta_{\lft}^{(1)}(y_{1}^{\lft};x)=\; & (x-y_{1}^{\lft})^{4/\kappa}\int_{y_{1}^{\lft}}^{x}dw\,(w-y_{1}^{\lft})^{-8/\kappa}(x-w)^{-4/\kappa}=B_{2}\,(x-y_{1}^{\lft})^{1-8/\kappa}.
\end{align*}

The results for the left and right side visits can be collected in
the (well known) $N=1$ probability amplitude already stated in Equation~\eqref{eq: known 1-pt result},
\begin{align*}
\zeta^{(1)}(x;y_{1})=\; & \chi^{(1)}(x;y_{1})=\; B_{2}\,|y_{1}-x|^{1-\frac{8}{\kappa}},
\end{align*}
with our multiplicative normalization convention resulting in $B_{2}$
given in~\eqref{eq: B2 formula}.

\subsubsection{Two-point solutions}

Let us start the discussion of the two-point solutions from the case
where both visits take place on the right hand side. The relevant
vector $v_{\rgt\rgt}^{(2)}\in\Wd_{3}\tens\Wd_{3}\tens\Wd_{2}$ reads
\begin{align*}
v_{\rgt\rgt}^{(2)}=\; & \frac{q^{4}(1+q^{2}+q^{4})}{(1-q^{4})^{2}(1+q^{4})}\Big((q^{2}+q^{4})\Wbas_{011}-\Wbas_{020}-(1+q^{2})\Wbas_{101}+(1-q^{2})\Wbas_{110}+\Wbas_{200}\Big),
\end{align*}
where $\Wbas_{t_{2}t_{1}d}\equiv\Wbas_{t_{2}}\tens\Wbas_{t_{1}}\tens\Wbas_{d}$.
Thus the probability amplitude is 
\begin{align*}
\Ampl_{\rgt\rgt}^{(2)}(x;y_{1}^{\rgt},y_{2}^{\rgt})=\; & \frac{q^{4}(1+q^{2}+q^{4})}{(1-q^{4})^{2}(1+q^{4})}\Big((q^{2}+q^{4})\varphi_{;0;1,1}(x;y_{1}^{\rgt},y_{2}^{\rgt})-\varphi_{;0;2,0}(x;y_{1}^{\rgt},y_{2}^{\rgt})\\
 & \qquad-(1+q^{2})\varphi_{;1;0,1}(x;y_{1}^{\rgt},y_{2}^{\rgt})+(1-q^{2})\varphi_{;1;1,0}(x;y_{1}^{\rgt},y_{2}^{\rgt})\\
 & \qquad+\varphi_{;2;0,0}(x;y_{1}^{\rgt},y_{2}^{\rgt})\Big).
\end{align*}
The transformation to real integrals is still straightforward albeit
more involved, as one needs to take into account the phases related
to the order of the integration variables. The number of terms is
also larger, e.g., the integral $\varphi_{;0;2,0}$ breaks into $16$
different terms (some of which immediately cancel against each other).

Collecting the results in the expression for the probability amplitude,
however, there are again lots of simplifications: 
\begin{align*}
\Ampl_{\rgt\rgt}^{(2)}(x;y_{1}^{\rgt},y_{2}^{\rgt})=\;\frac{q^{-2}+1+q^{2}}{q^{-2}+q^{2}}\Big(\rho_{;0;0,2}(x;y_{1}^{\rgt},y_{2}^{\rgt})+\rho_{;0;1,1}(x;y_{1}^{\rgt},y_{2}^{\rgt})\Big).
\end{align*}
Again we notice that as the first index of all remaining real integrals is zero,
the integration contour is closed. The probability amplitude is also
real.

The amplitudes with other orderings of visits can be calculated similarly.
The results can be collected as 
\begin{align*}
\Ampl_{\lft\lft}^{(2)}(y_{1}^{\lft},y_{2}^{\lft};x)=\; & \frac{q^{-2}+1+q^{2}}{q^{-2}+q^{2}}\Big(\rho_{0,2;0;}(y_{1}^{\lft},y_{2}^{\lft};x)+\rho_{0,1;1;}(y_{1}^{\lft},y_{2}^{\lft};x)\Big)\\
\Ampl_{\lft\rgt}^{(2)}(y_{1}^{\lft};x;y_{1}^{\rgt})=\; & \frac{q^{-2}+1+q^{2}}{q^{-3}+q^{-1}+q+q^{3}}\Big((q^{-2}+1+q^{2})\rho_{0;2;0}(y_{1}^{\lft};x,y_{1}^{\rgt})\\
 & +\frac{q^{-3}+q^{-1}+q+q^{3}}{q^{-2}+q^{2}}\rho_{0;1;1}(y_{1}^{\lft};x;y_{1}^{\rgt})+\rho_{0;0;2}(y_{1}^{\lft};x;y_{1}^{\rgt})\Big)\\
\Ampl_{\rgt\lft}^{(2)}(y_{1}^{\lft};x;y_{1}^{\rgt})=\; & \frac{q^{-2}+1+q^{2}}{q^{-3}+q^{-1}+q+q^{3}}\Big((q^{-2}+1+q^{2})\rho_{0;0;2}(y_{1}^{\lft};x,y_{1}^{\rgt})\\
 & +\frac{q^{-3}+q^{-1}+q+q^{3}}{q^{-2}+q^{2}}\rho_{0;1;1}(y_{1}^{\lft};x;y_{1}^{\rgt})+\rho_{0;2;0}(y_{1}^{\lft};x;y_{1}^{\rgt})\Big)\\
\Ampl_{\rgt\rgt}^{(2)}(x;y_{1}^{\rgt},y_{2}^{\rgt})=\; & \frac{q^{-2}+1+q^{2}}{q^{-2}+q^{2}}\Big(\rho_{;0;0,2}(x;y_{1}^{\rgt},y_{2}^{\rgt})+\rho_{;0;1,1}(x;y_{1}^{\rgt},y_{2}^{\rgt})\Big).
\end{align*}
One can check that 
\begin{align*}
\Ampl_{++}^{(2)}(x;y^+_{1},y^+_{2})=\; & B_{2}^{2}\,\frac{\Gamma(\frac{16-\kappa}{\kappa})\,\Gamma(\frac{4}{\kappa})}{\Gamma(\frac{12-\kappa}{\kappa})\,\Gamma(\frac{8}{\kappa})}
    \;(y^+_{1}-x)^{1-\frac{8}{\kappa}}(y^+_{2}-y^+_{1})^{1-\frac{8}{\kappa}}\\
 & \qquad\times\phantom{}_{2}F_{1}\left(\frac{4}{\kappa},\frac{\kappa-8}{\kappa};\frac{8}{\kappa};\frac{y^+_{2}-y^+_{1}}{y^+_{2}-x}\right)
\end{align*}
and 
\begin{align*}
\Ampl_{+-}^{(2)}(y^-_{1};x;y^+_{1})=\; & B_{2}^{2}\,\frac{\Gamma(\frac{16-\kappa}{\kappa})\,\Gamma(\frac{8}{\kappa})}{\Gamma(\frac{12-\kappa}{\kappa})\,\Gamma(\frac{12}{\kappa})}\;
    (x-y^-_{1})^{\frac{4}{\kappa}}(y^+_{1}-x)^{-\frac{4}{\kappa}}(y^+_{1}-y^-_{1})^{2-\frac{16}{\kappa}}\\
 & \qquad\times\phantom{}_{2}F_{1}\left(\frac{8}{\kappa},\frac{\kappa-4}{\kappa};\frac{12}{\kappa};-\frac{x-y^-_{1}}{y^+_{1}-x}\right),
\end{align*}
and that $\Ampl_{--}^{(2)}(y^-_{2},y^-_{1};x)$ and $\Ampl_{-+}^{(2)}(y^-_{1};x;y^+_{1})$
are given by the obvious reflection in the above formulas. In particular
our formulas for $\Ampl_{++}^{(2)}$ and $\Ampl_{--}^{(2)}$ agree
up to the choice of normalization with those given in \cite{SZ-boundary_proximity_of_SLE}.

\subsection{Divergences of the real integrals\label{sec: divergences of integrals}}

As we mentioned above, the integrals over the real line contain divergences.
The integrals converge for $\kappa>8$, but diverge when $0<\kappa\le8$,
which is the range of the most interesting values of $\kappa$. There
are several strategies to tame the divergences, of which we emphasize two.

First, by construction, the spin chain - Coulomb gas method will result in
formulas that are analytic in $\kappa$, and the fundamental way of regularizing the
divergences of the integrals therefore is:
\begin{itemize}
\item \textbf{Analytic continuation.} We can first restrict to $\kappa>8$,
where the integrals converge, and analytically continue the final
expressions to smaller values of $\kappa$. 
\end{itemize}
As usual, analytic continuation can be done in several ways. The basis functions
$\varphi$, defined as integrals as in Figure~\ref{fig: FW integration contour},
themselves converge for all values of $\kappa$ and their suitable linear combinations
are thus already the analytic answer that we are looking for. Whenever possible,
it is nevertheless desirable to have explicit expressions for the answer
in terms of known analytic functions. This is, in fact, essentially what we have
been doing so far. Already the multiplicative constants $B_1, B_3, B_2$
appearing in the asymptotics properties in Section~\ref{sec: asymptotics via correspondence}
were a priori defined as real integrals convergent only for $\kappa>8$, but they
were expressible in terms of Gamma-functions which readily
provide their analytic continuation, e.g.,
$ B_{2} = \int_{0}^{1}\ud w\; w^{-\frac{4}{\kappa}}(1-w)^{-\frac{8}{\kappa}}
= \Gamma(\frac{\kappa-4}{\kappa}) \Gamma(\frac{\kappa-8}{\kappa}) / \Gamma(2\frac{\kappa-6}{\kappa}) $.
Furthermore, in Section~\ref{sec: transformation to real integrals} we gave
formulas for the final answers for the zig-zag amplitudes $\Ampl^{(1)}_\pm$ and $\Ampl^{(2)}_\omega$,
in terms of for example hypergeometric functions which also have known analytic continuations.

Although the real integrals always remain in essence similar to the cases considered above,
we can not in general reduce the answers to such well known special functions. It is therefore useful
to have a direct procedure to regularize the divergent real integrals generally, in a way that
provides their analytic continuation, has transparent properties, and can be used for their numerical
evaluation. We focus on one such procedure:
\begin{itemize}
\item \textbf{Cutoff regularization.} We can start from the final expressions
involving real integrals, and introduce a small cutoff $\eps$ to
regularize all divergent integrals. More precisely, we require that
all integration variables are further away than $\eps$ from any of
the points $x$ or $y_{k}$. With this prescription, the results diverge
as $\eps\searrow0$. All divergent terms are powers of $\eps$, with
the exponents depending on $\kappa$. They can be subtracted unambiguously
at least for irrational values of $\kappa$. The final result is then
obtained by taking $\eps\searrow0$ after subtracting the divergent
counterterms. We will discuss the details below. 
\end{itemize}

Let us now sketch how to prove that this regularization leads to the correct final result.

First, the loop integrals $\varphi$ of Figure~\ref{fig: FW integration contour}
converge for all values of $\kappa$ and thus can
be used to define the analytic continuation of the result from $\kappa>8$
to $0<\kappa\le8$.
We can then do the transformation
to real integrals, which was described in Section~\ref{sec: transformation to real integrals},
in a way that avoids the divergences. We first choose $\eps$ which
is smaller than half of the separation of any two of the points $y_{k}$
or $x$. When deforming the loops into integrals over the real line,
we replace the sections of contours on the real line, which are closer
than $\eps$ to the points $y_{k}$ or $x$, by (semi-)circles having
radii $\eps$. This approach results in a higher dimensional analogue
of the usual Pochhammer contour.
In this way a regularization is obtained by modification
of the contours, and no terms are dropped. Therefore it also gives
the analytic continuation of the results to small values of $\kappa$,
independently of the value of $\eps$.

Second, the pieces of the above contour on the real line 
equal the cutoff regularized integrals. The integrals
over the (semi-)circles can be expanded around $\eps=0$, and the
terms which are divergent as $\eps\searrow0$ provide the counterterms
for the cutoff regularization. For generic irrational $\kappa$ the
expansions contain no constant term. Therefore, taking $\eps\searrow0$,
the analytically continued result matches with the cutoff-regularized
one for all values of $\kappa$ for which the cutoff procedure could
be defined unambiguously.%
\footnote{For the specific values of $\kappa$, where the counterterms of the
cutoff procedure involve constants, cutoff regularization can be defined
such that it matches with the other schemes. Equivalently we can,
e.g., require that the counterterms are analytic in $\kappa$.%
}

Let us then work out the details of the cutoff regularization, i.e.,
find a method to calculate the counterterms. We already pointed out
that this can be done by studying the expansion of the contributions
from the (semi-)circles to the integrals above, but tracking
the phases of these integrals is quite involved. It turns out to be
easier to read off the divergent terms from the real integrals directly.
We can first take $\kappa>8$ and start from the integrals without
any cutoff. Then we separate the ``divergent'' terms by dividing
the integrations into several pieces, effectively introducing a ``cutoff''.

Let us first discuss the generic framework in more detail. We shall
also give an example below. We start from the integral $\rho$ where
all integrals are along the real line and the integrand is real. We
divide the integrals over each of the real intervals into two pieces:
the ``regular'' one where all integration variables are further
away than $\eps$ from the endpoints, and the ``divergent'' one
where one of the variables (either the first or the last one) is within
$\eps$ from the endpoints. The basic idea is then to develop the
divergent pieces as series at $\eps=0$.

For an $N$-point function, the highest possible divergence appears
when all integration variables are within $\eps$ from different points
$y_{j}$. Taking into account the behavior of the integrand and the
integration measure, such contribution is $\sim\eps^{N(1-8/\kappa)}$.
Developing the integrand as series at $\eps=0$, and taking into account
the contributions having divergent terms from $n<N$ integrations,
the generic divergent contribution has the power behavior
\begin{align}
\sim\; & \eps^{n(1-8/\kappa)}\eps^{k},\qquad\text{where}\; n=1,2,\ldots,N\;\text{and}\; k=0,1,2,\ldots.\label{eq: generic divergent power behavior}
\end{align}
All such terms can be in principle calculated by analyzing the divergent
terms. Analytically continuing to $\kappa<8$, terms with small $k$
will be divergent as $\eps\searrow0$. (Alternatively, we could keep
$\kappa<8$ fixed from the start and work with two cutoffs.) Since
we started from an integral that was independent of $\eps$, these
terms must cancel when all divergent and regular pieces are summed,
and they are thus the required counterterms. How all of this works
is best illustrated by considering an example.

Let us discuss the $N=2$ integral 
\begin{align*}
\rho_{;0;0,2}(x;y_{1},y_{2})=\; & \int_{y_{1}}^{y_{2}}\int_{w_{1}}^{y_{2}}dw_{1}dw_{2}\left[\frac{(w_{2}-w_{1})(y_{2}-y_{1})}{(y_{2}-w_{1})(y_{2}-w_{2})(w_{2}-y_{1})(w_{1}-y_{1})}\right]^{\frac{8}{\kappa}}\times F(w_{1},w_{2};x;y_{1},y_{2}),
\end{align*}
where $x<y_{1}<y_{2}$ and we denoted by 
\begin{align*}
F(w_{1},w_{2};x;y_{1},y_{2})=\left[\frac{(y_{2}-x)(y_{1}-x)}{(w_{2}-x)(w_{1}-x)}\right]^{\frac{4}{\kappa}}
\end{align*}
the part which would be replaced by a more complicated function for
a higher point integral having a similar structure, i.e., integral
of two variables between consecutive points $y_{j}$. The regular
term is 
\begin{align*}
R=\; & \int_{y_{1}+\eps}^{y_{2}-\eps}\!\int_{w_{1}}^{y_{2}-\eps}\! dw_{1}dw_{2}\left[\frac{(w_{2}-w_{1})(y_{2}-y_{1})}{(y_{2}-w_{1})(y_{2}-w_{2})(w_{2}-y_{1})(w_{1}-y_{1})}\right]^{\frac{8}{\kappa}}\times F(w_{1},w_{2};x;y_{1},y_{2})\\
\end{align*}
and the divergent terms can be written as 
\begin{align*}
 & D_{1}+D_{2}+D_{3}+D_{4}+D_{5}\\
= & \left(\int_{y_{1}}^{y_{1}+\eps}\!\int_{y_{1}+\eps}^{y_{2}-\eps}\!+\int_{y_{1}+\eps}^{y_{2}-\eps}\!\int_{y_{2}-\eps}^{y_{2}}\!+\int_{y_{1}}^{y_{1}+\eps}\!\int_{y_{2}-\eps}^{y_{2}}\!+\int_{y_{1}}^{y_{1}+\eps}\!\int_{w_{1}}^{y_{1}+\eps}\!+\int_{y_{2}-\eps}^{y_{2}}\!\int_{w_{1}}^{y_{2}}\!\right)dw_{1}dw_{2}\\
 & \times\left[\frac{(w_{2}-w_{1})(y_{2}-y_{1})}{(y_{2}-w_{1})(y_{2}-w_{2})(w_{2}-y_{1})(w_{1}-y_{1})}\right]^{\frac{8}{\kappa}}F(w_{1},w_{2};x;y_{1},y_{2}),
\end{align*}
where the first two terms include one divergent piece of integration,
and the last three include two pieces.

The leading contribution from the divergent pieces is contained in
the third term $D_{3}$, where $|w_{1}-y_{1}|<\eps$ and $|w_{2}-y_{2}|<\eps$,
as the terms $D_{4}$ and $D_{5}$ are suppressed by the factor $(w_{2}-w_{1})^{8/\kappa}$.
We denote the $\mathcal{O}(\eps)$ integration variables as $\hat{w}_{1}=w_{1}-y_{1}$
and $\hat{w}_{2}=y_{2}-w_{2}$. Developing at $\eps=0$ we find 
\begin{align*}
D_{3}=\; & \int_{0}^{\eps}\!\int_{0}^{\eps}d\hat{w}_{1}d\hat{w}_{2}\hat{w}_{1}^{-8/\kappa}\hat{w}_{2}^{-8/\kappa}\Big[F(y_{1},y_{2};x;y_{1},y_{2})\\
 & \!\!+\hat{w}_{1}\frac{\partial}{\partial{w_{1}}}F(w_{1},y_{2};x;y_{1},y_{2})\Big|_{w_{1}=y_{1}}\!\!-\hat{w}_{2}\frac{\partial}{\partial{w_{2}}}F(y_{1},w_{2};x;y_{1},y_{2})\Big|_{w_{2}=y_{2}}\!\!+\mathcal{O}(\eps^{2})\Big],
\end{align*}
where we wrote the terms of the expansions up to next-to-leading order,
corresponding to $k=1$ in \eqref{eq: generic divergent power behavior}.
Doing the integrals gives the counterterms 
\begin{align*}
D_{3}=\; & \frac{\eps^{2(1-8/\kappa)}}{(1-8/\kappa)^{2}}\Bigg[F(y_{1},y_{2};x;y_{1},y_{2})+\frac{\eps(1-8/\kappa)}{2(1-4/\kappa)}\\
 & \times\left(\frac{\partial}{\partial{w_{1}}}F(w_{1},y_{2};x;y_{1},y_{2})\Big|_{w_{1}=y_{1}}\!-\frac{\partial}{\partial{w_{2}}}F(y_{1},w_{2};x;y_{1},y_{2})\Big|_{w_{2}=y_{2}}\right)+\mathcal{O}(\eps^{2})\Bigg]\\
=\; & \frac{\eps^{2(1-8/\kappa)}}{(1-8/\kappa)^{2}}\Bigg[1-\frac{2\eps(1-8/\kappa)(y_{2}-y_{1})}{\kappa(1-4/\kappa)(y_{2}-x)(y_{1}-x)}+\mathcal{O}(\eps^{2})\Bigg].
\end{align*}

As another example, let us consider the term $D_{1}$. Denoting again
$\hat{w}_{1}=w_{1}-y_{1}$, we find 
\begin{align*}
D_{1}=\; & \int_{0}^{\eps}\!\int_{y_{1}+\eps}^{y_{2}-\eps}d\hat{w}_{1}dw_{2}\hat{w}_{1}^{-8/\kappa}(y_{2}-w_{2})^{-8/\kappa}\Bigg[F(y_{1},w_{2};x;y_{1},y_{2})\\
 & \quad+\hat{w}_{1}\bigg(\frac{\partial}{\partial{w_{1}}}F(w_{1},w_{2};x;y_{1},y_{2})\Big|_{w_{1}=y_{1}}-\frac{8(y_{2}-w_{2})}{\kappa(w_{2}-y_{1})(y_{2}-y_{1})}F(y_{1},w_{2};x;y_{1},y_{2})\bigg)+\mathcal{O}(\eps^{2})\Bigg]\\
=\; & \frac{\eps^{1-8/\kappa}}{1-\frac{8}{\kappa}}\int_{y_{1}+\eps}^{y_{2}-\eps}dw_{2}(y_{2}-w_{2})^{-8/\kappa}\Bigg[F(y_{1},w_{2};x;y_{1},y_{2})\\
 & \quad+\frac{\eps(1-8/\kappa)}{2(1-4/\kappa)}\bigg(\frac{\partial}{\partial{w_{1}}}F(w_{1},w_{2};x;y_{1},y_{2})\Big|_{w_{1}=y_{1}}-\frac{8(y_{2}-w_{2})}{\kappa(w_{2}-y_{1})(y_{2}-y_{1})}F(y_{1},w_{2};x;y_{1},y_{2})\bigg)+\mathcal{O}(\eps^{2})\Bigg].
\end{align*}
Thus rather nontrivial integrals remain in these counterterms. Notice
that even though the explicit $\eps$-factor which arises from the
divergent pieces is of lower order than in $D_{3}$, the overall divergence
is of the same order as the integral over $w_{2}$ also diverges for
$\eps\searrow0$.

The calculation for $D_{2}$ is similar as for $D_{1}$. The terms
$D_{4}$ and $D_{5}$ only contribute at $\mathcal{O}(\eps^{2-8/\kappa})$,
and their calculation is rather involved. Actually we slightly cheated
in the calculation of next-to-leading order terms for $D_{1}$: we
replaced $w_{2}-w_{1}$ by $w_{2}-y_{1}$ even though this approximation
fails when $w_{2}$ is close to the lower bound of its integration
range. Corrections due to this approximation can be combined with
the contributions from $D_{4}$.

In Appendix~\ref{sub: numerical evaluation} we discuss how the regularized integrals are used to
numerically compute the SLE boundary visit amplitudes.

\bigskip{}

\section{Notions of SLE boundary visits and applications\label{sec: interpretations and applications}}

In this section we give the definition of chordal SLE in the upper
half-plane $\bH$, and give the conformal covariance rule to transport
the boundary visit amplitudes from the half-plane to any other domain.
We then consider alternative definitions of SLE boundary visits, and
discuss applications of our main result.

\subsection{Definition of chordal SLE in half-plane\label{sub: def chordal SLE}}

By conformal invariance, it is sufficient to define the chordal $\SLEkappa{\kappa}$
in one reference domain with marked points. The upper half-plane $\bH$
with the starting point of the curve at $0$ and the end point of
the curve at $\infty$ is the most common choice. The following definition
also gives a convenient time parametrization for the curve. To define
the chordal $\SLEkappa{\kappa}$ in $(\bH;0,\infty)$, consider the
Loewner chain
\begin{align}
g_{0}(z)=\; & z, & \qquad\der tg_{t}(z)=\; & \frac{2}{g_{t}(z)-X_{t}} &  & \text{(for \ensuremath{z\in\bH})}\label{eq: Loewner chain in H}
\end{align}
where the driving process $(X_{t})_{t\geq0}$ is taken to be
\begin{align*}
X_{t}=\; & \sqrt{\kappa}\, B_{t}
\end{align*}
a multiple of the standard Brownian motion $(B_{t})_{t\geq0}$ on
the real line --- the parameter $\kappa$ gives the variance increment
per unit time.

The hull $K_{t}$ of the chordal $\SLEk$ at time $t$ is the closure
of the set of points $z\in\bH$ for which the solution to the Loewner
differential equation, Equation \eqref{eq: Loewner chain in H}, has
ceased to exist by time $t$. The hulls are growing compacts, $K_{s}\subset K_{t}$
for $s\leq t$. It can be shown \cite{RS-basic_properties} that the
hulls are generated by a continuous curve $\gamma:[0,\infty)\rightarrow\bH$
in the sense that the unbounded component of the complement $\bH\setminus\gamma[0,t]$
of an initial segment up to time $t$ coincides with the complement
$\bH\setminus K_{t}$ of the hull. We think of the chordal $\SLEk$
simply as this random curve $\gamma$.

\subsection{Conformal covariance of boundary visit amplitudes\label{sub: conformal covariance}}

We content ourselves to writing down the solutions to the boundary
visit question in the upper half-plane $\bH$ for a chordal $\SLEk$
from $x$ to $\infty$. The answer can be transported to other domains
by conformal covariance as follows.

Let us denote by $\Ampl_{(\domain;a,b)}^{(N)}(y_{1},\ldots,y_{N})$ 
the boundary zig-zag amplitude for chordal $\SLEk$
in domain $\domain$ from $a$ to $b$,
defined in a similar manner as in the half-plane, when the points
$y_{1},\ldots,y_{N}\in\bdry\domain$ are on smooth parts of the boundary
of the domain. Consider the chordal $\SLEk$ curve $\gamma$ in $(\domain;a,b)$,
and a conformal map $f\colon\domain\rightarrow f(\domain)$. For boundary
points $y\in\bdry\domain$ at which $f'(y)$ exists, a neighborhood
of $y$ of radius $\eps$ is approximately mapped to a neighborhood
of the image $f(y)$ and having radius $\eps\times|f'(y)|$. The SLE
curve itself is conformally invariant, that is, $f(\gamma)$ has the
law of a chordal $\SLEk$ in $(f(\domain);f(a),f(b))$. Correspondingly,
after passing to the limit of small radii in the definition of the
amplitude 
\begin{align*}
 & \lim_{\eps\searrow0}\left(\frac{1}{\prod_{j}\eps_{j}^{h}}\times\PR\left[\SLEk\text{\,\ visits neighborhoods of \ensuremath{y_{j}}\,\ of radii \ensuremath{\eps_{j}}}\right]\right),
\end{align*}
we get that the boundary zig-zag amplitudes satisfy the following
conformal covariance rule
\begin{align}
\Ampl_{(\domain;a,b)}^{(N)}(y_{1},\ldots,y_{N})=\; & \left(\prod_{j=1}^{N}|f'(y_{j})|^{h}\right)\times\Ampl_{(f(\domain);f(a),f(b))}^{(N)}\left(f(y_{1}),\ldots,f(y_{N})\right),\label{eq: conformal covariance formula}
\end{align}
and similarly for the complete correlation functions $\Corr_{(\domain;a,b)}^{(N)}$.

Appendix \ref{sub: conformal covariance from CFT} discusses this
conformal covariance from the viewpoint of conformal field theory.

\subsection{Different definitions of SLE boundary visits\label{sub: different definitions of boundary visit}}

There are several formulations of boundary visits, and one expects
many limits of the types of Equations \eqref{eq: full correlation with balls}
or \eqref{eq: zig-zag proba with balls} to exist. Consider for example
the following alternative formulations:
\begin{itemize}
\item \emph{Touching small boundary intervals (for $\kappa>4$):} In the
phase $\kappa>4$, where the curve $\gamma$ can touch the boundary
of the domain, a natural notion of reaching a neighborhood of a point
$y_{j}\in\bR\setminus\set x\subset\bdry\bH$ is that the curve $\gamma$
touches the boundary between the point $y_{j}$ and a point which
is $\eps_{j}$ further away from the starting point $x$ of the curve.
If $y_{j}>x$ set $I_{\eps_{j}}(y_{j})=[y_{j},\, y_{j}+\eps_{j}]$
and if $y_{j}<x$ set $I_{\eps_{j}}(y_{j})=[y_{j}-\eps_{j},\, y_{j}]$.
The corresponding boundary visit amplitude is given by the limit of
\emph{
\begin{align}
 & \eps_{1}^{-h}\cdots\eps_{N}^{-h}\;\PR\Big[\gamma\cap I_{\eps_{j}}(y_{j})\neq\emptyset\quad\forall j=1,2,\ldots,N\Big]\label{eq: interval visit amplitude}
\end{align}
}as $\eps_{1},\ldots,\eps_{N}\searrow0$.
\item \emph{Reaching small conformal distances from the boundary points:}
For $\Lambda\subsetneq\bC$ a simply connected open domain and $z\in\Lambda$,
define the conformal radius $\rho_{\Lambda}(z)$ such that if $f:\bD\rightarrow\Lambda$
is a conformal map with $f(0)=z$, then $\rho_{\Lambda}(z)=|f'(0)|$.
By Schwarz lemma and Köbe $\frac{1}{4}$-theorem, $\rho_{\Lambda}(z)$
is comparable to the distance of $z$ to $\bdry\Lambda$:
\begin{align*}
\frac{1}{4}\rho_{\Lambda}(z)\leq\; & \dist\left(z,\,\bdry\Lambda\right)\leq\rho_{\Lambda}(z).
\end{align*}
Now for $y_{j}\in\bR\setminus\set x\subset\bdry\bH$, let $U_{j}$
be the (unique) connected component of $\bH\setminus\gamma$ such
that $y_{j}\in\bdry U_{j}$. Join to $U_{j}$ its reflection across
the real axis, to obtain a larger domain in which $y_{j}$ is an interior
point --- more precisely, let $V_{j}$ be the interior 
of $U_{j} \cup \bR \cup U_{j}^{*}$, where $U_{j}^{*}=\setcond{\bar{z}}{z\in U_{j}}$.
The quantity $\rho_{\bH\setminus\gamma}(y_{j})=\rho_{V_{j}}(y_{j})$
gives a conformally covariant notion of the distance of $y_{j}$ to
$\gamma$ --- recall that $\frac{1}{4}\rho_{\bH\setminus\gamma}(y_{j})\leq\;\dist\left(y_{j},\,\gamma\right)\leq\rho_{\bH\setminus\gamma}(y_{j}).$
The corresponding boundary visit amplitude is given by the limit of
\emph{
\begin{align}
 & \eps_{1}^{-h}\cdots\eps_{N}^{-h}\;\PR\Big[\rho_{\bH\setminus\gamma}(y_{j})<\eps_{j}\quad\forall j=1,2,\ldots,N\Big]\label{eq: conformal visit amplitude}
\end{align}
}as $\eps_{1},\ldots,\eps_{N}\searrow0$.
\end{itemize}
One could give an endless list of possible formulations: it is essentially
possible to define the notion of a boundary visit as the intersection
of the curve with a small neighborhood of any imaginable shape. Each
of the different formulations admits both a complete correlation function
analogous to Equation \eqref{eq: full correlation with balls} as
exemplified in the two cases above, and an ordered zig-zag amplitude
analogous to Equation \eqref{eq: zig-zag proba with balls}. The formulations
\eqref{eq: interval visit amplitude} and \eqref{eq: conformal visit amplitude}
are convenient for various reasons. In Appendix \ref{sec: derivations of the value of h}
we in particular present a derivation of the correct value of the
scaling exponent $h=\frac{8-\kappa}{\kappa}$ given in \eqref{eq: h13}
based on each of them.

\subsection{Applications of the results and universal and non-universal aspects\label{sub: applications and universality}}

In Section \ref{sub: different definitions of boundary visit} we
have argued that the SLE boundary visit amplitudes describe the probabilities
of events where the SLE trace comes close to marked boundary points,
independent of the details of the definition of these events. In this
section we mention further applications.

First, however, we emphasize that the details of the formulation or
application affect a multiplicative constant in the answer, but not
the functional shape of the zig-zag amplitude $\Ampl^{(N)}(x;y_{1},\ldots,y_{N})$
or the correlation function $\Corr^{(N)}(x;y_{1},\ldots,y_{N})$.
For example, visiting small neighborhoods of different shapes should
happen with comparable but not necessarily equal probabilities.
In renormalization group language, the multiplicative constants are
non-universal, whereas the functions $\Ampl^{(N)}(x;y_{1},\ldots,y_{N})$
and $\Corr^{(N)}(x;y_{1},\ldots,y_{N})$ are universal as scaling
functions (correlation functions). Also some ratios of the multiplicative
constants are universal: the most immediate example comes from considering
the formula $\chi^{(N)}(x;y_{1},\ldots,y_{N})=\sum_{\sigma\in \SymmGrp_{N}}\Ampl^{(N)}(x;y_{\sigma(1)},\ldots,y_{\sigma(N)})$
for the complete correlation function as a sum over different orders
of visits --- for the formula to be meaningful, the ratios of the
different multiplicative constants for a given $N$ have to be independent
of the formulation.

A slightly trivial but nevertheless illuminating example of the
universality of the functional shape and non-universality of the constant factor is
to imagine what would have happened in Equation \eqref{eq: full correlation with balls}
had we chosen to measure the size of the semi-disk neighborhoods with
diameter $\eps$ instead of radius $\eps$ --- the limit would obviously
have been a factor $2^{Nh}$ smaller. As a nontrivial example, note that
the literature contains two definitions of the SLE Green's function at interior points:
one for neighborhoods defined with usual Euclidean distances, and another
with conformal radius. It has been shown
in~\cite{LR-Minkowski_content_and_natural_parametrization_for_SLE} that the two
Green's functions are the same up to a multiplicative constant (whose value is not
explicitly known). In fact, the idea used
in~\cite{LR-Minkowski_content_and_natural_parametrization_for_SLE} 
is the correct explanation with SLE analysis of the universality of the functional shape of
$\Ampl^{(N)}_\omega$, and of the non-universality of the multiplicative constant.
Roughly, if the SLE curve is conditioned to approach a point $y$, and
one considers the curve locally near $y$, then in small scale the curve
will look like it is drawn from a certain stationary distribution which is
independent of what other far away points the curve is conditioned to visit.
The curve with stationary law has certain non-zero probabilities of hitting
a half-disk, boundary interval, or some other shape, and the ratios of these
probabilities give the ratios of the amplitudes in the respective formulations.

In most cases, an exact formula for the non-universal multiplicative constants
would be too much to hope for.
However, Appendix \ref{sub: touching boundary interval} contains
one concrete example in which the multiplicative constant is explicit:
the $N=1$ case in the ``touching small boundary intervals'' formulation
is Equation \eqref{eq: touching small intervals 1pt function}.

In Section \ref{sec: the problem} we argued that the amplitudes $\Ampl^{(N)}$
and $\Corr^{(N)}$ are obtained as solutions to a system of linear
partial differential equations and boundary conditions. Solutions
to this linear homogeneous problem are at best fixed up to a multiplicative
constant, and the above considerations explain that this is only natural.

\subsubsection{Boundary visit probabilities for interfaces in lattice models\label{sub: lattice model boundary visit application}}

The principal motivation for the introduction and study of SLEs is
that these random curves are the scaling limits of interfaces in lattice
models of statistical mechanics at criticality. The SLE zig-zag probabilities
are closely related to the probabilities for an interface in a lattice
model to pass through given boundary points. For some models these
probabilities in turn have direct physical interpretations, for example
the boundary visit probability of interface in $Q$-random cluster
model ($Q$-FK model) gives a boundary magnetization in the $Q$-Potts
model via the Edwards-Sokal coupling \cite{ES-generalization_of_FKSW_representation_and_Monte_Carlo}.

For lattice model interfaces, too, the exact meaning of passing through
a boundary point involves some choices, and different choices lead
to different non-universal constant factors. The idea, however, always
is to consider the model on a lattice domain $\domain_{\delta}$ of
small lattice mesh size $\delta$ so that $\domain_{\delta}$ approximates
a given planar domain $\domain\subset\bC$ as $\delta\searrow0$.
One defines a boundary visit locally by requiring the lattice model
interface to use for example a given edge or a given vertex near a
marked point $y\in\bdry\domain$ on the boundary. The probabilities
of thus visiting $N$ marked points on smooth parts of the boundary
$\bdry\domain$ are of order $\delta^{Nh}$, provided that also the
lattice approximations to the boundary have a regular and consistent
local structure as $\delta\searrow0$. Thus the lattice mesh $\delta$
serves as a measure of the neighborhood size, and much like in \eqref{eq: full correlation with balls},
the limit of the lattice model interface probability renormalized
by $\delta^{-Nh}$ should be given by $\Ampl^{(N)}$ or $\Corr^{(N)}$,
correctly conformally transported to the domain $\domain$ by the
conformal covariance rule of Section \ref{sub: conformal covariance}.

In Section \ref{sec: lattice models and numerics} we discuss in more
detail a few well-known lattice models and the details of the question
of boundary visits of interfaces for them. We find that our formulas
for $\Ampl^{(N)}$ and $\Corr^{(N)}$ are in very good agreement with
the probabilities obtained from numerical simulations of these lattice
models.

\subsubsection{Covariant measure of SLE on the boundary\label{sub: covariant boundary measure application}}

For lattice models, the most natural way of quantifying boundary proximity
of an interface is by counting the number of boundary points visited
by it, e.g., within a given boundary segment. In the scaling limit,
the count must be renormalized properly by a power of the lattice
spacing $\delta$: the probability to visit a given boundary point
is of order $\delta^{h}$ and the expected number of boundary points
visited in a segment is of order $\delta^{h-1}$ (which diverges for
$\kappa>4$ and tends to zero for $\kappa<4$).

The article \cite{AS-covariant_measure_of_SLE_on_boundary} presents
a construction of a covariant measure of SLEs on the boundary, which
is the analogous boundary proximity count in the continuum. Roughly,
this SLE boundary measure $\mu_{\domain;a,b}$, associated to the
chordal $\SLEk$ in domain $\domain$ from $a$ to $b$, is a random
locally finite measure $\mu_{\domain;a,b}$ on $\bdry\domain$, supported
on the set where the chordal $\SLEk$ curve $\gamma_{\Lambda;a,b}$
from $a$ to $b$ in $\domain$ touches the boundary $\bdry\domain$.
This measure is conformally covariant with exponent $h$, i.e., if
$f\colon\domain\rightarrow\domain'$ is a conformal map, then $\mu_{\domain;a,b}(\ud x)=|f'(x)|^{h}\,\mu_{f(\domain);f(a),f(b)}(\ud f(x))$
in law. The domain Markov property for the measure states that conditionally
on an initial segment of the chordal $\SLEk$ curve in $(\domain;a,b)$,
the measure $\mu_{\domain;a,b}$ restricted to a set $A\subset\bdry\domain$
away from the initial segment has the same law as $\mu_{\domain\setminus{\rm segment};{\rm tip},b}$
restricted to the same set. These properties characterize the family
of measures $\mu_{\domain;a,b}$ up to a multiplicative constant.

The SLE boundary measure is constructed by studying a local martingale
associated to the correlation function $\chi^{(1)}$. By construction
this function $\chi^{(1)}$ then gives the density of the expectation
of $\mu=\mu_{\bH;0,\infty}$ with respect to the Lebesgue measure
on $\bR$. The higher complete correlation functions $\chi^{(N)}$
of the present article should be the integral kernels for moments
of the SLE boundary measure
\begin{align*}
\eps^{-N}\,\EX\left[\prod_{j=1}^{N}\mu([y_{j},y_{j}+\eps])\right]\sim\; & \const\times\Corr^{(N)}(0;y_{1},\ldots,y_{N}).
\end{align*}
In fact the proof \cite{AS-covariant_measure_of_SLE_on_boundary}
of non-triviality of the constructed SLE boundary measure employs
the two-point function $\Corr^{(2)}$, which had been found in \cite{SZ-boundary_proximity_of_SLE}.

A convenient way to explicitly characterize a random measure is to
give its Laplace transform. Denote briefly $\mu=\mu_{\bH;0,\infty}$.
For a test function $\phi\colon\bR\setminus\set 0\rightarrow\bR$
let
\begin{align*}
L(\phi):=\; & \EX\left[e^{-\int_{\bR}\phi\ud\mu}\right]
\end{align*}
be the Laplace transform of $\mu$ at $\phi$. For the sake of concreteness,
consider $\phi$ supported on the positive real axis. Then the expansion
of the Laplace transform around the zero function is given by
\begin{align*}
L(\eps\phi)=\; & \EX\left[e^{-\eps\int_{\bR}\phi\ud\mu}\right]\\
=\; & 1-\eps\EX\left[\int_{\bR}\phi(y)\ud\mu(y)\right]+\frac{\eps^{2}}{2}\EX\left[\iint\phi(y_{1})\phi(y_{2})\ud\mu(y_{1})\ud\mu(y_{2})\right]+\cdots\\
=\; & 1+\sum_{N=1}^{\infty}(-\eps)^{N}c_{N}\idotsint_{\set{y_{1}<y_{2}<\cdots<y_{N}}}\phi(y_{1})\cdots\phi(y_{N})\,\chi^{(N)}(0;y_{1},\ldots,y_{N})\,\ud y_{1}\cdots\ud y_{N},
\end{align*}
where $c_N$ are non-universal multiplicative constants.

The construction of \cite{AS-covariant_measure_of_SLE_on_boundary}
establishes that a unique (up to normalization) random measure satisfying
the required abstract properties exists. The results of this article
in principle give explicit formulas for the random measure in terms
of integral kernels for its moments or the power series expansion
of its Laplace transform.

\subsubsection{Conditioned SLE and first visit point recursion for the zig-zag amplitudes\label{sub: conditioning application and proof strategy}}

Let us discuss one more interpretation of the results, which in fact
also suggests a natural strategy of rigorous proof that our formulas
give the order refined SLE Green's functions on the boundary, as defined
in Section \ref{sec: Introduction} or alternatively in Section \ref{sub: different definitions of boundary visit}.

Consider conditioning the chordal $\SLEk$ curve $\gamma$ to visit
a boundary point $y$, for definiteness in $(\bH;x,\infty)$ again.
As such, this is a zero-probability event (for $\kappa<8$), and one
must perform a limiting procedure to properly define the conditioning:
first condition on visiting $B_{\eps}(y)$ and then let $\eps\searrow0$.
The conditioned curve can be described explicitly: its Radon-Nikodym
derivative with respect to the ordinary chordal SLE is proportional
to the indicator of the event of the visit, and in the limit $\eps\searrow0$
we get a Girsanov transform of the ordinary chordal SLE
\begin{align*}
\frac{\ud\PR_{(\bH;x,y,\infty)}^{{\rm cond.}}}{\ud\PR_{(\bH;x,\infty)}}\Big|_{\mathcal{F}_{t}}\propto\; &
\Corr_{\bH\setminus K_{t}}^{(1)}(\gamma(t);y)\,=\,|g_{t}'(y)|^{h}\;\Corr^{(1)}(X_t;g_{t}(y)).
\end{align*}
This description of the conditioned curve is equivalent to the more
familiar $\SLE_{\kappa}(\rho)$ with $\rho=\kappa-8$, i.e., the random
Loewner chain \eqref{eq: Loewner chain in H} with driving process
given by 
\begin{align*}
X_{0}=x,\qquad\ud X_{t}=\; & \sqrt{\kappa}\,\ud B_{t}+\frac{\rho}{X_{t}-g_{t}(y)}\ud t, & \text{where }\rho=\; & \kappa-8.
\end{align*}
After the random time when the conditioned curve reaches $y$ (i.e.,
when $|X_{t}-g_{t}(y)|\to0$), the curve will continue like an ordinary
chordal SLE in the complement of the initial segment of the curve
up to that time.

Using the one-point function $\Corr^{(1)}$, one may thus describe
the SLE conditioned to visit a given boundary point. Conditioning
on visiting several points could be similarly done with our functions
$\Corr^{(N)}$ or $\Ampl^{(N)}$. Below we will however turn the logic
around, and see how our formulas could be rigorously proved using
this conditioning.

The idea is to use the conditioning to reduce the $N$-point function
question to an $(N-1)$-point question. Namely, for the SLE curve
$\gamma$ to make visits to $B_{\eps_{1}}(y_{1}),\ldots,B_{\eps_{N}}(y_{N})$
in this order, it needs to make the first visit to $y_{1}$ by definition,
and we may proceed by conditioning on this. We know, for example by
considerations similar to Appendices \ref{sub: touching boundary interval}
or \ref{sub: reaching small conformal radius}, that the probability
of this first visit is of order $\eps_{1}^{h}\,\Corr^{(1)}(x;y_{1})$,
and we can describe the conditional law of the curve given this first
visit essentially by the $\SLE_{\kappa}(\rho)$ process above. After
the time $\tau$ of the first visit, the curve is again a chordal
SLE in the random domain $\bH\setminus K_{\tau}$ at that time, and
we would like it to visit the neighborhoods of the $N-1$ remaining
points $y_{2},\ldots,y_{N}$. We may inductively assume that the $(N-1)$-point
visit formulas $\Ampl^{(N-1)}$ for chordal SLE have been established.
Thus we need to be able to average the $(N-1)$-point zig-zag amplitude
$\Ampl_{\bH\setminus K_{\tau}}^{(N-1)}(\gamma_{\tau};y_{2},\ldots,y_{N})$
over the randomness of the domain $(\bH\setminus K_{\tau};\gamma_{\tau},\infty)$
that remains after the first visit. That will be achieved if we can
construct a martingale for the conditioned SLE, whose value at the
time $\tau$ is $\Ampl_{\bH\setminus K_{\tau}}^{(N-1)}(\gamma_{\tau};y_{2},\ldots,y_{N})$.
The key point is that such a martingale is constructed using the formula
for $\Ampl^{(N)}$ that we find in the present work --- namely we
set
\begin{align*}
M_{t}=\; & \prod_{j=2}^{N}|g_{t}'(y_{j})|^{h}\times\frac{\Ampl^{(N)}(X_{t};g_{t}(y_{1}),\ldots,g_{t}(y_{N}))}{\Corr^{(1)}(X_{t};g_{t}(y_{1}))}.
\end{align*}
This is a local martingale by the differential equations \eqref{sec: 2nd order PDE}
that our $\Ampl^{(N)}$ satisfies, and its value at time $\tau$ is
the desired $(N-1)$-point zig-zag amplitude in the random domain
$\bH\setminus K_{\tau}$ essentially by the asymptotics conditions
\eqref{eq: the very first visited point} we impose on $\Ampl^{(N)}$.
What remains is to show that $(M_{t})_{t\in[0,\tau]}$ is a uniformly
integrable martingale. This relies partly on a priori estimates of
SLE probabilities \cite{Beffara-dimension_of_the_SLE_curves,LW-multi_point_Greens_functions_for_SLE}
and on careful control of the functions appearing in the spin chain
- Coulomb gas correspondence of the present article and in \cite{KP-covariant_boundary_correlations}.
One also needs to control some approximations made, but roughly speaking
at this stage optional stopping for the martingale $(M_{t})$ proves
that $\Ampl^{(N)}$ gives the $N$-point boundary zig-zag amplitude
or $N$-point order refined SLE Green's function on the boundary.

Carrying out the proof with this strategy is the topic of a subsequent
work in collaboration with Konstantin Izyurov.

\bigskip{}

\section{Comparisons with lattice model simulations\label{sec: lattice models and numerics}}

It is somewhat intricate and computationally demanding to obtain satisfactory
computer simulations of SLE curves \cite{Kennedy-fast_algorithm_for_simulating_the_chordal_SLE}.
Therefore, comparing our results with direct numerics of SLEs would
be difficult. A more practical alternative is to simulate lattice
models whose interfaces tend to SLEs in the scaling limit. The boundary
visits in such lattice models indeed constitute a natural interpretation
and an important physical application of our results, as discussed
in Section \ref{sub: lattice model boundary visit application}. In
the present section we elaborate on the idea in the context of various
lattice models. We discuss simulation of these models and their interfaces
and boundary visits of the interfaces. Finally, we compare the numerical
results obtained from these simulations to our solution presented
in Sections \ref{sec: quantum group and Coulomb gas} and \ref{sec: regularized real integrals and evaluation}.

On physical grounds it is completely natural to expect that the scaling
limit of renormalized lattice interface visit probabilities is proportional
to the SLE Green's functions $\Corr^{(N)}$ and $\Ampl^{(N)}$. We
nevertheless remark that even in models whose interface is rigorously
known to converge to a chordal SLE in the scaling limit (e.g., Sections
\ref{sub: LERW model}, \ref{sub: percolation model}, \ref{sub: FK model}
below), highly nontrivial additional mathematical work would be needed
to establish this. Actually, the validity of the physically unsurprising
equivalence is highly sensitive to the details of the lattice approximation
of the domain boundary, and again even valid approximation schemes
lead to different non-universal proportionality constants. Incidentally,
the equivalence of the two formulations has been rigorously established
for one case: one and two-point boundary visits of the FK-Ising model
interface (Section \ref{sub: FK model} below) on boundary segments
parallel to coordinate axes --- the boundary visit probabilities (or
equivalent boundary spin correlation functions) were used in \cite{HK-Ising_interfaces_and_free_boundary_conditions}
as a technique to control the scaling limit of an interface in a dual
model (the Ising model with particular boundary conditions). Our simulation
results below of course show a good match to our analytical solution,
and thus clearly support the physically expected equivalence of the
formulations.

Let us still make general comments about the numerical comparison
of simulation data with our main results. Small lattice mesh sizes
$\delta$ are of course desirable to reduce finite size scaling effects,
i.e., to obtain better approximations to the conformally invariant
scaling limit situation. As always, however, small mesh size $\delta$
or corresponding large size of the simulated system quickly increases
needed computational resources, particularly so in critical models
that we are interested in. For our question, there is yet another
difficulty. With lattice mesh $\delta$, the probability of having
$N$ boundary visits by the interface is of order $\delta^{Nh}$,
where $h=h_{1,3}(\kappa)=\frac{8-\kappa}{\kappa}>0$ and $\kappa$
depends on the model. We are thus interested in rare events, whose
probability further decreases with mesh size $\delta$ and number
of visit points $N$, so in order to obtain acceptable statistics,
we need increasingly large numbers of samples. The trade-off between
reducing finite size effects and improving statistics is therefore
a major issue. High values of the exponent $h_{1,3}(\kappa)$, or
correspondingly models with small $\kappa$ are the most problematic.
We have simulated models corresponding to $\kappa=2$ (LERW, Section
\ref{sub: LERW model}), $\kappa=\frac{24}{5}$, and $\kappa=\frac{16}{3}$
(different FK-models, Section \ref{sub: FK model}), and $\kappa=6$
(percolation, Section \ref{sub: percolation model}). In the most
difficult case $\kappa=2$ we are essentially limited to $N\leq2$,
and significant finite size effects still remain in the data (see
Figure \ref{fig: LERW 2pt data}). In the least problematic case $\kappa=6$,
finite size effects can be made reasonably small up to $N=4$ (see
Figure \ref{fig: perco 4pt data}). The issues in numerical evaluation
of our analytical results have been separately discussed in Appendix
\ref{sub: numerical evaluation}, and we note that besides large $N$,
difficulties also arise due to small $\kappa$.

\subsection{Lattice model interfaces}

\subsubsection{Relevant domains and conformal maps\label{sub: lattice domains and conformal maps}}

We have simulated different statistical models in lattice approximations
of domains of the simplest possible shapes: the square and the equilateral
triangle. The frequencies of boundary visits of interfaces have been
collected, and for comparison with our formulas they need to be transported
to the half-plane $\bH$ by conformal maps. The domains, lattice approximations,
and the conformal maps are described below.

The unit square

\begin{align*}
S=\; & \set{z\in\bC\;\Big|\;0<\re(z)<1,\;0<\im(z)<1}
\end{align*}
will be discretized by a square lattice of small mesh size $\delta$:
the vertex set is $S_{\delta}=\delta\bZ^{2}\cap\overline{S}$ and
edges connect vertices at distance $\delta$. A conformal map $f_{S}\colon S\rightarrow\bH$
from the square to the half-plane is the Jacobi elliptic sine function
$\mathrm{sn}$ composed with a Möbius transform, and our choice is
\begin{align*}
f_{S}(u)=\; & \frac{\mathrm{sn}\left((2u-1)K;m\right)+1}{\mathrm{sn}\left((2u-1)K;m\right)-1/\sqrt{m}}\;\frac{\mathrm{sn}\left(K;m\right)-1/\sqrt{m}}{\mathrm{sn}\left(K;m\right)+1},
\end{align*}
where $m$ is the elliptic modulus of square and $K=K(m)$ is the
corresponding complete elliptic integral of the first kind. This choice
is such that the lower left corner is mapped to the origin, the top
right corner to infinity, and the bottom right and top left corners
to $+1$ and $-1$, respectively.

The unit equilateral triangle
\begin{align*}
T=\; & \set{z\in\bC\;\Bigg|\;-\half<\re(z)<\half,\;0<\im(z)<\frac{\sqrt{3}}{2}-\sqrt{3}\,|\re(z)|}
\end{align*}
will be discretized by a fine triangular lattice. The small mesh size
$\delta$ is the distance between its neighboring vertices, and $T_{\delta}$
denotes the set of such triangular lattice vertices in $\overline{T}$.
A conformal map $f_{T}\colon T\rightarrow\bH$ from the triangle to
the half-plane is the inverse of a Schwarz-Christoffel map,
\begin{align*}
f_{T}^{-1}(z)=\; & \frac{\Gamma(\frac{5}{6})}{\sqrt{\pi}\,\Gamma(\frac{1}{3})}\times\int_{0}^{z}(1-w)^{-2/3}(1+w)^{-2/3}\ud w.
\end{align*}
The choice is such that $f_{T}$ maps the midpoint of the bottom side
to the origin, and the left and right bottom corners to $-1$ and
$+1$, respectively.

\subsubsection{Loop-erased random walk\label{sub: LERW model}}

The loop-erased random walk (LERW) is a path obtained by performing
loop erasure to a finite piece of a simple random walk. The conformal
invariance of the scaling limit of interior-to-boundary LERW was shown
in \cite{LSW-LERW_and_UST}. Different LERW variants, including the
one we study here, were proven to have conformally invariant scaling
limits in \cite{Zhan-scaling_limits_of_planar_LERW}. The scaling
limit of the path we describe below is chordal $\SLE_{2}$.

\begin{figure}
\begin{centering}
\includegraphics[width=0.6\textwidth]{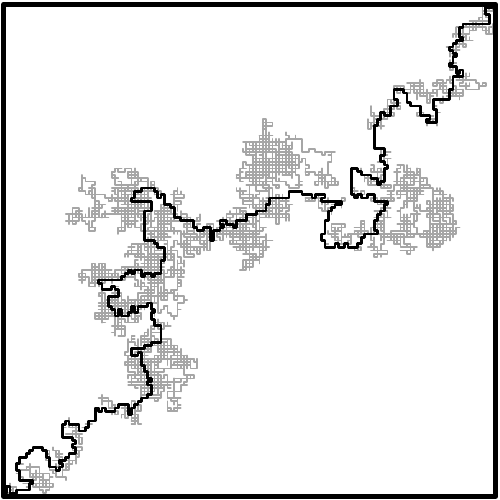}
\par\end{centering}

\caption{\emph{A loop-erasure of a random walk in a box, from the bottom-left
corner to the top-right corner.\label{fig: LERW}}}
\end{figure}

We consider the square lattice domain $S_{\delta}$, which approximates
the unit square, as in Section \ref{sub: lattice domains and conformal maps}.
We send a simple random walk $(W_{n})_{n=0}^{\infty}$ at the
lower left corner $W_{0}=\delta+\ii\delta$. We condition on the event
that the walk exits the square via the upper right corner, and we
denote the time of exit by $\tau$. The loop-erased random walk is
the simple path $\gamma_{\delta}$ which is obtained from $(W_{n})_{n=0}^{\tau-1}$
by chronologically erasing all loops (sequences of consecutive steps
which start and end at the same vertex). Figure \ref{fig: LERW} shows
a realization of a LERW in $S_{\delta}$ with lattice mesh $\delta=\frac{1}{150}$.
The figure also suggests that the loop-erased path is unlikely to
come close to the boundary except at the two end points, indicating
the difficulties of sampling boundary visits of this model with fine
lattice mesh.

We define boundary visit as the event that the path $\gamma_{\delta}$
passes through a vertex $x$ at distance $\delta$ from the boundary
$\bdry S$ of the square. The behavior of the boundary visit probabilities
should be
\begin{align}
\PR[\gamma_{\delta}\text{ visits }x_{1},x_{2},\ldots,x_{N}]\approx\; & \const\times\prod_{j=1}^{N}\left(|f'(x_{j})|\,\delta\right)^{h}\times\Ampl^{(N)}(0;f(x_{1}),\ldots,f(x_{N})),\label{eq: lattice interface boundary visit approximation}
\end{align}
where $h=h_{1,3}(2)=3$ and $f=f_{S}:S\rightarrow\bH$ is the conformal
map from the unit square to the half-plane given in Section \ref{sub: lattice domains and conformal maps}.

The simulation is done as follows: we sample a conditioned random
walk using explicitly calculated transition probabilities, then perform
the loop erasure of the random walk, and collect data of visited boundary
points of the loop erasure. We correct the boundary visit frequencies
obtained from the simulations by dividing by the factor $\prod_{j=1}^{N}\left(|f_{S}'(x_{j})|\,\delta\right)^{h}$
that appears in \eqref{eq: lattice interface boundary visit approximation},
and then compare with our SLE boundary visit amplitude $\Ampl^{(N)}$
at $\kappa=2$. Note that the probabilities decay as $\delta^{Nh}$
and due to the high value of the exponent $h=h_{1,3}(2)=3$ it is
very hard to obtain good statistics with a small mesh size, especially
for higher $N$. Figures \ref{fig: 1pt data} and \ref{fig: LERW 2pt data}
present data from simulations with lattice mesh $\delta=\frac{1}{120}$
and $10^{7}$ realizations and with lattice mesh $\delta=\frac{1}{60}$
and $10^{8}$ realizations, respectively. The agreement with our analytical
results is reasonable. The otherwise difficult small $\kappa$ turns
out to have one advantage: the orders of magnitude of the visits in
different pieces of the plot are rather different, and one notes in
particular that the universal ratio of the boundary visit amplitudes with $y_{2}<x=0$
and $y_{2}>y_{1}=1$ obtained by our method is undeniably correct
--- a single multiplicative constant has been fitted for the two pieces
$\zeta_{++}^{(2)}$ and $\zeta_{+-}^{(2)}$ in Figure \ref{fig: LERW 2pt data}.

\subsubsection{Percolation\label{sub: percolation model}}

Percolation is an easily defined model of statistical physics, showing
nevertheless interesting critical behavior. Its conformal invariance
had been predicted in \cite{LPPS-crossing_probabilities_in_percolation},
and impressive exact results had been predicted using conformal field
theory. The proof of conformal invariance of scaling limit of site
percolation on triangular lattice was obtained by Smirnov in \cite{Smirnov-critical_percolation},
based on a formula found by Cardy \cite{Cardy-percolation}. The interface
that we define below converges in the scaling limit to chordal $\SLE_{6}$,
see \cite{Smirnov-critical_percolation,CN-critical_percolation_exploration_path}.

\begin{figure}
\begin{centering}
\includegraphics[width=0.7\textwidth]{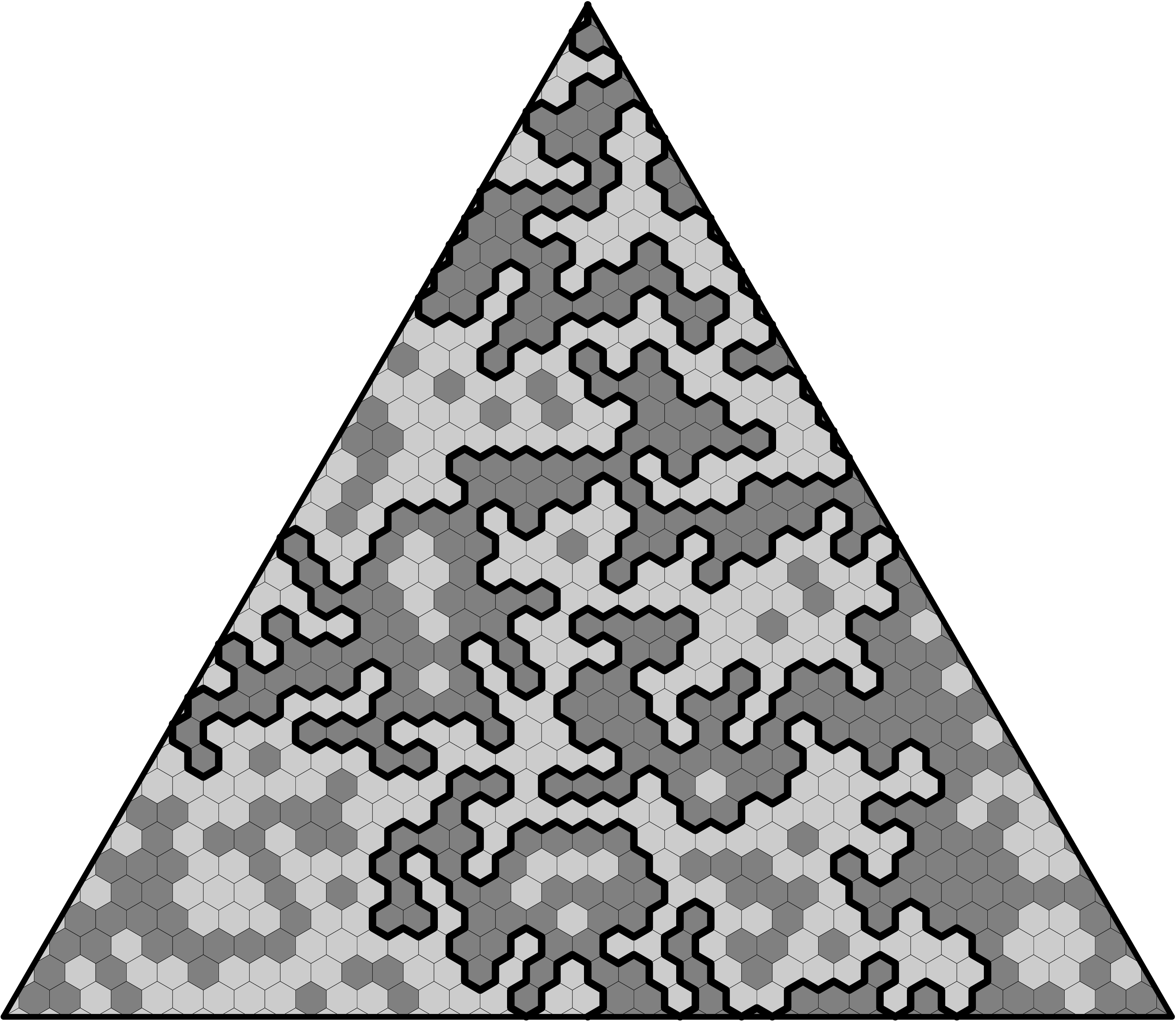}
\par\end{centering}

\caption{\emph{Critical percolation in a triangle, the exploration path starting
from the middle of the bottom side leaves white hexagons to its left
and black hexagons to its right.\label{fig: perco}}}
\end{figure}

We take a domain $T_{\delta}$ which is a triangular lattice approximation
of an equilateral triangle as in Section \ref{sub: lattice domains and conformal maps}.
Triangular lattice site percolation with parameter $p\in(0,1)$ associates
to each vertex of a domain in the triangular lattice (which we portray
as a hexagon, a face of the dual lattice) a color: white with probability
$p$ and black with probability $1-p$, independently. One studies
questions concerning connected components of sites of one color at
the critical parameter value $p=p_{c}=\half$. We impose white boundary
conditions on the left half of the boundary $\bdry T_{\delta}\cap\set{\re(z)<0}$,
and black on the right half $\bdry T_{\delta}\cap\set{\re(z)>0}$.
There is a unique path $\gamma_{\delta}$ on the dual lattice from
the midpoint of the bottom side of the triangle to the top vertex
of the triangle, leaving white vertices on the left and black vertices
on the right. This path, commonly called the percolation exploration
path, is our interface. Figure \ref{fig: perco} shows a realization
of the exploration path in $T_{\delta}$ with lattice mesh $\delta=\frac{1}{40}$.
Quite the contrary to Figure \ref{fig: LERW}, here there is no shortage
of places on the boundary that are visited by the path.

We define boundary visit as the event that the path $\gamma_{\delta}$
passes through the exteriormost corner $x$ of a hexagon next to the
boundary layer. The behavior of the boundary visit probabilities should
be given by Equation \eqref{eq: lattice interface boundary visit approximation},
where now $h=h_{1,3}(6)=\frac{1}{3}$ and $f=f_{T}:T\rightarrow\bH$
is a conformal map from the triangle to the half-plane given in Section
\ref{sub: lattice domains and conformal maps}.

\begin{figure}
\begin{centering}
\includegraphics[width=0.9\textwidth]{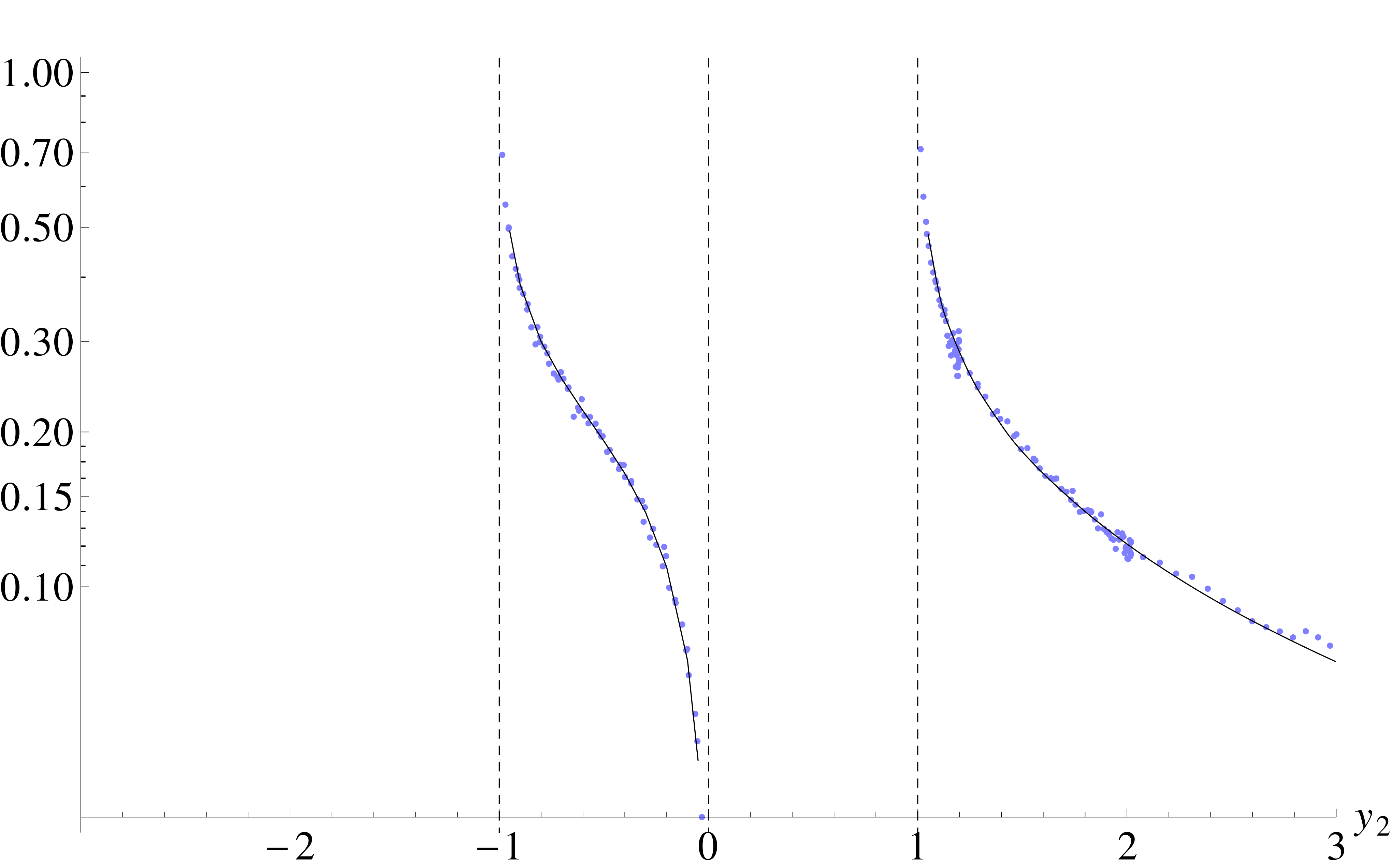}\\
\includegraphics[width=0.9\textwidth]{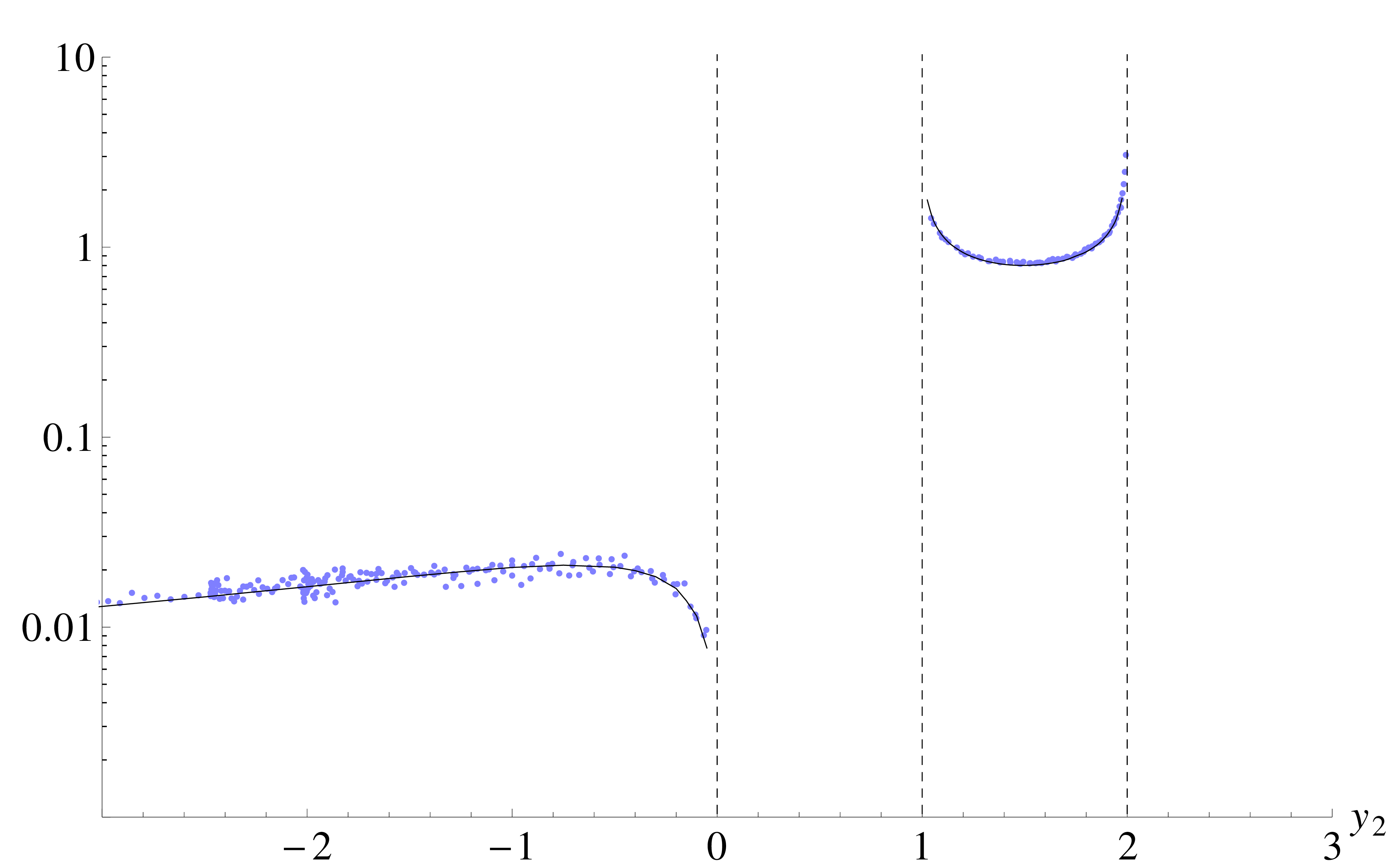}
\par\end{centering}

\caption{\emph{Data of three-point boundary visit frequencies collected from
simulations of critical percolation. In the upper plot we set $x=0$,
$y_{1}=1$, $y_{3}=-1$, and in the lower plot we set $x=0$, $y_{1}=1$,
$y_{3}=2$. In both plots the conformally corrected frequency is shown
as a function of $y_{2}$ on logarithmic scale. The solid curves are
multiples of the three-point boundary visit amplitudes $\zeta^{(3)}(x;y_{1},y_{2},y_{3})$
(that is, combinations of $\Ampl_{+--}$ and $\Ampl_{++-}$ on the
upper and of $\Ampl_{+-+}$ and $\Ampl_{+++}$ on the lower plot).
The fitted multiplicative constant is again the same for all the different
pieces.\label{fig: perco 3pt data}}}
\end{figure}

\begin{figure}
\begin{centering}
\includegraphics[width=0.9\textwidth]{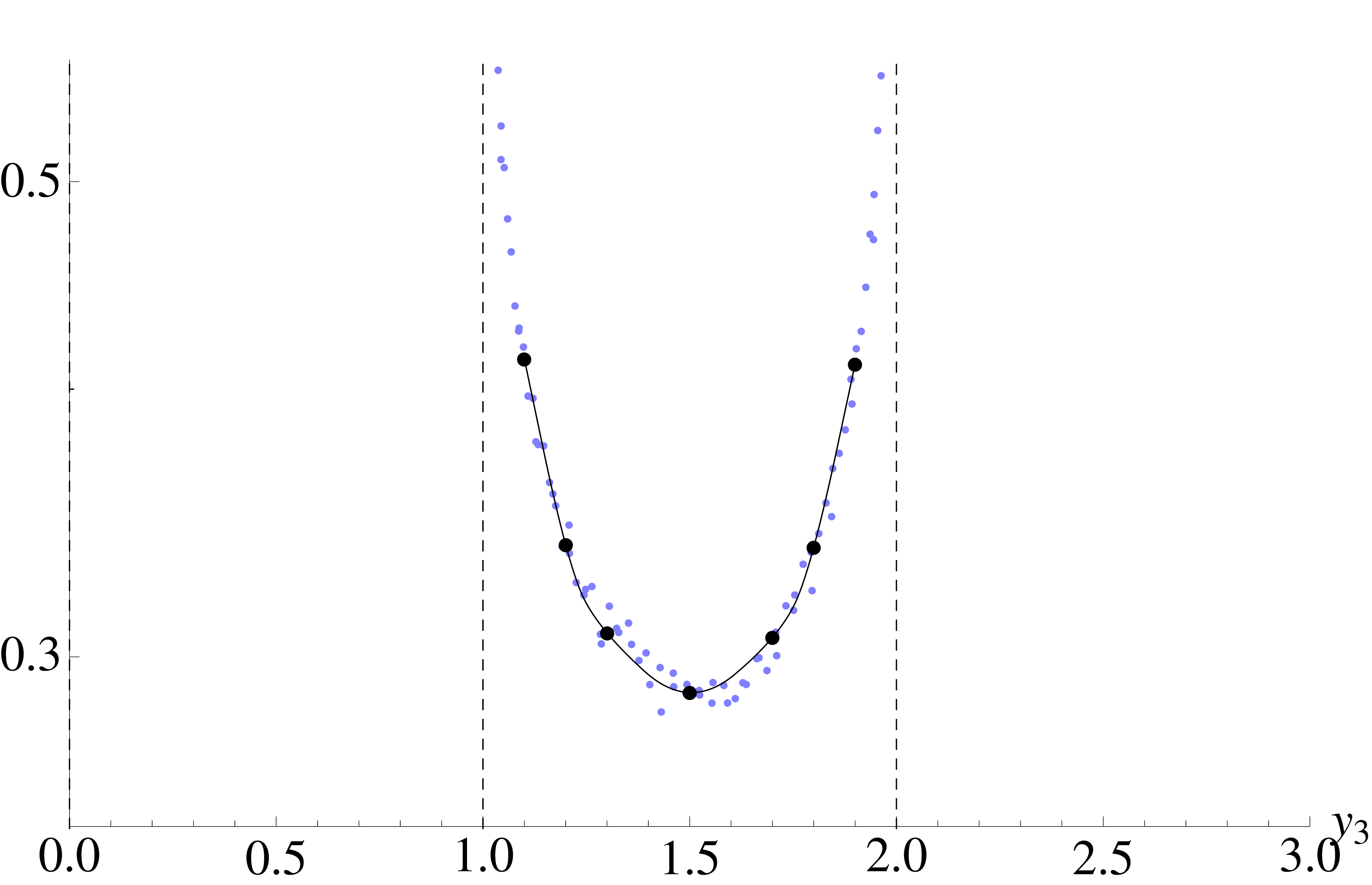}\\
\includegraphics[width=0.9\textwidth]{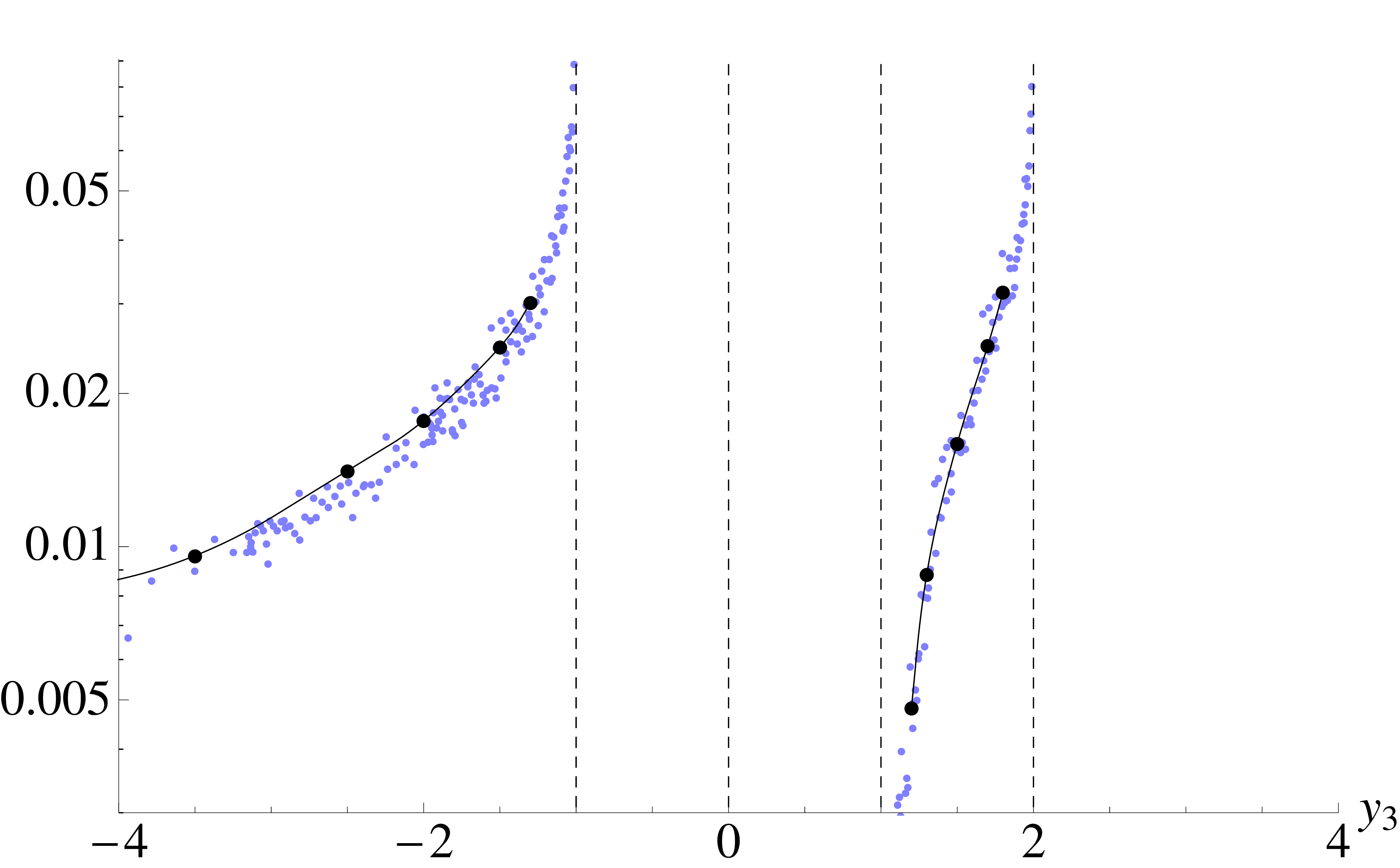}
\par\end{centering}

\caption{\emph{Data of four-point boundary visit frequencies collected from
simulations of critical percolation. On the upper plot we set $x=0$,
$y_{1}=-1$, $y_{2}=1$, $y_{4}=2$ and plot as a function of $y_{3}$.
On the lower plot we set $x=0$, $y_{1}=1$, $y_{2}=-1$, $y_{4}=2$
and plot as a function of $y_{3}$. The conformally corrected frequencies
in both plots are on a logarithmic scale. The solid curves are multiples
of the four-point boundary visit amplitudes $\zeta^{(3)}(x;y_{1},y_{2},y_{3},y_{4})$
(that is, $\Ampl_{-+++}$ on the upper and combinations of $\Ampl_{+-++}$
and $\Ampl_{+--+}$ on the lower plot). The fitted multiplicative
constant is again the same for all the different pieces.\label{fig: perco 4pt data}}}
\end{figure}

The simulation of percolation configurations hardly requires any comments.
The only computationally intensive step is to extract the interface
from the configuration. Another practical issue for high $N$, small
$\delta$ and large number of samples is the storage of the obtained
data of boundary visits. Once the data of boundary visit frequencies
is collected, we again correct them by dividing by the factor $\prod_{j=1}^{N}\left(|f_{T}'(x_{j})|\,\delta\right)^{h}$,
and then compare with our SLE boundary visit amplitude $\Ampl^{(N)}$
at $\kappa=6$. Figures \ref{fig: 1pt data}, \ref{fig: high kappa 2pt data},
\ref{fig: perco 3pt data}, and \ref{fig: perco 4pt data} present
data for $N=1,2,3,4$, respectively, obtained from simulations with
lattice mesh $\delta=\frac{1}{500}$ and $10^{5}$ realizations, with
lattice mesh $\delta=\frac{1}{300}$ and $2\times10^{6}$ realizations,
with lattice mesh $\delta=\frac{1}{80}$ and $10^{6}$ realizations,
and with lattice mesh $\delta=\frac{1}{160}$ and $2\times10^{8}$
realizations, respectively. The agreement with our analytical results
is nearly perfect. Note again that for any fixed $N$, only one multiplicative
constant has been fitted, and the ratios of the magnitudes of boundary
visit frequencies in different pieces of the plots are obtained from
our results.

\subsubsection{FK-model\label{sub: FK model}}

\begin{figure}
\begin{centering}
\includegraphics[width=0.6\textwidth]{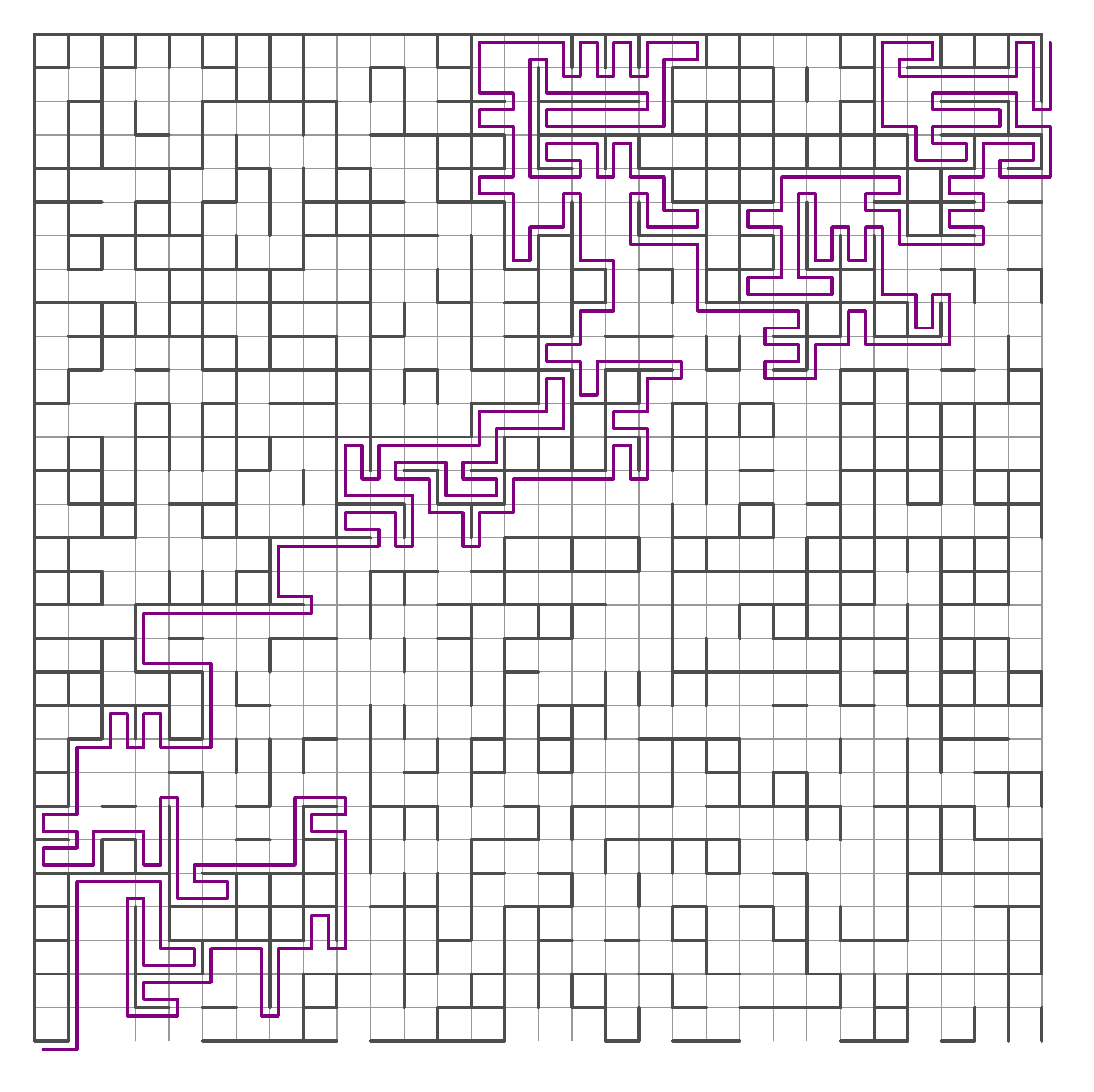}
\par\end{centering}

\caption{\emph{FK-model (random cluster model) interface closely follows the
outer boundary of the cluster connected to the wired part of the boundary:
the left and top sides.\label{fig: FK model}}}
\end{figure}

The random cluster model (also called FK-model, named after Fortuin
and Kasteleyn \cite{FK-on_the_random_cluster_model}) with parameters
$(p,Q)$ is a generalization of bond percolation, which for integer
values of $Q$ is closely related to the $Q$-Potts model. For $Q\in[0,4]$
it is expected to undergo a continuous phase transition at the critical
value $p=p_{c}(Q)=\frac{\sqrt{Q}}{1+\sqrt{Q}}$,%
\footnote{That this self-dual value is critical has been established in \cite{BD-the_self_dual_point_of_2d_RCM_is_critical}
for $Q\geq1$.%
} and behave conformally invariantly at the critical point. With Dobrushin
boundary conditions, there is an interface somewhat analogous to the
exploration path of percolation, which at the critical point is expected
to converge in the scaling limit to (chordal) $\SLEk$, where $\kappa=\kappa(Q)=\frac{4\pi}{\arccos(-\sqrt{Q}/2)}$.
The SLE scaling limit is rigorously known in two special cases: the
case $Q=2$ is known as the FK-Ising model and the techniques of \cite{Smirnov-towards_conformal_invariance,Smirnov-conformal_invariance_in_RCM_1}
led to a proof \cite{CDHKS-convergence_of_Ising_interfaces_to_SLE},
and the limiting case $Q=0$ corresponds to the uniform spanning tree
treated in \cite{LSW-LERW_and_UST}. Figure \ref{fig: FK model} shows
a realization of $Q=4$ FK-model interface with lattice mesh $\delta=\frac{1}{30}$,
together with the interface.

It is worth noticing that the probabilities of boundary visits of
the interface can be used to express the boundary magnetization, and
more generally boundary spin correlation functions of the Potts model,
with one of the boundary arcs having fixed spin. These exemplify some
of the physical applications of the boundary visit problem.

For simulations in this article we restrict our attention to the values
$Q=2$ and $Q=3$. Integer values of $Q$ are convenient because there
exists a Monte Carlo Markov chain by Swendsen and Wang, which does
not suffer as much of critical slowing down as the more common Markov
chains based on local updates \cite{SW-nonuniversal_critical_dynamics_in_Monte_Carlo_simulations}.
This efficiency of simulation is important, because we need good statistics
to get accurate information about the small probability events of
multiple boundary visits. Swendsen-Wang algorithm works for all integer
$Q$, but for $Q>4$ the model has a first order phase transition
and does not exhibit conformal invariance. For $Q=4$ the finite size
corrections scale too badly for reliable simulations.

We define the model in the lattice approximation $S_{\delta}$ of
the unit square $S$ given in Section \ref{sub: lattice domains and conformal maps}.
The random cluster model is a random subset $\omega$ of edges of
$S_{\delta}$, with probability proportional to
\begin{align*}
\PR_{(p,Q)}[\set{\omega}]\propto & \left(\frac{p}{1-p}\right)^{|\omega|}Q^{k(\omega)},
\end{align*}
where $k(\omega)$ denotes the number of connected components (``clusters'')
of the subgraph of $S_{\delta}$ defined by all vertices and the edges
$\omega$. The appropriate Dobrushin boundary conditions amount to
conditioning on the event that all edges of the left and top boundaries
of the square are in $\omega$. The interface $\gamma_{\delta}$ is
the path obtained as the boundary of the $\frac{\delta}{4}$-thickening
of the component connected to the left and top, i.e., a path closely
surrounding the ``wired cluster'', see Figure \ref{fig: FK model}.

The interface being defined on a lattice different from the square
lattice, it is now natural to define boundary visits to points with
half-lattice-unit coordinates. Moreover, the wiring of the boundary
introduces some asymmetry in the definition. On the bottom we say
that $(x+\half)\delta$ is visited if the path goes outside the domain
at $(x+\half)\delta-\frac{\delta}{4}\ii$, and on the right a similar
definition is used. On the left we say that $\ii(y+\half)\delta$
is visited if the path comes to the point $\ii(y+\half)+\frac{\delta}{4}$,
and on the top a similar definition is used. These definitions are
natural, as is illustrated by the figure of the interface. The behavior
of the boundary visit probabilities should again be given by Equation
\eqref{eq: lattice interface boundary visit approximation}, where
now $h=h_{1,3}(\frac{16}{3})=\frac{1}{2}$ for $Q=2$ and $h=h_{1,3}(\frac{24}{5})=\frac{2}{3}$
for $Q=3$, and $f=f_{S}:S\rightarrow\bH$ is the conformal map from
the unit square to the half-plane as in Section \ref{sub: lattice domains and conformal maps}.

\begin{figure}
\begin{centering}
\includegraphics[width=0.8\textwidth]{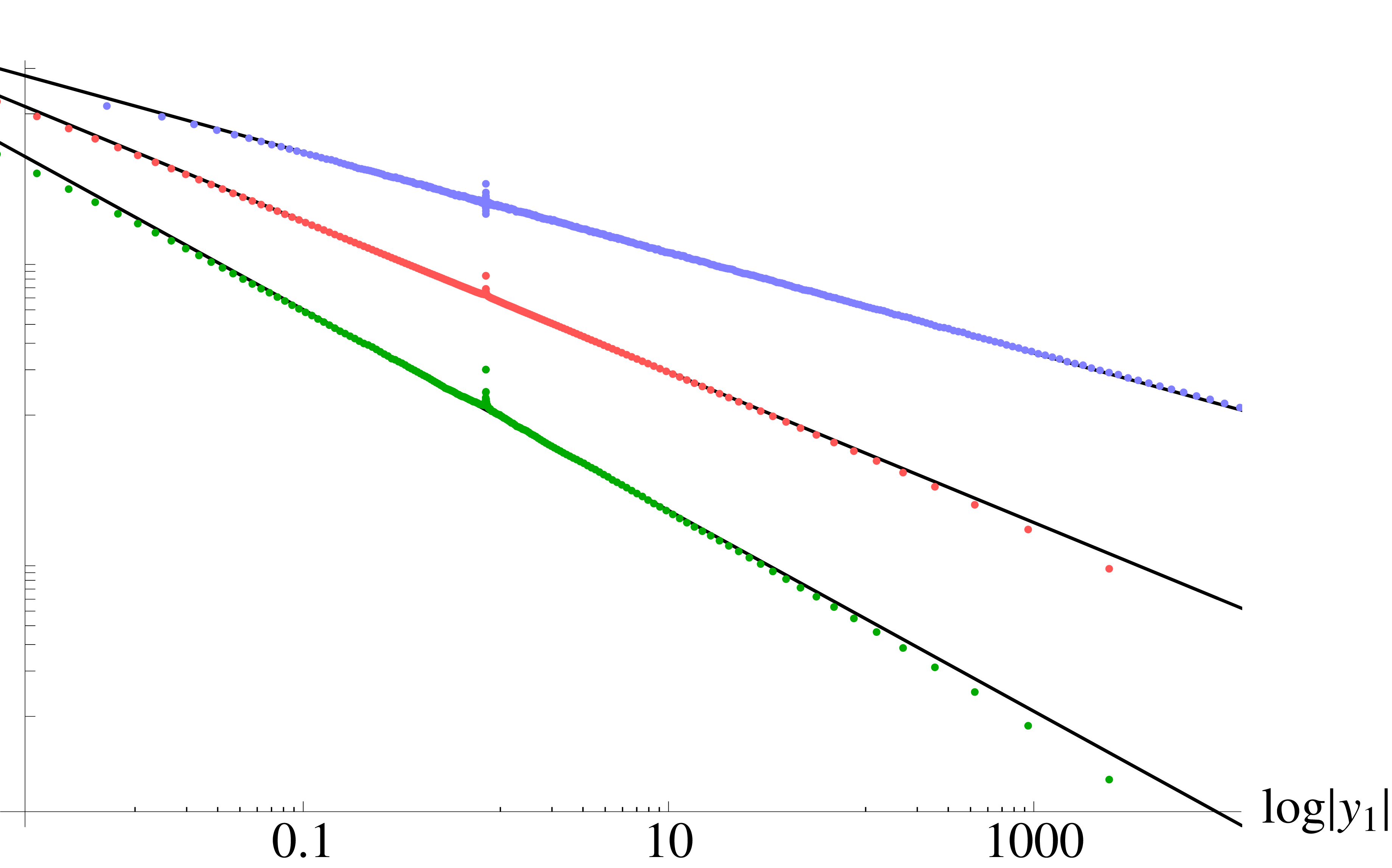}\\
\includegraphics[width=0.8\textwidth]{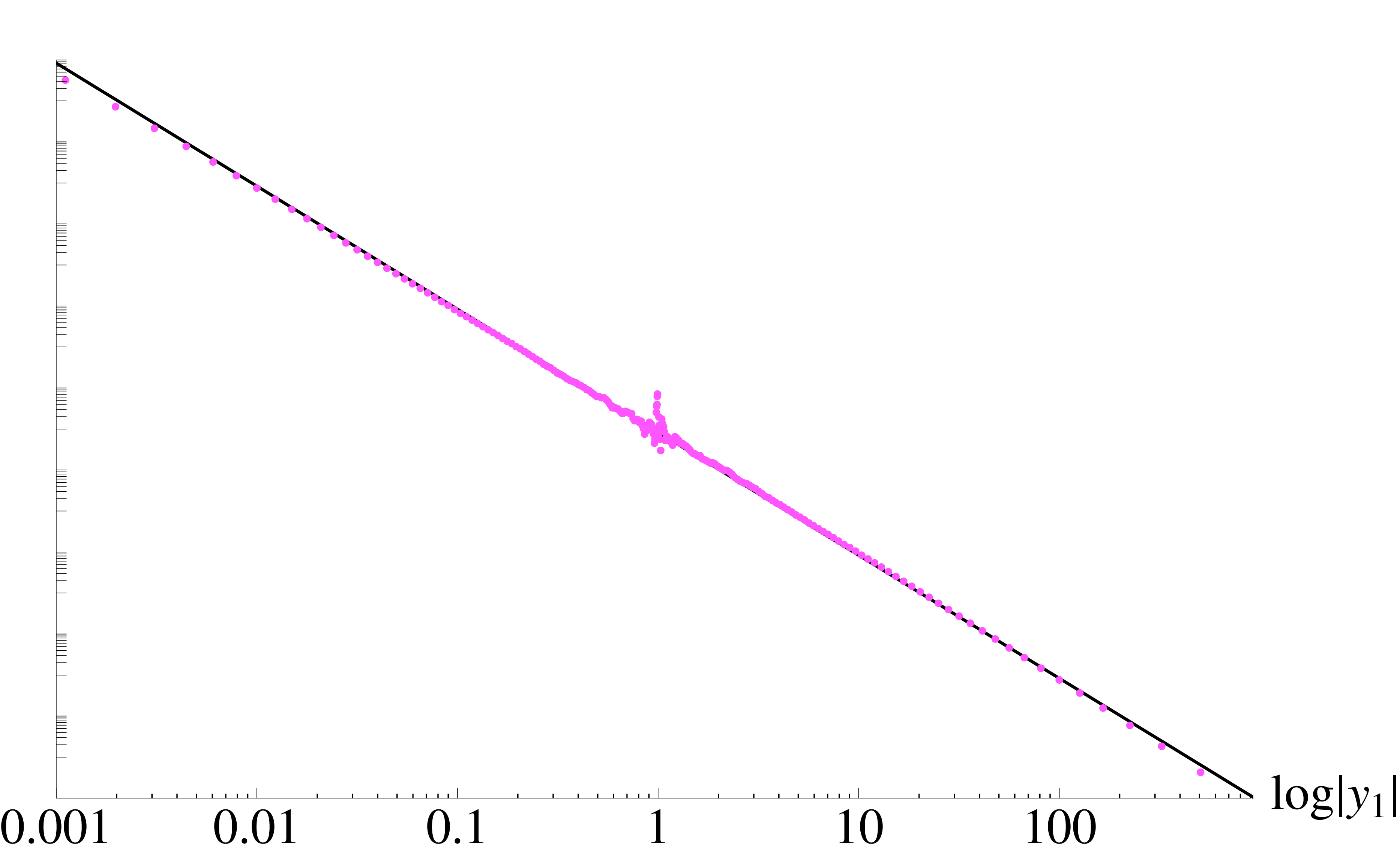}
\par\end{centering}

\caption{\emph{Data of one-point boundary visit frequencies collected from
simulations of lattice models. We have set $x=0$ and plotted the
conformally corrected frequency of visits as a function of $y_{1}$
on log-log scale. The solid lines are fitted power laws, in accordance
with $\Ampl^{(1)}(x,y_{1})\propto|y_{1}-x|^{-h}$. 
The simulations
are done in polygonal domains (triangle for percolation and square
for the other models), and the bumps in the data in the middle of
the plots are due to a corner of the polygonal domain. }\protect \\
Upper plot\emph{: percolation (top, blue), FK-Ising model (middle,
red), FK model with $Q=3$ (bottom, green)}\protect \\
Lower plot\emph{: loop-erased random walk\label{fig: 1pt data}}}
\end{figure}

\begin{figure}
\begin{centering}
\includegraphics[width=0.75\textwidth]{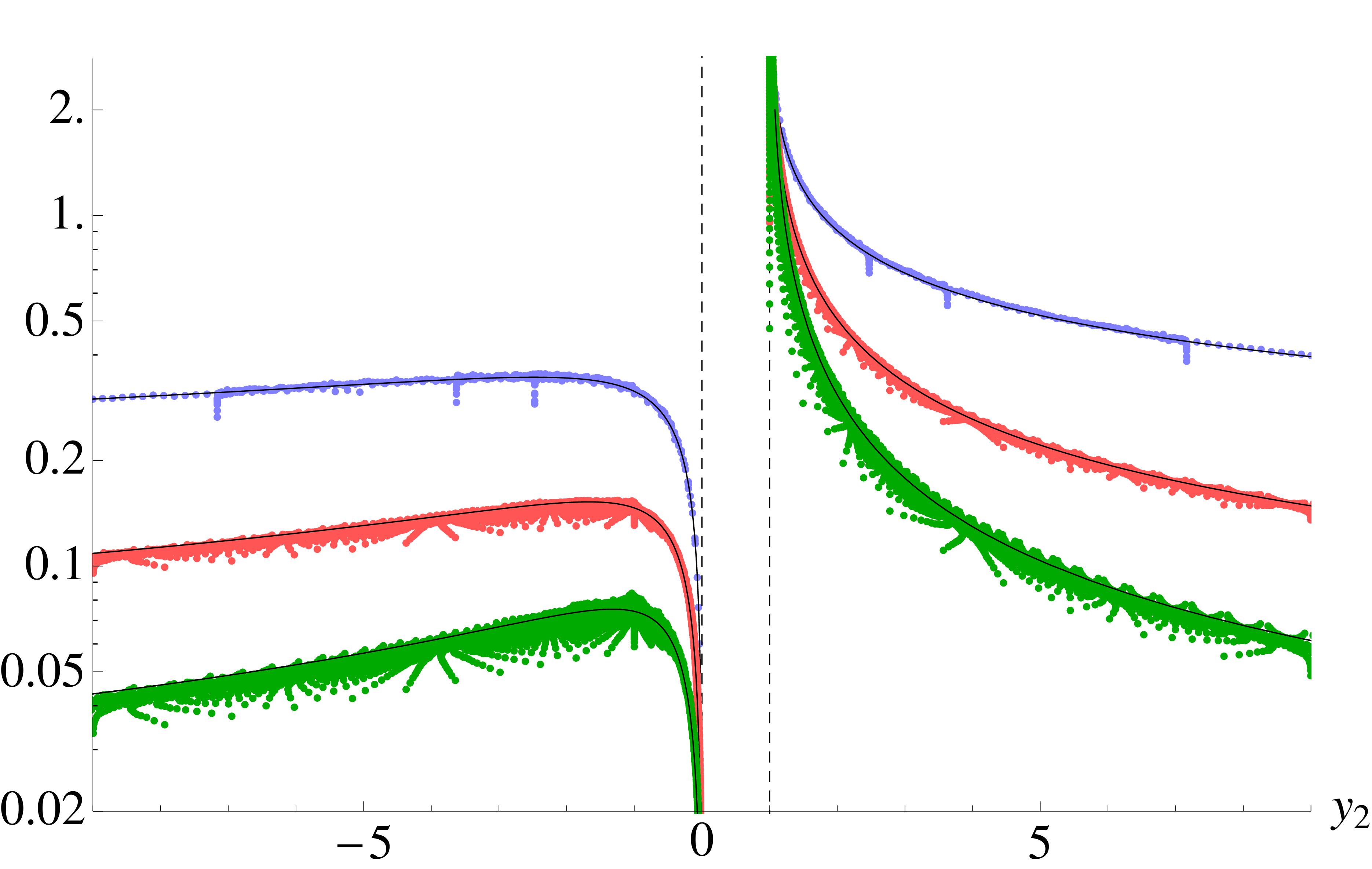}
\par\end{centering}

\caption{\emph{Data of two-point boundary visit frequencies collected from
simulations of lattice models: percolation (top, blue), FK-Ising model
(middle, red), FK-model with $Q=3$ (bottom, green). We set $x=0$,
$y_{1}=1$ and plot the conformally corrected frequency as a function
of $y_{2}$ on logarithmic scale. The solid curves are multiples of
the two-point boundary visit amplitudes $\zeta^{(2)}(x;y_{1},y_{2})$,
with the same multiplicative constant used for the two pieces: $\zeta_{++}(x;y_{1},y_{2})$
when $y_{2}>1$ and $\zeta_{+-}(y_{2};x;y_{1})$ when $y_{2}<0$.
For FK-Ising we have used the known exact multiplicative constant
from \cite{HK-Ising_interfaces_and_free_boundary_conditions}, for
other models this non-universal constant is fitted to data.\label{fig: high kappa 2pt data}}}
\end{figure}

\begin{figure}
\begin{centering}
\includegraphics[width=0.9\textwidth]{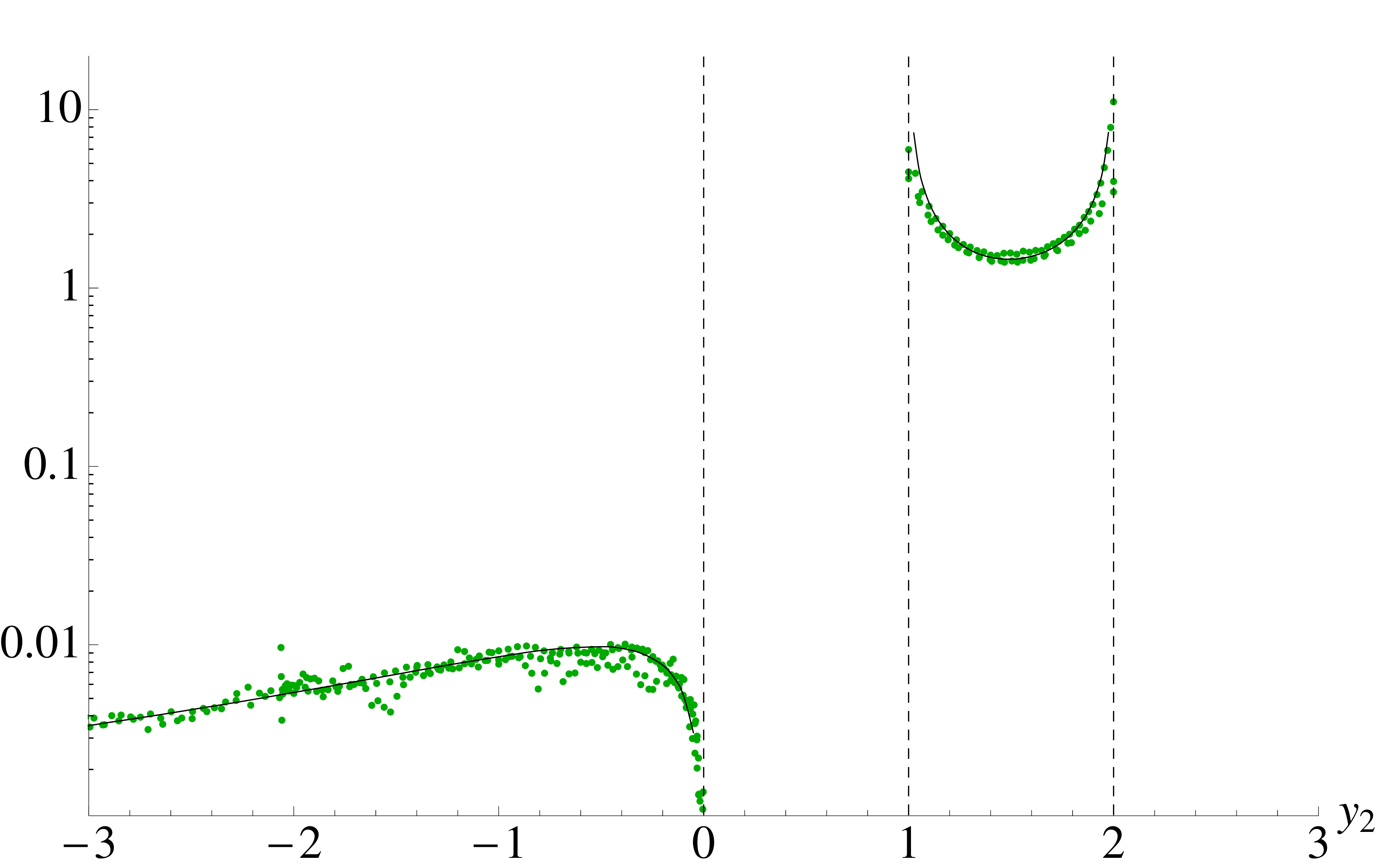}\\

\par\end{centering}

\caption{\emph{Data of three-point boundary visit frequencies collected from
simulations of FK random cluster model with $Q=3$. In this plot we
set $x=0$, $y_{1}=1$, $y_{3}=2$. The plot shows conformally corrected
frequency as a function of $y_{2}$ on logarithmic scale. The solid
curves are multiples of the three-point boundary visit amplitudes
$\zeta^{(3)}(x;y_{1},y_{2},y_{3})$ (that is, combinations of $\Ampl_{+-+}$
and $\Ampl_{+++}$). The fitted multiplicative constant is again the
same for the different pieces.\label{fig: FK Q3 3pt data}}}
\end{figure}

Our simulation runs the Swendsen-Wang Monte Carlo Markov chain and
collects time averages of the boundary visiting events. Neither the
initial transient nor the autocorrelation time at the stationary distribution
cause any noticeable statistical errors --- the inevitable trade-off
between finite size effects and computational time is the main source
of numerical error. We correct the boundary visit frequencies obtained
from the simulations by dividing by the factor $\prod_{j=1}^{N}\left(|f_{S}'(x_{j})|\,\delta\right)^{h}$,
and then compare with our SLE boundary visit amplitude $\Ampl^{(N)}$
at $\kappa=\kappa(Q)$. For $N\leq3$ we get good enough statistics
and the agreement with our analytical results is very good: Figures
\ref{fig: 1pt data} and \ref{fig: high kappa 2pt data} show $N=1$
and $N=2$ data for both $Q=2$ and $Q=3$, with $\delta=\frac{1}{100}$
and $10^{7}$ samples in each case. We have included the plot of three-point
boundary visit data in Figure \ref{fig: FK Q3 3pt data} only for
$Q=3$ because the value of $\kappa$ ($\kappa=\frac{24}{5}$) is
sufficiently different from the case of percolation ($\kappa=6$)
so that the shapes of the functions are clearly distinct (for this
we use $\delta=\frac{1}{100}$ and $5\times10^{6}$ samples).

We still point out how remarkably much is known of the FK-Ising case
$Q=2$, largely owing to the techniques of discrete complex analysis
\cite{Smirnov-conformal_invariance_in_RCM_1,CS-discrete_complex_analysis_on_isoradial_graphs,CS-universality_in_2d_Ising,Smirnov-discrete_complex_analysis_and_probability}.
This is the only lattice model for which the scaling limit of renormalized
boundary visiting probabilities has in fact been proven to exist,
and even the corresponding non-universal constants for $N=1$ and
$N=2$ have been found explicitly \cite{HK-Ising_interfaces_and_free_boundary_conditions}.
The exact $N=1$ formula reads for $x$ away from the corners
\begin{align*}
\frac{1}{\sqrt{\delta}}\;\frac{1}{\sqrt{|f_{S}'(x)|}}\;\PR_{\mathrm{FK-Ising}}\left[\gamma_{\delta}\text{ visits }x\right]\;\underset{\delta\searrow0}{\longrightarrow}\quad & \sqrt{\frac{1+\sqrt{2}}{2\pi}}\times|f_{S}(x)|^{-1/2}.
\end{align*}
We find excellent numerical agreement of the exponent value (best
fit gives $0.499872$ instead of $\half$) and the non-universal multiplicative
constant (best fit gives $0.618241$ instead of $\sqrt{\frac{1+\sqrt{2}}{2\pi}}\approx0.619866$).
The exact $N=2$ formula reads for $x_{1},x_{2}$ away from corners
and on the same side
\begin{align*}
 & \frac{1}{\delta}\;\frac{1}{\sqrt{|f_{S}'(x_{1})|\,|f_{S}'(x_{2})|}}\;\PR_{\mathrm{FK-Ising}}\left[\gamma_{\delta}\text{ visits }x_{1}\text{ then }x_{2}\right]\\
\underset{\delta\searrow0}{\longrightarrow}\quad & \frac{(4+2\sqrt{2})\,\Gamma(\frac{3}{4})^{2}}{\pi^{5/2}}\,\times\,\frac{\phantom{}_{2}F_{1}\left(\frac{-1}{2},\frac{3}{4};\frac{3}{2};1-\frac{f_{Q}(x_{1})}{f_{Q}(x_{2})}\right)}{\sqrt{f_{Q}(x_{1})}\;\sqrt{f_{Q}(x_{2})-f_{Q}(x_{1})}}.
\end{align*}
The solid line in the middle plot in Figure \ref{fig: high kappa 2pt data}
uses this explicit non-universal multiplicative constant. This comparison
to an exact scaling limit result gives a fair idea of the finite size
effects present in the simulation data of the FK-Ising model, but
one must remember that the finite size corrections scale differently
for other models.

\noindent 

\noindent 

\noindent 
\begin{figure}
\begin{centering}
\includegraphics[width=0.75\textwidth]{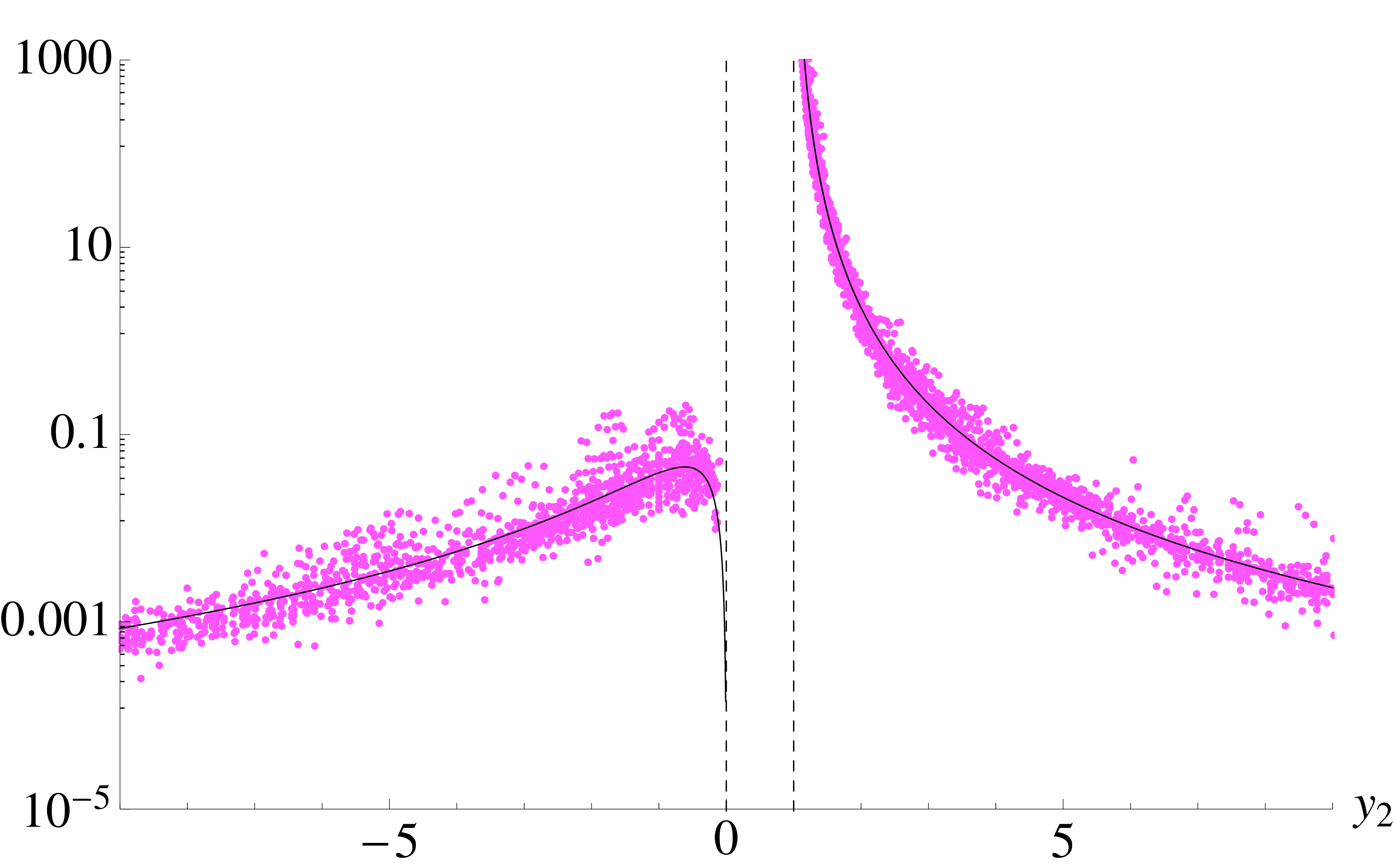}
\par\end{centering}

\caption{\emph{Data of two-point boundary visit frequencies collected from
simulations of LERW. We set $x=0$, $y_{1}=1$ and plot the conformally
corrected frequency as a function of $y_{2}$ on logarithmic scale.
The solid curves are multiples of the two-point boundary visit amplitudes
$\zeta^{(2)}(x;y_{1},y_{2})$, with again the fitted multiplicative
constant being the same for the two pieces.\label{fig: LERW 2pt data}}}
\end{figure}

\noindent 

\noindent 

\noindent 

\subsection{Simulation data and results of the comparison}

Simulation data and corresponding plots of our analytical results
are presented in Figures \ref{fig: 1pt data}~---~\ref{fig: perco 4pt data}.
The general conclusion is that the boundary visit probabilities of
lattice model interfaces are in agreement with the predictions of
type \eqref{eq: lattice interface boundary visit approximation},
where the amplitudes $\Ampl^{(N)}$ are given by our main results.
The main source of numerical error is finite size effects.

Figure \ref{fig: 1pt data} shows one-point visit amplitudes on a
$\log$-$\log$ scale. The data from all models follows the power
law $\Ampl^{(1)}(x;y)=|y-x|^{-h}$ over a range of scales. The slope
$h$ is so different for $\kappa=2$ that we have included a separate
plot for the LERW case. Particular finite size effects caused by error
near the corners of the polygonal domain (triangle or square) are
seen as bumps in the data. This effect diminishes for smaller $\delta$,
but it is visibly present in our data for all $N$. We have centered
the $N=1$ data so that the bump appears in the middle of the plot.
For $N\geq2$ this error affects a part of the data points across
the whole range of the plot, resulting in an apparent failure of a
perfect data collapse seen as thickness of the data point cloud.

Figures \ref{fig: high kappa 2pt data} and \ref{fig: LERW 2pt data}
show two-point boundary visit data on a logarithmic scale both in
the case where the points $y_{1},y_{2}$ to be visited are on the
same side and in the case where they are on different sides. We have
scaled to the case $y_{1}=1$ and plotted as a function of $y_{2}$,
so that ideally all data from a given model should collapse on the
curve constructed from the two pieces $\Ampl_{++}^{(2)}(0;1,y_{2})$
(for $y_{2}>1$) and $\Ampl_{+-}^{(2)}(0;1,y_{2})$ (for $y_{2}<0$).
The same fitted multiplicative constant is used on both pieces for
each model, and a clear agreement is observed in all cases. For the
FK-Ising model case we have even avoided fitting, as we have been
able to use the rare known explicit non-universal constant mentioned
in Section \ref{sub: FK model}. Data from all models show some finite
size effects, and roughly these are worse for smaller $\kappa$. The
functional shape of all plots is nevertheless clearly correct. Again
the shape for $\kappa=2$ is so different from others that we have
plotted it separately.

Figures \ref{fig: perco 3pt data} and \ref{fig: FK Q3 3pt data}
show three-point boundary visit data on a logarithmic scale for critical
percolation and the critical $Q=3$ FK-model, respectively. Data from
percolation are still very well on the curves of our analytical results.
In the $Q=3$ FK-model the finite size effects are more apparent.
Again, a single fitted multiplicative constant has been used for all
pieces. In particular the several orders of magnitude difference of
the boundary visit frequencies on the two sides of Figure \ref{fig: FK Q3 3pt data}
is in excellent agreement with our analytical results, even if, due
to finite size effects, the data points otherwise only serve to give
a sketchy idea of the shape of the function here.

Figure \ref{fig: perco 4pt data} shows four-point boundary visit
data on a logarithmic scale for critical percolation. Both the numerical
evaluation of our results $\Ampl^{(4)}$ and decent simulation results
are starting to be computationally very heavy --- we have had to interpolate
the analytical result from the calculations at the points shown on
the plots. Nevertheless, the plot shows agreement of simulation data
with our result.

\bigskip{}

\section{Conclusions and outlook\label{sec: conclusions}}

We have presented a method based on quantum group calculations, which
gives explicit solutions of the chordal $\SLEk$ boundary visit probability
amplitudes $\Ampl^{(N)}$ and $\Corr^{(N)}$ for arbitrary numbers
$N$ of marked boundary points. The answers are expressed in terms
of linear combinations of Coulomb gas integrals, and can be transformed
to regularized real integrals. They give the universal answer to various
formulations of the SLE boundary visit question, up to an overall
non-universal constant, which depends on the formulation. In particular,
they give the renormalized scaling limit boundary visit probabilities
for lattice model interfaces.

Our results are obtained by solving a partial differential equation
system with boundary conditions given recursively by the solutions
with smaller number $N$ of marked points. The system is suggested
by plausible considerations of asymptotics, but we have not fully
justified the use of this procedure. In an ongoing work with Konstantin
Izyurov we plan to implement the strategy outlined in Section \ref{sub: conditioning application and proof strategy}
to prove rigorously that the formulas obtained in the present article
indeed give the SLE multi-point Green's functions on the boundary.

The method we have used is an application of the spin chain - Coulomb
gas correspondence presented in a more general setup in \cite{KP-covariant_boundary_correlations},
and applied to the problem of multiple SLE pure geometries and
crossing probabilities in \cite{KP-pure_partition_functions_of_multiple_SLEs}. The method provides a systematic approach
to a class of SLE and CFT problems depending on arbitrary numbers
of marked points. It works directly only for irrational values of
$\kappa$, but for questions such as boundary visit amplitudes, one
can naturally extend the final results to all $\kappa$ by requiring
continuity. It would be interesting to generalize the spin chain -
Coulomb gas correspondence itself to rational values of $\kappa$.
This would presumably involve non-semisimple representation theory
of the corresponding quantum group as well as results that correspond
to logarithmic conformal field theory correlation functions.

It would be interesting to find also formulas for boundary visit probabilities
for other variants of SLE, such as the radial $\SLEk$ and dipolar
$\SLEk$, $\SLEk(\rho)$, or even more general variants. Finally,
one of the most natural remaining open questions about Schramm-Loewner
evolutions is the bulk analogue of the question answered in the present
article: finding a formula for the multi-point Green's function of
the chordal SLE (for recent progress on this, see \cite{RS-basic_properties,Beffara-dimension_of_the_SLE_curves,LS-natural_parametrization_of_SLE,LW-multi_point_Greens_functions_for_SLE}).
\begin{quotation}
\vfill{}

\end{quotation}
\textbf{\emph{Acknowledgments}}\emph{:}

\smallskip{}

Konstantin Izyurov and Eveliina Peltola have shared with us many of
their insights during related collaborations and discussions. We also
thank Dmitry Beliaev, Denis Bernard, Steven Flores, Christian Hagendorf,
Clément Hongler, Peter Kleban, Antti Kupiainen, Greg Lawler, Jacob
Simmons, and Stanislav Smirnov for interesting discussions and helpful
comments. We also thank the anonymous referees for useful comments.
This work was initiated in the ISF workshop ``Random matrices
and integrability: from theory to applications'' in Yad Hashmona,
and parts of it were carried out at Technion and University of Haifa
at Oranim, at the University of Southern Denmark in Odense, and at
the University of Geneva --- we gratefully acknowledge the hospitality.
We also acknowledge the CESGA (Centro de Supercomputación de Galicia)
Supercomputing Center for computational time. It would be impossible
to list the innumerable breweries which provided constant inspiration
during the course of this work.

N.J. is funded in part by the Spanish grant FPA2011-22594, by Xunta
de Galicia (Conselleria de Educación, grant INCITE09-206-121-PR and
grant PGIDIT10PXIB206075PR), by the Consolider-Ingenio 2010 Programme
CPAN (CSD2007-00042), and by FEDER. N.J. is also supported by the
Juan de la Cierva program.

M.J. was supported in part by grants PERG07-GA-2010-268246, PIF-GA-2011-300984,
the EU program ``Thales'' and ``HERAKLEITOS II'' ESF/NSRF 2007-2013
and was also co-financed by the European Union (European Social Fund,
ESF) and Greek national funds through the Operational Program ``Education
and Lifelong Learning'' of the National Strategic Reference Framework
(NSRF) under ``Funding of proposals that have received a positive
evaluation in the 3rd and 4th Call of ERC Grant Schemes''.

K.K. is supported by the Academy of Finland grant ``Conformally invariant
random geometry and representations of infinite dimensional Lie algebras''.

\bigskip{}

\appendix

\section{SLE derivations of the exponent and a PDE\label{sec: derivations of the value of h}}

This appendix provides SLE calculations for the $N=1$ case,
to establish the same value of the exponent $h$ with the two alternative notions of
boundary visits given in Section~\ref{sub: different definitions of boundary visit}.
Visits to small boundary intervals are treated in \ref{sub: touching boundary interval},
and visits to small conformal distance neighborhoods in \ref{sub: reaching small conformal radius}.
The latter implies up to constant bounds for the probabilities of boundary visits with the notion
used in the introduction, since the conformal distance $\rho_{\bH\setminus \gamma}(y)$ is
proportional to the ordinary distance $d(\gamma,y)$. This up to constants estimate had
also been derived differently in \cite{AK-intersection_probabilities_for_SLE_and_semicircle}.
The work \cite{Lawler-Minkowski_SLE_real_line} establishes
the existence of the SLE boundary Green's function in complete generality.

In \ref{sec: 2nd order PDE} we relate the second order differential
equation \eqref{eq: second order differential equation} to a martingale for the chordal SLE.

We do not provide a direct justification of the third order differential
equations \eqref{eq: third order differential equations} for the boundary visit amplitudes
with SLE analysis, but instead only discuss them from the point of view of conformal field theory
in Appendix~\ref{sub: singular vectors}.
We nevertheless note that in
\cite{Dubedat-SLE_and_Virasoro_representations_localization,Dubedat-SLE_and_Virasoro_representations_fusion} and
\cite{KP-covariant_boundary_correlations,KP-pure_partition_functions_of_multiple_SLEs} these
equations were shown to hold for limiting cases of multiple SLE partition functions,
and it is natural to interpret the boundary visiting SLE as a degeneration of such
multiple SLEs.

\subsection{Touching a small boundary interval\label{sub: touching boundary interval}}

One can write down the exact solution for the probability of a chordal
SLE to hit a boundary interval $[y,y+\eps]$ (for $y>x$) and do the
asymptotics as $\eps\searrow0$, see, e.g., \cite{BB-zig_zag,AS-Hausdorff_dimension_of_SLE_real_line}.
We include the argument briefly here.

Assume that $x<l<r$ and let $P(x,l,r)$ be the probability that a
chordal $\SLEk$ in the half-plane $\bH$ from $x$ to $\infty$ touches
the interval $[l,r]$, and note that by translation and scaling invariance
it can be reduced to a function of one variable,
\begin{align*}
P(x,l,r):=\; & \PR_{(\bH;x,\infty)}[\gamma\cap[l,r]\neq\emptyset], & P(x,l,r)=\; & p\left(\frac{l-x}{r-x}\right).
\end{align*}
By domain Markov property we create a martingale $(M_{t})_{t\geq0}$:
we define $M_{t}$ as the above probability conditionally on the knowledge
of an initial segment $\gamma[0,t]$
\begin{align*}
M_{t}=\; & \PR_{(\bH;x,\infty)}\left[\gamma\cap[l,r]\neq\emptyset\;\big|\;\mathcal{F}_{t}\right]=\PR_{(H_{t};\gamma(t),\infty)}\left[\gamma\cap[l,r]\neq\emptyset\right].
\end{align*}
By conformal invariance under the map $g_{t}$ in \eqref{eq: Loewner chain in H}
this can be written as
\begin{align*}
M_{t}=\; & \PR_{(\bH;X_{t},\infty)}\left[\gamma\cap[g_{t}(l),g_{t}(r)]\neq\emptyset\right]=P\left(X_{t},\, g_{t}(l),\, g_{t}(r)\right).
\end{align*}
Stochastic calculus tells that for this to be a martingale, the drift
term
\begin{align*}
 & \frac{\kappa}{2}\frac{\partial^{2}}{\partial x^{2}}P+\frac{2}{l-x}\pder lP+\frac{2}{r-x}\pder rP
\end{align*}
in the Itô derivative must vanish. This is an ordinary differential
equation for $p$,
\begin{align*}
p''(u)+\frac{-4+(2\kappa-4)u}{\kappa u(1-u)}\, p'(u)=\; & 0.
\end{align*}
Integrating with the boundary conditions $p(0)=1$, $p(1)=0$ we obtain
that (for $4<\kappa<8$)

\begin{align*}
\PR_{(\bH;x,\infty)}\Big[\gamma\cap[l,r]\neq\emptyset\Big]=\; & \frac{4\,\sqrt{\pi}}{2^{8/\kappa}\,\Gamma(\frac{8-\kappa}{2\kappa})\,\Gamma(\frac{\kappa-4}{\kappa})}\;\int_{\frac{l-x}{r-x}}^{1}u^{-\frac{4}{\kappa}}\;(1-u)^{2\frac{4-\kappa}{\kappa}}\;\ud u.
\end{align*}
From this exact answer we find that the probability of hitting a small
interval of size $\eps$ at $y$ scales as $\eps^{h}$ with amplitude
$|y-x|^{-h}$

\begin{align}
\PR_{(\bH;x,\infty)}\Big[\gamma\cap[y,\, y+\eps]\neq\emptyset\Big]\sim\; & \eps^{\frac{8-\kappa}{\kappa}}\,\frac{4\,\sqrt{\pi}\,\kappa}{(8-\kappa)\,2^{8/\kappa}\,\Gamma(\frac{8-\kappa}{2\kappa})\,\Gamma(\frac{\kappa-4}{\kappa})\,}\,(y-x)^{\frac{\kappa-8}{\kappa}}.\label{eq: touching small intervals 1pt function}
\end{align}
Also the multiplicative constant in
\begin{align*}
\lim_{\eps\searrow0}\left(\eps^{-h}\times\PR[\gamma\cap I_{\eps}(y)\neq\emptyset]\right)=\; & \const\times\Ampl^{(1)}(x;y)
\end{align*}
is explicit here, but it is given by a somewhat complicated expression,
and such constants are in any case non-universal.

\subsection{Reaching a small conformal distance from boundary point\label{sub: reaching small conformal radius}}

Another derivation of the scaling exponent is based on the notion
of boundary visit defined in terms of conformal distance. Namely,
one can find explicitly the asymptotics of the probability that the
chordal SLE reaches a small conformal distance from a marked boundary
point. The strategy is similar to the above, but the martingale argument
leads to a parabolic partial differential equation, which we do not
solve explicitly, but instead we just find the leading eigenvector
and eigenvalue of the generator, and hence deduce the small neighborhood
size asymptotics of solutions.

For the martingale argument we need to keep track of one more point,
the rightmost point $r$ in the image of the SLE hull. Choose therefore
$x<r<y$ and let $Q(x,r,y,s)$ be the probability that for a chordal
$\SLEk$ $\gamma$ in the half-plane $\bH$ from $x$ to $\infty$
the conformal radius of $y$ in $\bH\setminus(\gamma\cup(-\infty,r])$
(with a Schwarz reflection as before) is at most $e^{-s}$. In the
limit $r\searrow x$ this correctly measures the conformal distance
to the curve $\gamma$ only. By translation and scaling invariance
$Q$ can be reduced to a function of two variables,
\begin{align*}
Q(x,r,y,s):=\; & \PR_{(\bH;x,\infty)}[\rho_{\bH\setminus(\gamma\cup(-\infty,r])}(y)\leq e^{-s}], & Q(x,r,y,s)=\; & q\left(\frac{r-x}{y-r},\; s+\log(y-r)\right).
\end{align*}
By domain Markov property we again create a martingale $(M_{t})_{t\geq0}$
\begin{align*}
M_{t}=\; & \PR_{(\bH;x,\infty)}\left[\rho_{\bH\setminus(\gamma\cup(-\infty,r])}(y)\leq e^{-s}\;\big|\;\mathcal{F}_{t}\right],
\end{align*}
and by conformal invariance we write it as
\begin{align*}
M_{t}=\; & \PR_{(\bH;X_{t},\infty)}\left[\rho_{\bH\setminus(\gamma\cup(-\infty,g_{t}(r)])}(g_{t}(y))\leq e^{-s+\log|g_{t}'(y)|}\right]=Q\left(X_{t},\, g_{t}(r),\, g_{t}(y),\, s-\log|g_{t}'(y)|\right).
\end{align*}
For this to be a martingale, the Itô derivative drift term
\begin{align*}
 & \frac{\kappa}{2}\frac{\partial^{2}}{\partial x^{2}}Q+\frac{2}{r-x}\pder rQ+\frac{2}{y-x}\pder yQ+\frac{2}{(y-x)^{2}}\pder sQ
\end{align*}
must vanish. This is a parabolic partial differential equation for
$q$,
\begin{align*}
\left[\pder{\sigma}-\mathcal{G}\right]q(\theta,\sigma)=\; & 0\qquad\text{with generator} & \mathcal{G}=\; & \frac{\kappa}{4}\theta(1+\theta)^{2}\frac{\partial^{2}}{\partial\theta^{2}}+(1+\theta)(1+2\theta)\pder{\theta}.
\end{align*}
The asymptotics of small neighborhood size $\eps=e^{-s}\to0$ correspond
to $s\to+\infty$ and therefore $\sigma\to+\infty$ in the above parabolic
equation. In this limit the solution behaves like $q(\theta,\sigma)\sim e^{\lambda_{0}\sigma}q_{0}(\theta)$,
where $q_{0}$ is the positive eigenvector and $\lambda_{0}$ the
corresponding leading eigenvalue of the generator $\mathcal{G}$.
One finds explicitly 
\begin{align*}
q_{0}(\theta)=\; & (1+\theta)^{1-\frac{8}{\kappa}}, & [\mathcal{G}q_{0}](\theta)= & \left(1-\frac{8}{\kappa}\right)q_{0}(\theta),\qquad\text{i.e., }\lambda_{0}=1-\frac{8}{\kappa}.
\end{align*}
From this asymptotic we find that the probability of reaching a small
conformal distance $e^{-s}=\eps$ at $y$ scales as $e^{\lambda_{0}s}=\eps^{h}$
with the correct scaling exponent $h=-\lambda_{0}=\frac{8-\kappa}{\kappa}$.

\subsection{The second order PDE from stochastic calculus\label{sec: 2nd order PDE}}

Let $\gamma$ be the chordal $\SLEk$ curve in $(\bH;x,\infty)$ parametrized
as in Section \ref{sub: def chordal SLE}. By the domain Markov property,
conditionally on an initial segment $\gamma^{-}=\gamma\big|_{[0,T]}$
of the curve up to a stopping time $T$, the rest of the curve $\gamma^{+}=\gamma\big|_{[T,\infty)}$
is a chordal $\SLEk$ in the domain $\bH\setminus K_{T}$ from the
tip $\gamma(T)$ of the initial segment to $\infty$. Consider stopping
times $T$ smaller than the time at which any boundary visit happens.
Then, conditionally on the initial segment $\gamma^{-}$, the contribution
to the boundary visit amplitude $\Ampl^{(N)}(x;y_{1},\ldots,y_{N})$
is $\Ampl_{(\bH\setminus K_{T};\gamma(T),\infty)}^{(N)}(y_{1},y_{2},\ldots,y_{N})$.
Using the conformal map $g_{T}\colon\bH\setminus K_{T}\rightarrow\bH$
and conformal covariance of $\Ampl_{(\Lambda;a,b)}^{(N)}$, the conditional
contribution equals
\begin{align}
M_{T}=\; & \left(\prod_{j=1}^{N}g_{T}'(y_{j})^{h}\right)\times\Ampl^{(N)}(X_{T};g_{T}(y_{1}),\ldots,g_{T}(y_{N})).\label{eq: limit local martingale}
\end{align}
By construction, then, $(M_{t})_{t\geq0}$ is a local martingale.
We can compute the Itô derivative of $M_{t}$, and require that the
drift term in it vanishes, leading to the second order partial differential
equation

\begin{align*}
\left[\frac{\kappa}{2}\frac{\partial^{2}}{\partial x^{2}}+\sum_{j=1}^{N}\left(\frac{2}{y_{j}-x}\pder{y_{j}}-\frac{2h}{(y_{j}-x)^{2}}\right)\right]\Ampl^{(N)}(x;y_{1},\ldots,y_{N})=\; & 0,
\end{align*}
which is Equation \eqref{eq: second order differential equation}
in the PDE system of Section \ref{sub: differential equations}. The
alternative explanation of this equation by conformal field theory
is given in Appendix \ref{sub: singular vectors}.

\bigskip{}

\section{Conformal field theory considerations\label{sec: conformal field theory}}

\subsection{Boundary visit amplitudes as conformal field theory correlation functions\label{sub: conformal covariance from CFT}}

From conformal field theory point of view, the boundary visit amplitudes
are essentially correlation functions of boundary primary fields of
conformal weights $h$ in a conformal field theory with central charge
$c(\kappa)=\frac{(3\kappa-8)(6-\kappa)}{2\kappa}$, see \cite{BB-zig_zag}.
We remark that the value \eqref{eq: h13} is a conformal weight in
the Kac table, $h=h_{1,3}(\kappa)=\frac{8-\kappa}{\kappa}$. This
suggests the possibility of a degeneracy at grade three, which we
argue to give rise
to the third order PDEs \eqref{eq: third order differential equations}
below in Appendix \ref{sub: singular vectors}.

The covariance rule \eqref{eq: conformal covariance formula} reflects
the conformal transformation properties of primary fields. More precisely,
the boundary zig-zag amplitude should be thought of as a ratio
\begin{align*}
\Ampl^{(N)}(x;y_{1},y_{2},\ldots,y_{N})=\; & \frac{\bra\psi_{1,2}(x)\,\psi_{1,3}(y_{1})\cdots\psi_{1,3}(y_{N})\,\psi_{1,2}(\infty)\ket}{\bra\psi_{1,2}(x)\,\psi_{1,2}(\infty)\ket},
\end{align*}
where:
\begin{itemize}
\item The numerator $\bra\psi_{1,2}(x)\,\psi_{1,3}(y_{1})\cdots\psi_{1,3}(y_{N})\,\psi_{1,2}(\infty)\ket$
is a correlation function of $N$ boundary primary fields $\psi_{1,3}$
of conformal weight $h=h_{1,3}(\kappa)=\frac{8-\kappa}{\kappa}$ located
at $y_{1},y_{2},\ldots,y_{N}$, and two boundary primary fields $\psi_{1,2}$
of conformal weight $\delta=h_{1,2}(\kappa)=\frac{6-\kappa}{2\kappa}$
located at $x$ and $\infty$.
\item The denominator $\bra\psi_{1,2}(x)\,\psi_{1,2}(\infty)\ket$ is the
correlation function of two boundary primary fields $\psi_{1,2}$
located at $x$ and $\infty$. This correlation function is in fact
just a constant (independent of $x$), but the presence of the fields
$\psi_{1,2}(x)$ both in the numerator and denominator is the reason
why the conformal covariance rule \eqref{eq: conformal covariance formula}
does not contain a Jacobian factor $|f'(x)|^{\delta}$.
\end{itemize}

\subsection{Singular vectors and differential equations\label{sub: singular vectors}}

From the point of view of conformal field theory, partial differential
equations such as \eqref{eq: second order differential equation}
and \eqref{eq: third order differential equations} are consequences
of conformal Ward identities if the relevant boundary primary fields
have vanishing descendants.

At the tip of the SLE curve, the boundary changing field is a primary
field $\ketvec{\psi_{1,2}}$ of conformal weight $\delta=h_{1,2}(\kappa)=\frac{6-\kappa}{2\kappa}$,
which has a vanishing descendant $\left(L_{-1}^{2}-\frac{4}{\kappa}L_{-2}\right)\ketvec{\psi_{1,2}}=0$
at level $2$ \cite{BB-SLE_martingales,BB-CFTs_of_SLEs,BB-conformal_transformations}.
The associated conformal Ward identity is the second order PDE \eqref{eq: second order differential equation}.

At the points to be visited by the SLE curve, the boundary fields
are primaries $\ketvec{\psi_{1,3}}$ of conformal weights $h=h_{1,3}(\kappa)=\frac{8-\kappa}{\kappa}$,
and they have vanishing descendants
\begin{align*}
\left(L_{-1}^{3}-\frac{16}{\kappa}L_{-2}L_{-1}+\frac{8(8-\kappa)}{\kappa^{2}}L_{-3}\right)|\psi_{1,3}\rangle=\; & 0
\end{align*}
at level $3$. The associated conformal Ward identities are the third
order PDEs \eqref{eq: third order differential equations}.

\subsection{Asymptotics from operator product expansions\label{sec: CFT asymptotics}}

Conformal field theory allows a finite number of different asymptotics
as the distance of any two arguments of $\Ampl^{(N)}$ or $\Corr^{(N)}$
tends to zero. The reason is that the boundary primary field $\psi_{1,2}(x)$
is degenerate at level two \cite{BB-SLE_martingales,BB-CFTs_of_SLEs,BB-conformal_transformations},
and similarly the boundary primary fields $\psi_{1,3}(y_{j})$ are
degenerate at level three \cite{BB-zig_zag} (this level three degeneracy
is not a priori granted, but it is suggested by known $N=1$ and $N=2$
cases and justified a posteriori by a proof of our formula). The degeneracies
imply selection rules for the fusion of the corresponding fields.
A fusion of primary fields located at $z$ and $w$, with respective
conformal weights $h^{(z)}$ and $h^{(w)}$, to a field of conformal
weight $h^{(\infty)}$ and its descendants, leads to terms of the
form
\begin{align*}
 & (z-w)^{h^{(\infty)}-h^{(z)}-h^{(w)}}\times\reg.
\end{align*}
in the operator product expansion. Here and below, $\reg.$ stands
for functions that are holomorphic and non-vanishing on the ``diagonal''
$z=w$. Taking into account the selection rules, conformal field theory
suggests the following:
\begin{itemize}
\item \emph{Possible asymptotics as two visit points approach each other}:
The fusion of the fields at $y_{j}$ and $y_{k}$ may contain primary
fields of weights $h_{1,1}=0$, $h_{1,3}=\frac{8-\kappa}{\kappa}$,
$h_{1,5}=\frac{2(12-\kappa)}{\kappa}$. Correspondingly the functions
$\Ampl^{(N)}$ and $\Corr^{(N)}$ have the form 
\begin{align}
 & \quad(y_{j}-y_{k})^{2(1-\frac{8}{\kappa})}\times\reg.+(y_{j}-y_{k})^{1-\frac{8}{\kappa}}\times\reg.+(y_{j}-y_{k})^{\frac{8}{\kappa}}\times\reg.\label{eq: possible asymptotics y-y}
\end{align}
as $|y_{k}-y_{j}|\rightarrow0$.
\item \emph{Possible asymptotics as the starting point and a visit point
approach each other}: The fusion of the fields at $x$ and $y_{j}$
may contain primary fields of weights $h_{1,2}=\frac{6-\kappa}{2\kappa}$,
$h_{1,4}=\frac{3(10-\kappa)}{2\kappa}$. Correspondingly the functions
$\Ampl^{(N)}$ and $\Corr^{(N)}$ have the form 
\begin{align}
 & (x-y_{k})^{1-\frac{8}{\kappa}}\times\reg.+(x-y_{k})^{\frac{4}{\kappa}}\times\reg.\label{eq: possible asymptotics x-y}
\end{align}
as $|y_{j}-x|\rightarrow0$.
\end{itemize}

The possible asymptotics above can also be viewed directly as resulting from the
indicial equations for the Frobenius series solutions to the
system of partial differential equations given in Section~\ref{sec: the problem}.
This point of view to fusion is adopted in the article \cite{Dubedat-SLE_and_Virasoro_representations_fusion},
where also the justification of Frobenius
series ansatz and more profound consequences are studied.

\bigskip{}

\section{Some explicit quantum group formulas\label{sec: explicit quantum group formulas}}

\subsection{Explicit normalization conventions for subrepresentations\label{sec: explicit submodules}}

In the spin chain~-~Coulomb gas correspondence, the asymptotics
of the functions may be read off from projections to irreducible subrepresentations
in consecutive tensorands.
We specifically make use of the tensor products
\begin{align*}
\Wd_{3}\tens\Wd_{3}\isom\; & \Wd_{1}\oplus\Wd_{3}\oplus\Wd_{5}
\end{align*}
and
\begin{align*}
\Wd_{2}\tens\Wd_{3}\isom\; & \Wd_{2}\oplus\Wd_{4},\qquad & \Wd_{3}\tens\Wd_{2}\isom\;\Wd_{2}\oplus\Wd_{4} & .
\end{align*}
We will need projections to the irreducible subrepresentations. Note
that if we want to identify the subrepresentations concretely with
the irreducibles described in Section \ref{sec: irreducible representations},
we have to fix normalization factors. This corresponds to a choice
of embedding of the irreducibles to the tensor products as subrepresentations.
Our normalization conventions given below are specializations of \cite[Lemma 2.4]{KP-covariant_boundary_correlations}.

For the former tensor product representation, $\Wd_{3}\tens\Wd_{3}$,
we denote the projections to the three irreducible subrepresentations
by $\pi^{(d)}:\Wd_{3}\tens\Wd_{3}\rightarrow\Wd_{d}\subset\Wd_{3}\tens\Wd_{3}$,
where $d\in\set{1,3,5}$. For the latter two, $\Wd_{2}\tens\Wd_{3}$
and $\Wd_{3}\tens\Wd_{2}$, we denote the projections to the two irreducible
subrepresentations by $\pi^{(d)}:\Wd_{2}\tens\Wd_{3}\rightarrow\Wd_{d}\subset\Wd_{2}\tens\Wd_{3}$
and $\pi^{(d)}:\Wd_{3}\tens\Wd_{2}\rightarrow\Wd_{d}\subset\Wd_{3}\tens\Wd_{2}$,
where $d\in\set{2,4}$. Although the same notation is used for these
latter two different projections, the meaning should always be clear
from the context.

Our embeddings of the irreducibles to the tensor products are the
following. It is enough to specify the image of the highest weight
vector $e_{0}$ in the tensor product, and our normalization choices
are 
\begin{align*}
\Wd_{1}\hookrightarrow\; & \Wd_{3}\tens\Wd_{3}:\qquad & \Wbas_{0}\mapsto\; & \frac{1}{(q^{2}-q^{-2})^{2}}\big(\Wbas_{0}\tens\Wbas_{2}-\Wbas_{1}\tens\Wbas_{1}+q^{-2}\Wbas_{2}\tens\Wbas_{0}\big)\\
\Wd_{3}\hookrightarrow\; & \Wd_{3}\tens\Wd_{3}:\qquad & \Wbas_{0}\mapsto\; & \frac{1}{q^{2}-q^{-2}}\big(-q^{2}\Wbas_{0}\tens\Wbas_{1}+\Wbas_{1}\tens\Wbas_{0}\big)\\
\Wd_{5}\hookrightarrow\; & \Wd_{3}\tens\Wd_{3}:\qquad & \Wbas_{0}\mapsto\; & \Wbas_{0}\tens\Wbas_{0}
\end{align*}
and 
\begin{align*}
\Wd_{2}\hookrightarrow\; & \Wd_{2}\tens\Wd_{3}:\qquad & \Wbas_{0}\mapsto\; & \frac{q^{4}}{1-q^{4}}\Wbas_{0}\tens\Wbas_{1}-\frac{q}{1-q^{2}}\Wbas_{1}\tens\Wbas_{0}\\
\Wd_{4}\hookrightarrow\; & \Wd_{2}\tens\Wd_{3}:\qquad & \Wbas_{0}\mapsto\; & \Wbas_{0}\tens\Wbas_{0}
\end{align*}
and 
\begin{align*}
\Wd_{2}\hookrightarrow\; & \Wd_{3}\tens\Wd_{2}:\qquad & \Wbas_{0}\mapsto\; & \frac{q^{2}}{1-q^{2}}\Wbas_{0}\tens\Wbas_{1}-\frac{q^{2}}{1-q^{4}}\Wbas_{1}\tens\Wbas_{0}\\
\Wd_{4}\hookrightarrow\; & \Wd_{3}\tens\Wd_{2}:\qquad & \Wbas_{0}\mapsto\; & \Wbas_{0}\tens\Wbas_{0}.
\end{align*}
These choices of normalizing constants strike a compromise between
simplicity of formulas for the quantum group representations and for
the asymptotics of the corresponding functions treated in Section
\ref{sec: asymptotics via correspondence}.

When an identification with a smaller tensor product is implied in
a projection to subrepresentation, we indicate this with a hat: we
thus define $\hat{\pi}^{(1)}\colon\Wd_{3}\tens\Wd_{3}\rightarrow\bC$,
$\hat{\pi}^{(3)}\colon\Wd_{3}\tens\Wd_{3}\rightarrow\Wd_{3}$, $\hat{\pi}^{(2)}\colon\Wd_{2}\tens\Wd_{3}\rightarrow\Wd_{2}$,
and $\hat{\pi}^{(2)}\colon\;\Wd_{3}\tens\Wd_{2}\rightarrow\Wd_{2}$
with the identifications of the subrepresentations given above. We
finally need to act on two consecutive components of the following
big tensor product
\begin{align*}
 & \Wd_{3}^{\tens R}\tens\Wd_{2}\tens\Wd_{3}^{\tens L}.
\end{align*}
We define the following projections to a doublet subrepresentation
in the tensor product of the doublet tensorand in the middle and a
triplet on either side of it, according to the ``$\pm$''-symbol
\begin{align*}
\hat{\pi}_{\rgt}^{(2)}:\; & \Wd_{3}^{\tens R}\tens\Wd_{2}\tens\Wd_{3}^{\tens L}\rightarrow\Wd_{3}^{\tens(R-1)}\tens\Wd_{2}\tens\Wd_{3}^{\tens L}\\
\hat{\pi}_{\rgt}^{(2)}=\; & (\id_{\Wd_{3}})^{\tens(R-1)}\tens\hat{\pi}^{(d)}\tens(\id_{\Wd_{3}})^{\tens L}\\
\hat{\pi}_{\lft}^{(2)}:\; & \Wd_{3}^{\tens R}\tens\Wd_{2}\tens\Wd_{3}^{\tens L}\rightarrow\Wd_{3}^{\tens R}\tens\Wd_{2}\tens\Wd_{3}^{\tens(L-1)}\\
\hat{\pi}_{\lft}^{(2)}=\; & (\id_{\Wd_{3}})^{\tens R}\tens\hat{\pi}^{(2)}\tens(\id_{\Wd_{3}})^{\tens(L-1)}.
\end{align*}
Likewise, we define the following projections in two consecutive triplet
factors (in the $m$th and $(m+1)$st factors on the left or on the
right) 
\begin{align*}
\hat{\pi}_{\rgt;m}^{(3)}:\; & \Wd_{3}^{\tens R}\tens\Wd_{2}\tens\Wd_{3}^{\tens L}\rightarrow\Wd_{3}^{\tens(R-1)}\tens\Wd_{2}\tens\Wd_{3}^{\tens L}\\
\hat{\pi}_{\rgt;m}^{(3)}=\; & (\id_{\Wd_{3}})^{\tens(R-m-1)}\tens\hat{\pi}^{(3)}\tens(\id_{\Wd_{3}})^{\tens(m-1)}\tens\id_{\Wd_{2}}\tens(\id_{\Wd_{3}})^{\tens L}\\
\hat{\pi}_{\lft;m}^{(3)}:\; & \Wd_{3}^{\tens R}\tens\Wd_{2}\tens\Wd_{3}^{\tens L}\rightarrow\Wd_{3}^{\tens R}\tens\Wd_{2}\tens\Wd_{3}^{\tens(L-1)}\\
\hat{\pi}_{\lft;m}^{(3)}=\; & (\id_{\Wd_{3}})^{\tens R}\tens\id_{\Wd_{2}}\tens(\id_{\Wd_{3}})^{\tens(m-1)}\tens\hat{\pi}^{(3)}\tens(\id_{\Wd_{3}})^{\tens(L-m-1)}.
\end{align*}
Finally, we also define the following projections in two consecutive
triplet factors (in the $m$th and $(m+1)$st factors on the left
or on the right)
\begin{align*}
\hat{\pi}_{\rgt;m}^{(1)}:\; & \Wd_{3}^{\tens R}\tens\Wd_{2}\tens\Wd_{3}^{\tens L}\rightarrow\Wd_{3}^{\tens(R-2)}\tens\Wd_{2}\tens\Wd_{3}^{\tens L}\\
\hat{\pi}_{\rgt;m}^{(1)}=\; & (\id_{\Wd_{3}})^{\tens(R-m-1)}\tens\hat{\pi}^{(1)}\tens(\id_{\Wd_{3}})^{\tens(m-1)}\tens\id_{\Wd_{2}}\tens(\id_{\Wd_{3}})^{\tens L}\\
\hat{\pi}_{\lft;m}^{(1)}:\; & \Wd_{3}^{\tens R}\tens\Wd_{2}\tens\Wd_{3}^{\tens L}\rightarrow\Wd_{3}^{\tens R}\tens\Wd_{2}\tens\Wd_{3}^{\tens(L-2)}\\
\hat{\pi}_{\lft;m}^{(1)}=\; & (\id_{\Wd_{3}})^{\tens R}\tens\id_{\Wd_{2}}\tens(\id_{\Wd_{3}})^{\tens(m-1)}\tens\hat{\pi}^{(1)}\tens(\id_{\Wd_{3}})^{\tens(L-m-1)}.
\end{align*}

Additionally, we denote by $\pi_{\pm}^{(2)},\pi_{\pm}^{(4)},\pi_{\pm;m}^{(1)},\pi_{\pm;m}^{(3)},\pi_{\pm;m}^{(5)}$ the projections
\[ \Wd_{3}^{\tens R}\tens\Wd_{2}\tens\Wd_{3}^{\tens L} \to \Wd_{3}^{\tens R}\tens\Wd_{2}\tens\Wd_{3}^{\tens L} \]
analogous to the hatted counterparts $\hat{\pi}_{\pm}^{(2)},\hat{\pi}_{\pm}^{(4)},\hat{\pi}_{\pm;m}^{(1)},\hat{\pi}_{\pm;m}^{(3)},\hat{\pi}_{\pm;m}^{(5)}$,
respectively, but without the identification of the submodule with a shorter tensor product.

\subsection{The quantum group solutions for some 4-point visits\label{sub: 4-point QG solutions}}

For brevity, we factor out the constant
\begin{align*}
C_{4}=\; & \frac{q^{7}\left(q^{4}+q^{2}+1\right)^{3}}{\left(q^{2}-1\right)^{4}\left(q^{2}+1\right)^{5}\left(\left(q^{12}+2q^{8}+q^{6}+2q^{4}+q^{2}+2\right)q^{4}+1\right)}.
\end{align*}
Then, with a shorthand notation similar to that in Sections \ref{sec: 2pt quantum group solution}
and \ref{sec: 3pt quantum group solution}, the normalized solutions
for the cases needed for Figure~\ref{fig: perco 4pt data} are
\begin{align*}
v_{+-++}^{(4)}=\; & C_{4}\times\Bigg(\left(q^{2}+1\right)\left(q^{4}+1\right)q^{6}\Wbas_{00112}-q^{4}\Wbas_{00202}-\left(q^{2}+1\right)q^{5}\Wbas_{00211}+q^{4}\Wbas_{22000}\\
 & \qquad+\left(q^{2}+1\right)\left(q^{4}+1\right)q^{8}\Wbas_{01012}-\left(q^{2}+1\right)q^{6}\Wbas_{01102}-\left(q^{2}+1\right)^{2}q^{7}\Wbas_{01111}\\
 & \qquad+\left(q^{5}+q^{3}\right)\Wbas_{01201}+\left(q^{6}+q^{4}\right)\Wbas_{01210}-q^{8}\Wbas_{02002}-\left(q^{2}+1\right)q^{9}\Wbas_{02011}+\Wbas_{20200}\\
 & \qquad+\left(q^{7}+q^{5}\right)\Wbas_{02101}+\left(q^{8}+q^{6}\right)\Wbas_{02110}+\left(-q^{4}-1\right)\Wbas_{02200}+\left(q^{4}-1\right)q^{5}\Wbas_{12001}\\
 & \qquad-\left(q^{2}+1\right)\left(q^{5}+q\right)^{2}\Wbas_{10012}+\left(-q^{8}+q^{6}+q^{2}\right)\Wbas_{10102}+\left(q^{4}-1\right)q^{6}\Wbas_{12010}\\
 & \qquad+\left(-q^{11}+q^{7}+q^{5}+q^{3}\right)\Wbas_{10111}+\left(q^{4}-1\right)q\Wbas_{10201}+\left(q^{4}-1\right)q^{2}\Wbas_{10210}\\
 & \qquad+\left(-q^{10}+q^{8}+q^{4}\right)\Wbas_{11002}+\left(-q^{13}+q^{9}+q^{7}+q^{5}\right)\Wbas_{11011}+\left(q^{8}+q^{4}\right)\Wbas_{20002}\\
 & \qquad+\left(q^{2}-1\right)\left(q^{2}+1\right)^{2}q^{3}\Wbas_{11101}+\left(q^{2}-1\right)\left(q^{2}+1\right)^{2}q^{4}\Wbas_{11110}+\left(-q^{6}-q^{2}+1\right)\Wbas_{11200}\\
 & \qquad-\left(q^{6}+q^{2}-1\right)q^{2}\Wbas_{12100}+\left(q^{2}+1\right)\left(q^{4}+1\right)q^{5}\Wbas_{20011}-\left(q^{2}+1\right)q^{3}\Wbas_{20101}\\
 & \qquad-\left(q^{2}+1\right)q^{4}\Wbas_{20110}-\left(q^{2}+1\right)q^{5}\Wbas_{21001}-\left(q^{2}+1\right)q^{6}\Wbas_{21010}+\left(q^{4}+q^{2}\right)\Wbas_{21100}\Bigg)
\end{align*}
\begin{align*}
v_{+--+}^{(4)}=\; & C_{4}\times\Bigg(\frac{\left(q^{4}+1\right)\left(q^{4}+q^{2}+1\right)q^{5}\Wbas_{00022}}{q^{2}+1}+\left(q^{4}+1\right)\left(q^{4}+q^{2}+1\right)q^{6}\Wbas_{00112}+\left(q^{3}-q^{9}\right)\Wbas_{21100}\\
 & \qquad-\left(q^{4}+1\right)\left(q^{4}+q^{2}+1\right)q^{2}\Wbas_{00121}-\left(q^{4}+q^{2}+1\right)q^{4}\Wbas_{01012}+\left(q^{2}-q^{8}\right)\Wbas_{01021}\\
 & \qquad-\left(q^{4}+q^{2}+1\right)q^{5}\Wbas_{01102}+\left(-q^{11}-q^{9}+q^{5}+q^{3}\right)\Wbas_{01111}+\left(q^{7}+q^{5}+q^{3}\right)\Wbas_{01120}\\
 & \qquad+\left(q^{3}+\frac{q}{q^{2}+1}\right)\Wbas_{02002}+\left(q^{4}+q^{2}-1\right)q\Wbas_{02011}+\left(q^{2}+\frac{1}{q^{2}+1}-2\right)q\Wbas_{02020}\\
 & \qquad+\left(q^{4}+q^{2}-1\right)q^{2}\Wbas_{02101}+\left(q^{4}-q^{2}-1\right)q^{2}\Wbas_{02110}-\left(q^{4}+q^{2}+1\right)q^{6}\Wbas_{10012}\\
 & \qquad+\left(q^{4}-q^{10}\right)\Wbas_{10021}-\left(q^{4}+q^{2}+1\right)q^{7}\Wbas_{10102}+\left(q^{2}-q^{8}\right)\Wbas_{21010}+\left(q^{6}+2q^{4}-1\right)q^{3}\Wbas_{11011}\\
 & \qquad+\left(q^{9}+q^{7}+q^{5}\right)\Wbas_{10120}+\left(q^{7}+q^{5}+q^{3}\right)\Wbas_{11002}-\left(q^{8}+q^{6}-q^{2}-1\right)q^{5}\Wbas_{10111}\\
 & \qquad+\left(q^{4}-q^{2}-1\right)q^{3}\Wbas_{11020}+\left(q^{6}+2q^{4}-1\right)q^{4}\Wbas_{11101}+\left(q^{6}-2q^{2}-1\right)q^{4}\Wbas_{11110}\\
 & \qquad-\left(q^{4}+q^{2}+1\right)q^{2}\Wbas_{12001}+\left(1-q^{6}\right)\Wbas_{12010}+\left(q-q^{7}\right)\Wbas_{12100}+\frac{\left(q^{9}+q^{7}+q^{5}\right)\Wbas_{20002}}{q^{2}+1}\\
 & \qquad-\frac{\left(-q^{9}+q^{7}+q^{5}\right)\Wbas_{20020}}{q^{2}+1}-\left(q^{4}+q^{2}+1\right)q^{4}\Wbas_{21001}+\frac{\left(q^{4}+1\right)\left(q^{4}+q^{2}+1\right)q\Wbas_{22000}}{q^{2}+1}\\
 & \qquad+\left(q^{4}+q^{2}-1\right)q^{5}\Wbas_{20011}+\left(q^{4}-q^{2}-1\right)q^{6}\Wbas_{20110}+\left(q^{4}+q^{2}-1\right)q^{6}\Wbas_{20101}\Bigg)
\end{align*}
\begin{align*}
v_{-+++}^{(4)}=\; & C_{4}\times\Bigg(\left(q^{2}+1\right)\left(q^{4}+q^{2}+1\right)q^{6}\Wbas_{00112}+\left(q^{4}+1\right)\left(q^{4}+q^{2}+1\right)q^{2}\Wbas_{20002}\\
 & \qquad+\left(q^{8}+q^{6}-q^{2}-1\right)q^{4}\Wbas_{01012}+\left(q^{8}+q^{6}-q^{2}-1\right)q^{6}\Wbas_{01102}+\left(q^{6}+q^{4}+q^{2}\right)\Wbas_{22000}\\
 & \qquad-\left(q^{6}+2q^{4}-1\right)q^{5}\Wbas_{01201}+\left(q^{2}+1\right)\left(q^{4}+q^{2}+1\right)q^{2}\Wbas_{01210}-\left(q^{4}+q^{2}+1\right)q^{6}\Wbas_{02002}\\
 & \qquad+\left(-q^{9}+2q^{5}+q^{3}\right)\Wbas_{02011}+\left(-q^{11}+2q^{7}+q^{5}\right)\Wbas_{02101}+\left(q^{8}+q^{6}-q^{2}-1\right)\Wbas_{02110}\\
 & \qquad+\left(q^{6}-1\right)q^{2}\Wbas_{02200}-\left(q^{2}+1\right)\left(q^{4}+q^{2}+1\right)q^{4}\Wbas_{10012}-\left(q^{2}+1\right)\left(q^{4}+q^{2}+1\right)q^{6}\Wbas_{10102}\\
 & \qquad+\left(-q^{12}+2q^{6}+q^{4}+q^{2}\right)\Wbas_{11002}+\left(-q^{11}+2q^{7}+q^{5}\right)\Wbas_{10201}+\left(q^{8}+q^{6}-q^{2}-1\right)\Wbas_{10210}\\
 & \qquad-\left(q^{2}+1\right)^{2}\left(q^{4}-q^{2}-1\right)q^{3}\Wbas_{10111}-\left(q^{2}+1\right)^{2}\left(q^{4}+q^{2}-1\right)q^{3}\Wbas_{01111}\\
 & \qquad-\left(q^{2}+1\right)\left(q^{8}-2q^{6}-q^{4}+1\right)q^{3}\Wbas_{11101}+\left(q^{2}+1\right)\left(q^{8}-q^{4}-2q^{2}+1\right)\Wbas_{11110}\\
 & \qquad+\left(q^{10}-q^{6}-2q^{4}+q^{2}\right)\Wbas_{11200}+\left(q^{8}+q^{6}-q^{2}-1\right)q^{3}\Wbas_{12001}+\left(1-q^{4}\left(q^{2}+2\right)\right)\Wbas_{12010}\\
 & \qquad+\left(-q^{8}-2q^{6}+q^{2}\right)\Wbas_{12100}+\left(q^{8}+q^{6}-q^{2}-1\right)q\Wbas_{20011}+\left(q^{4}+q^{2}+1\right)q^{8}\Wbas_{00202}\\
 & \qquad+\left(q^{8}+q^{6}-q^{2}-1\right)q^{3}\Wbas_{20101}+\left(1-q^{4}\left(q^{2}+2\right)\right)\Wbas_{20110}-\left(q^{4}+q^{2}-1\right)q^{2}\Wbas_{20200}\\
 & \qquad-\left(q^{2}+1\right)\left(q^{4}+q^{2}+1\right)q^{3}\Wbas_{21001}+\left(-q^{6}+2q^{2}+1\right)\Wbas_{21010}+\left(-q^{8}+2q^{4}+q^{2}\right)\Wbas_{21100}\\
 & \qquad-\left(q^{2}+1\right)\left(q^{8}-2q^{6}-q^{4}+1\right)q\Wbas_{11011}-\left(q^{2}+1\right)\left(q^{4}+q^{2}+1\right)q^{3}\Wbas_{00211}\Bigg)
\end{align*}

\bigskip{}

\section{Numerical evaluation of the integrals\label{sub: numerical evaluation}}

Let us then describe how the integral expressions can be evaluated
numerically in practice. We have implemented two methods with symbolic
computation software: 
\begin{enumerate}
\item Direct evaluation of the complex loop integrals $\varphi_{t_{L}^{\lft},\ldots,t_{2}^{\lft},t_{1}^{\lft};d;t_{1}^{\rgt},t_{2}^{\rgt},\ldots,t_{R}^{\rgt}}$. 
\item Evaluation of the (real) integrals $\rho_{t_{L}^{\lft},\ldots,t_{2}^{\lft},t_{1}^{\lft};d;t_{1}^{\rgt},t_{2}^{\rgt},\ldots,t_{R}^{\rgt}}$
by using the $\eps$-regularization scheme described above. 
\end{enumerate}
Both of these approaches have advantages and disadvantages. The loop
integrals are well defined as such for all values of $\kappa$, but
involve complex integrands and complicated numerical contours which
slow down the integration. Real $\eps$-regularized integrals are
faster to evaluate, but one needs to add counterterms which also involve
integrals, thus increasing the total number of integrations. In addition,
the remaining $\eps$-dependence of the result needs to be controlled.

In both methods, low values of $\kappa$ are the most challenging.
In the loop integrals, the variations in the absolute value of the
integrand increase with decreasing $\kappa$, leading to more and
more precise cancellations between contributions from different sections
of the integrations contours. In order to make the $\eps$-regularization
work, a larger number of counterterms is necessary at small $\kappa$
than at values of $\kappa$ close to $\kappa=8$, which practically
limits this method to $\kappa\gtrsim4$. As it turns out, probability
amplitudes with $N=3$ boundary visits are still relatively fast to
evaluate, in particular when $\kappa$ is close to eight, whereas
it is already computationally demanding to evaluate the $N=4$ amplitudes.
For $N=3$ the calculation of the loop integrals is the faster method.
We have controlled the numerical errors by comparing the results obtained
by the two methods for the final result of the probability amplitude.

\subsection{Evaluation of the loop integrals}

\begin{figure}
\begin{centering}
\includegraphics[width=0.7\textwidth]{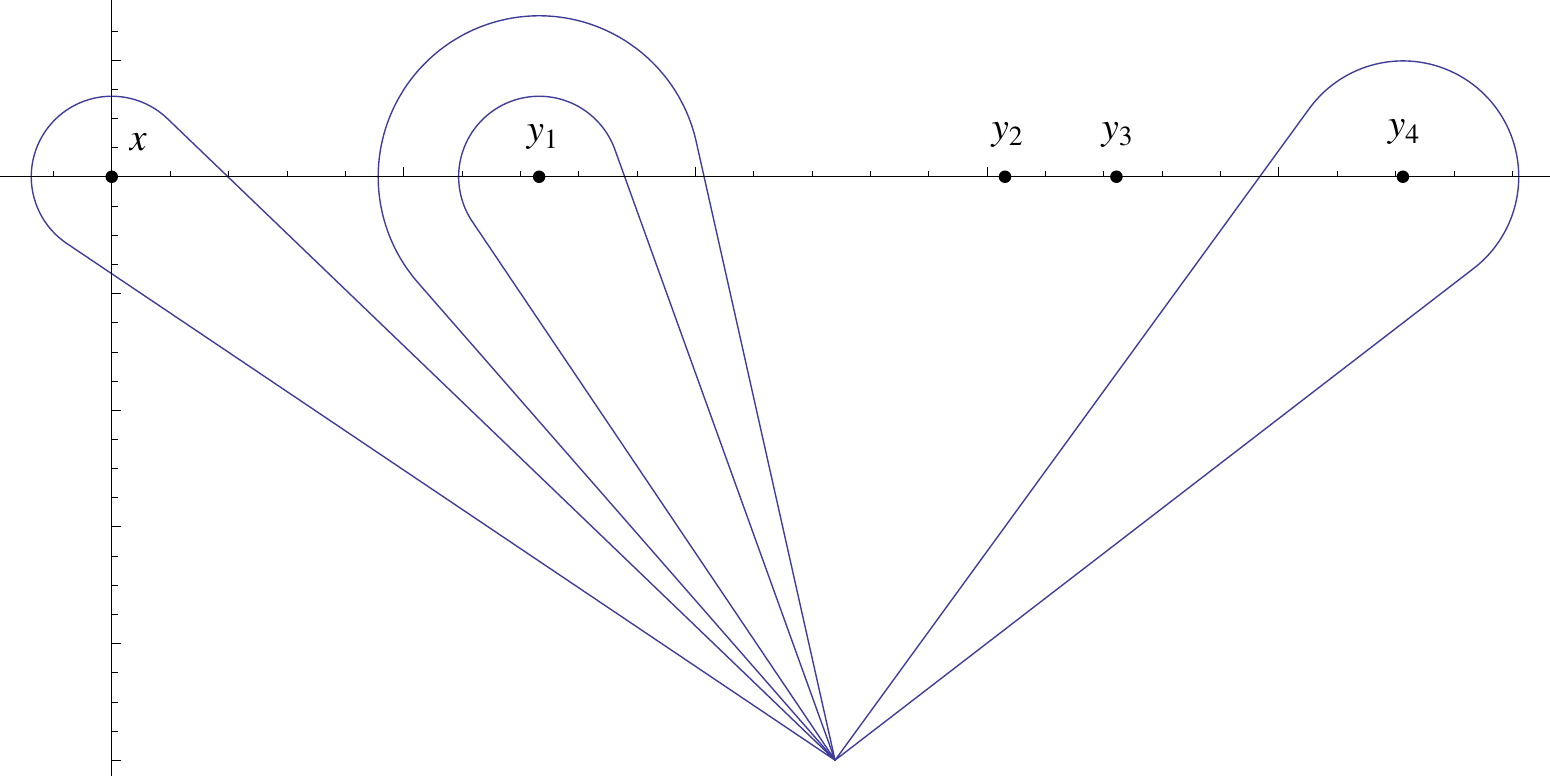}\\

\par\end{centering}

\caption{\emph{An example of the integration contours used for numerical evaluation
of our results.\label{fig: integration contour for numerics}}}
\end{figure}

In order to evaluate the loop integrals, we first need to specify
the integration contours. We choose the anchor point in the lower
half plane. Each contour is chosen to be a combination of two straight
lines and an arc of a circle, with the center of the circle located
at the encircled charge, and the lines being tangential to the circle
(see Fig.~\ref{fig: integration contour for numerics}). The radii
of the circles are chosen such that the minimum distance between any
pair of charges is (approximately) maximized. The contours $w_{k}=c_{k}(s_{k})$
are parametrized in terms of the real variables $s_{k}\in[0,1]$,
such that $w_{k}$ moves around the charge in the counterclockwise
direction with increasing $s_{k}$. The parametrization can be chosen
such that $c_{k}'(s_{k})$ is continuous at the points where the arc
joins with the lines.

\noindent 

The most tricky step is to write the multi-branched integrand in terms
of the principal branches of the power functions such that it is an
analytic function on the integration contours, and the phase convention
of Fig.~\ref{fig: FW integration contour} is realized. By the principal
branch we mean that 
\[
x^{y}=\exp\left(y\,\log(x)\right),
\]
where the principal branch of the logarithm satisfies $-\pi<\im\left(\log(x)\right)\le\pi$
for all complex $x\ne0$. Let us denote by $\hat{s}_{k}$ the value
of $s_{k}$ where $\im\, w_{k}$ takes its largest value. It is then
easy to check that the various terms of the integrand can be defined
as follows. 
\begin{itemize}
\item If the contours with indices $k_{1}$ and $k_{2}$ encircle two different
charges $y_{j_{1}}$ and $y_{j_{2}}$, with $y_{j_{1}}<y_{j_{2}}$,
we take 
\[
(w_{k_{2}}-w_{k_{1}})^{8/\kappa}=\exp\left(\frac{8}{\kappa}\,\log(w_{k_{2}}-w_{k_{1}})\right).
\]
Similar definition holds when either of the contours is around $x$. 
\item If the contours with indices $k_{1}$ and $k_{2}$ encircle the same
charge, with $c_{k_{1}}$ being the innermost contour, we take 
\begin{align*}
(w_{k_{2}}-w_{k_{1}})^{8/\kappa}=\; & \exp\left(\frac{8}{\kappa}\,\log(w_{k_{2}}-w_{k_{1}})\right) & \textrm{if}\quad0\le s_{k_{2}}\le\hat{s}_{k_{2}}\\
(w_{k_{2}}-w_{k_{1}})^{8/\kappa}=\; & \exp\left(\frac{8\pi}{\kappa}\ii+\frac{8}{\kappa}\,\log(w_{k_{1}}-w_{k_{2}})\right) & \textrm{if}\quad\hat{s}_{k_{2}}<s_{k_{2}}\le1.
\end{align*}

\item If the contour $c_{k}$ encircles $y_{j_{1}}$ we take for each $y_{j_{2}}\ne y_{j_{1}}$
\begin{align*}
(w_{k}-y_{j_{2}})^{-8/\kappa}=\; & \exp\left(-\frac{8}{\kappa}\,\log(w_{k}-y_{j_{2}})\right)\quad\textrm{if}\quad y_{j_{1}}>y_{j_{2}}\\
(y_{j_{2}}-w_{k})^{-8/\kappa}=\; & \exp\left(-\frac{8}{\kappa}\,\log(y_{j_{2}}-w_{k})\right)\quad\textrm{if}\quad y_{j_{1}}<y_{j_{2}},
\end{align*}
and for the contribution from the charge $y_{j_{1}}$ we use 
\begin{align*}
(w_{k}-y_{j_{1}})^{-8/\kappa}=\; & \exp\left(-\frac{8}{\kappa}\,\log(w_{k}-y_{j_{1}})\right) & \textrm{if}\quad0\le s_{k}\le\hat{s}_{k}\\
(w_{k}-y_{j_{1}})^{-8/\kappa}=\; & \exp\left(-\frac{8\pi}{\kappa}\ii-\frac{8}{\kappa}\,\log(y_{j_{1}}-w_{k})\right) & \textrm{if}\quad\hat{s}_{k}<s_{k}\le1.
\end{align*}
The terms involving $w_{k}$ and $x$ are treated analogously. 
\end{itemize}
The numerical integration can then be done after changing the integration
variables to $s_{k}$. It turns out that the integration on our symbolic
computation software is often faster, if each of the contours is explicitly
divided into the three pieces containing the two lines and the arc,
and the contributions are integrated separately.

The probability amplitudes $\Ampl^{(N)}$ often
have zeroes of poles at the rational values of $\kappa$ of interest
to us, but then one may just straightforwardly modify the normalizing
constants. For example at $\kappa=6$, $N=3$ we can add a normalization
factor $\propto1/(\kappa-6)$ and study $\Ampl^{(3)}/(\kappa-6)$
in the limit $\kappa\to6$. The numerical integration cannot be done,
however, arbitrary close to $\kappa=6$, because the integrals contributing
to $\Ampl^{(3)}$ do not vanish term by term, and noise due to the
limited numerical precision of such integrals will grow as $1/|\kappa-6|$
as $\kappa\to6$. We evaluated the amplitude for values of $\kappa$
near the critical one, say, at $\kappa=6.05$ and $\kappa=5.95$,
and estimated the amplitude at $\kappa=6$ as the average of the results.
More elaborate fitting, as a function of $\kappa$, can also be done.

\subsection{Evaluation of the $\eps$-regularized integrals}

The most involved step in the evaluation of the $\eps$-regularized
integrals is the identification of the counterterms. In Section~\ref{sec: divergences of integrals}
we already discussed how this can be done, and considered
explicitly a simple example. Computation of the terms at higher $N$
and to higher order in $\eps$ is in principle straightforward, but
the complexity of the expressions grows relatively fast. We have written
a code on symbolic computation software which automatically finds
the counterterms for a given integral. All leading order terms in
the expansion of the divergent terms at $\eps=0$ {[}i.e., the $k=0$
terms $\mathcal{O}\left(\eps^{-n(8/\kappa-1)}\right)$ in \eqref{eq: generic divergent power behavior},
with $n=1,2,\ldots,N${]} and at least the leading divergence from
the next-to-leading order term of the series {[}i.e., the terms $\mathcal{O}\left(\eps^{-(N-1)(8/\kappa-1)}\eps\right)${]}
are generated. Including these terms, the method converges for $N=2$
integrals when $\kappa>4$, and for $N=3$ integrals when $\kappa>16/3$.
In practice the limits can be somewhat higher due to limited numerical
precision.

After the counterterms have been identified, it is straightforward
to evaluate the sum of the regularized integral and all counterterms
for any fixed value of the cutoff $\eps$. Notice also that since
we are not able to subtract counterterms to all orders, some dependence
on $\eps$ remains, and we need to extrapolate the result down to
$\eps=0$. It is useful to calculate the amplitude at various values
of $\eps$, and fit the remaining $\eps$-dependence by using the
highest order term which was not subtracted. Moreover, a similar interpolation
as a function of $\kappa$, as was described above for the loop integrals,
is usually also required.

\end{document}